# Optical Photon Emission in Extended Airshowers

## - Hybrid computing in the context of CORSIKA 8 -

technische universität
dortmund

**Dominik Baack**
2023

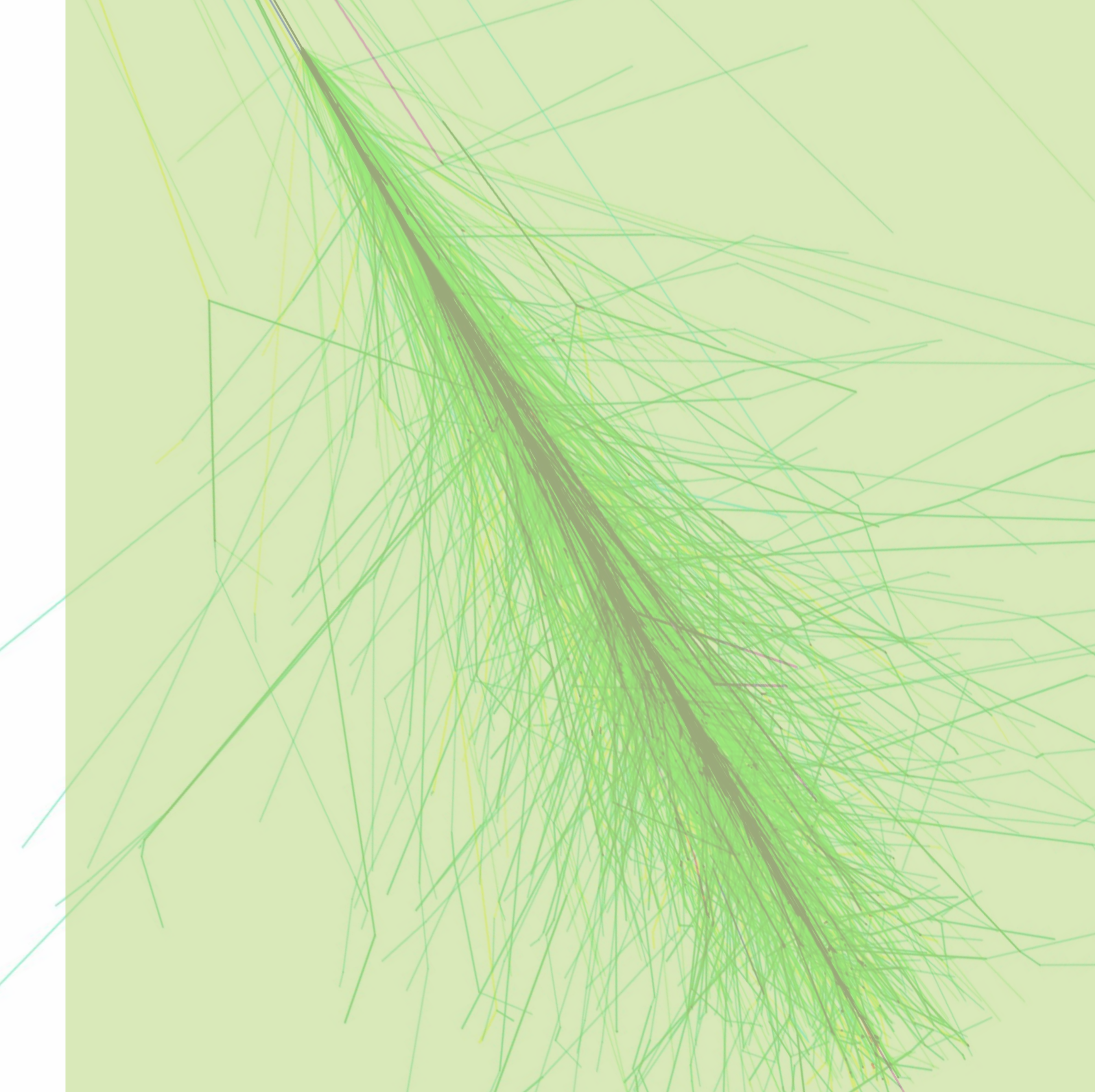

Dissertation
zur Erlangung des akademischen Grades
eines Doktors der Naturwissenschaften
(Dr. rer. nat.)

Fakultät Physik, Technische Universität Dortmund

Supervised by Prof. Dr. Dr. Wolfgang Rhode
&
Prof. Dr. Kevin Kröninger

# Abstract

## 1 Abstract

With the motivation to improve experimental gains and precision, established astroparticle experiments are currently undergoing massive upgrades. In addition, several new experiments are being built or planned. With the resulting gain in observational quality, the amount and accuracy of simulated data required for the analysis is also rising. In order to meet the increasing requirements and complexity due to the experiments' growth and to provide a unified software ecosystem, it was decided to re-develop the de facto standard extensive air shower simulation CORSIKA completely in C++ based on the original Fortran code. Since one of the largest runtime consumers is the propagation of millions of optical Cherenkov and fluorescence photons, and many experiments are starting to use them for measurements, it was decided to develop hardware-accelerated code to speed up the simulation. Specific methods have been developed to propagate photons on deep learning acceleration hardware similar to classical GPUs to take additional advantage of the current and future growth of the deep learning sector. In particular, Nvidia accelerators were tested.

## 2 Kurzfassung

Um den experimentellen Ertrag und die Präzision zu steigern, werden die etablierten Astroteilchenexperimente derzeit massiv aufgerüstet. Darüber hinaus sind mehrere neue Experimente im Bau oder in Planung. Mit der daraus resultierenden Verbesserung der Beobachtungsqualität steigt auch die Menge und Genauigkeit der benötigten simulierten Daten für die Analyse. Um den steigenden Anforderungen und der zunehmenden Komplexität durch das Wachstum der Experimente gerecht zu werden und ein einheitliches Software-Ökosystem zur Verfügung zu stellen, wurde beschlossen, die De-facto-Standardsimulation für Luftschauer CORSIKA auf Basis des ursprünglichen Fortran-Codes vollständig in C++ neu zu entwickeln. Da die Ausbreitung von Millionen optischer Cherenkov- und Fluoreszenzphotonen einer der größten Laufzeitverbraucher ist und viele Experimente beginnen, diese für Messungen zu verwenden, wurde beschlossen, einen hardwarebeschleunigten Code zu entwickeln, um die Simulation zu beschleunigen. Um das gegenwärtige und zukünftige Wachstum in der Verbreitung von Deep Learning zu nutzen, wurden spezielle Methoden entwickelt, um Photonen auf Deep Learning Beschleunigungshardware zu propagieren. Konkret wurden Beschleuniger des Herstellers Nvidia getestet.

# Contents

# 1 Introduction

With the increasing quality, size, and complexity of modern astroparticle experiments, such as CTA (Cherenkov Telescope Array - Large scale Cherenkov telescope array [7]), it is possible to observe cosmic rays in even greater detail than is possible with today's experiments. It is necessary to increase the amount and quality of the simulated data equally to enable the precise analysis of the observation data. These advances require new methods to improve physical correctness, which typically results in the slowdown of the simulation. In addition, a massive investment in computing time is needed to achieve the required magnitude of simulated data. With even more extensive experiments such as IceCube-Gen2 (IceCube Generation 2 - Upgraded in Ice Neutrino Detector [4]) or SKA (Square Kilometer Array - Large scale radio array that can observe cosmic air showers as side-project [25]) on the horizon, the computing power required to generate these simulations will become the limiting factor for comprehensive analysis. Even today, the difficulty of generating rare physical events with sufficient statistics in simulations is a major impediment to the study of rare physical events.

The only solution to achieve the required simulation statistics, apart from tailoring the simulation, which may introduce biases or hard-to-detect errors, is to speed up the simulation overall. Alternatively, investing the saved time in higher-quality simulation methods can improve the overall accuracy. The majority of the techniques outlined in this thesis have undergone testing in the latest version of CORSIKA 7, the widely utilised Monte Carlo simulation for atmospheric particle showers. The most effective methods were incorporated into the wholly redesigned CORSIKA 8. This new version of CORSIKA is written in modern C++ and follows a modular design approach, allowing new methods that the old and complex Fortran (Version 1977) code cannot support. The main optimisations explained in Sections 2 and 3 can be divided into two approaches. The first approach involves reducing the required computations through machine-learned cuts, while the second approach is to parallelise the workload generated during the simulation more efficiently.

This thesis focuses specifically on optical light transmission in the context of astroparticle physics. However, the results obtained are not exclusive to this field and are transferable to different applications, e.g. radio transmission.

The following paragraphs will provide an overview of the structure of this thesis:

**Chapter 1 - Introduction** The current section gives an outline and describes the structure of this thesis.

**Chapter 2 - Particle Physics** This section provides an outline of the physics of cosmic rays, including atmospheric cascades. A basic description of the relevant physical effects, such as Cherenkov and fluorescence light generation, is also given. Methods for calculating physically relevant properties, such as the refractive index, required in later methods extend this section.

**Chapter 3 - Simulation Fundamentals** Here an introduction of Monte Carlo Methods in general as well as the derivation of utilised methods and algorithm specifically developed for the simulation of optical light are explained.

**Chapter 4 - Methods and Implementation** This chapter focuses on the implementation and acceleration of the fundamental methods described in the previous chapter. Including comparing different approaches and evaluating their performance regarding runtime and quality of the results.

**Chapter 5 - GPU Computing** After the algorithmic optimisations in place, the focus shifts to the parallelisation of the simulation, where the evaluation of different parallelisation approaches and the implementation of the most promising one are in the foreground.

**Chapter 6 - Results and Comparison** The finished methods and algorithms are tested against theoretical models and the CORSIKA 7 implementation which acts as a reference. The results are compared predominantly in regard to physical correctness and runtime.

## 1.1 Disclaimer

Some parts of this thesis in the form of pictures or text segments were previously shown at conferences or published as ICRC proceedings ([17], [63], [8]). All numerical experiments and simulations were done on the LAMARR-Institute cluster in Dortmund, utilising a DGX A100 GPU.

AI-based spelling and grammar correction methods were used in the writing of this thesis. Specifically, Grammarly [1] and DeepL [2] writer.

## 1.2 Acknowledgements

First of all, I would like to thank Prof. Dr. Dr. Rhode for making it possible to work on this thesis topic and the various other projects I have been involved in over the past years.

Part of this work has been funded by the Deutsche Forschungsgemeinschaft (DFG) within the Collaborative Research Centre SFB 876 ”Providing Information by Resource-Constrained Analysis”, Project C3.

This research is partly funded by the German Federal Ministry of Education and Research and the State of North Rhine-Westphalia as part of the Lamarr Institute for Machine Learning and Artificial Intelligence.

# 2 Foundation

Experimental particle physics can be divided into two related disciplines: classic particle physics and astroparticle physics, indistinguishable in the early stages, later separating with different scientific goals. Both disciplines have their origins in the discovery of the first elementary particles, the electron in the late 19th century and the proton in the early 20th century, which are the building blocks of both fields. In the following years, knowledge was gained due to various experimental methods, which later led to the big artificial particle accelerators and detector systems used today. Over time, the numerous derived technologies for measurements on all scales establish the fundamental backbone of both branches. The scientific focus of the main branch of experimental elementary particle physics, still called so, is the precise study of particles and their interactions with each other, with the goal of a fundamental description of a theoretical model called the “Standard Model”. Therefore, the particles and the required development of the experimental technology have the highest importance.

Since its foundation at the beginning of the last century, so-called astroparticle physics has been using fundamentally similar detectors. Over the following decades, there was no separation between fields and the naturally occurring particles were used to advance fundamental particle knowledge. The multidisciplinary focus of today originated in the late 90s, where not the particles and interactions themselves were in the limelight, but the utilisation as a carrier of information from distant cosmic objects. These objects can function as natural particle accelerators, emitting particles of various types in a brought energy spectrum. Detecting this emission, often in combination with traditionally established methods of astrophysical observations like optical, X-ray, or radio telescopes, leads to a new understanding of the fundamental properties of those distant objects.

In the following sections, there will be a deeper description of cosmic rays’ origins, formation, and properties. Different methods of measurement and descriptions of physics-relevant properties follow.

## 2.1 Cosmic Rays Physics

The foundation stone for this area of physics was laid in the first half of the 20th century with the research on ionising radiation inside the atmosphere. Original beliefs stated that the Earth, e.g. the ground, itself could be responsible for the measurement of the 15 years earlier discovered radioactive particles. Therefore, a reduction in intensity with increasing distance from the potential source, the ground, was expected. Starting with 1911, the physicist Victor Hess [39] refutes this assumption by observing a rise in ionised particle levels with increasing height via a first survey done in multiple balloon flights. The following ascensions with improved equipment, e.g. electroscopes, enabled him to determine the first height distribution of radiation density inside the atmosphere and identify the origins as cosmic radiation, for which he was awarded the Physics Nobel Prize in 1936. The technical developments of the following decades allowed a more differentiated analysis of this radiation as part of particle physics, which led to the discovery of multiple new particle types, for example, the positron in 1932 [13], or the Muon in 1936 [60]. These advances progressed until the late 90s when modern astrophysics was established, and the focus shifted from the particle itself to the particle as messengers from distant sources. One of the earliest cosmic radiation sources was a neutrino, which could be associated with a supernova explosion in 1987 by the Kamioka experiment and shortly followed afterwards with the first dedicated gamma-ray measurement of the Supernovae remnant Crab Nebula in 1989 [72]. Both observations opened a new view into the depths of the universe.

### 2.1.1 Nature of Cosmic Rays

Cosmic rays are not a single homogeneous form of radiation but a broad spectrum of different types of hadronic, leptonic, and photonic emissions over a very wide energy range ranging over several orders of magnitude from $1\,\mathrm{keV}$ up to $10^{20}$ eV. The so-called GZK cutoff is the reason for an upper limit [36], where incoming particles start to interact with the ubiquitous cosmic microwave background [44] and produce multiple low-energy particles. Despite the limit, particles above the threshold were identified in different observations of various experiments. The exact origin is still unclear but can probably be attributed to heavier particles [49] or neutrinos [73].

A detailed analysis of the hadronic composition of the cosmic rays spectrum shows that all components are present in quantities similar to those found in the interstellar medium. However, the radiation itself is completely ionised. The flux distribution of particles arriving in the atmosphere is shown over the relevant Energy range in Figure 2.1 and starts with protons or hydrogen as the most common particle type.

Next is helium, with almost the same frequency, followed by heavier elements, which decrease significantly with increasing mass numbers.

The leptonic part consists mainly of electrons $e^-$, positrons $e^+$ and neutrinos $\nu_e, \nu_\mu, \nu_\tau$. The heavier leptons, muons, and tauons decay during their journey through interstellar space before they arrive at the Earth's atmosphere. As far as neutrinos are concerned, depending on their energy, the distance can be travelled almost unhindered through the interstellar medium.

The last components are highly energetic, therefore non-optical, photon emissions called gamma rays $\gamma$. For the analysis of specific sources, the photons and neutrinos make up the most relevant part of the particle flux since, unlike the other charged particles, they are not deflected by interstellar electric and magnetic fields. The weak gravitational influence acting on them, called gravitational bending or lensing, can be neglected for most sources of interest. Given the chaotic nature of the interstellar fields, the combination of the deflection of the charged particles and the large distances results in a de facto irreversible mixing, which leads to the fact that the charged particles have an almost isotropic distribution when they arrive at the Earth.

### 2.1.2 Origins of cosmic rays

Overall, charged cosmic rays cannot be traced directly to a single origin due to deflection. They are generally attributed to a superposition of individual sources distributed over the entire cosmos. Exceptions here are time-limited events of single sources, which can be verified by other experiments (e.g. optical) and be correlated with an increase in global flux. Further identification of entire source classes is possible from theoretical considerations, for example, based on the composition of cosmic rays measured by various experiments, shown in Figure 2.1. Here, it is of particular importance that the relative flux of the particle species shifts among themselves depending on their energy. In addition, special features, which can be seen in the combined flow shown in Figure 2.2, provide additional clues. The three most prominent features were named after an imaginary, "leg" which (with some imagination) can be interpreted into the contours form. The areas between the anatomic features correlate with three distinct acceleration areas of cosmic rays. The lower energy regime missing in this figure from $1\,\text{keV}$ to $100\,\text{MeV}$ originate from inside the solar system and are emitted by the sun in the form of solar winds. This origin can be verified by the measurement of fluctuation in the cosmic rays correlated with the solar cycles [69].
The second interval up to "knee" and "ankle" with the energy of $3 \times 10^{15}$ eV and $3 \times 10^{18}$ eV is formed by emission from sources in the local galaxy [33], the Milky

Way. This behaviour is caused by the fact that the particles have a gyro radius smaller than the thickness of the Milky Way due to the ubiquitous magnetic field, which effectively traps them. Sources can be credited to:

- **Stars** - Low energetic

- **Super Novae** and **remnants** - Higher Energies

In supernovae, the initial shock accelerates large amounts of matter, but even after the main explosion subsides, the entangled magnet fields bounce particles between shockfronts and accelerate them further. The highest energetic sources are extra-galactic and can be attributed to AGN - **A**ctive **G**alactic **N**ucleus or which are special “Black Holes”. The emitted jet and the expanding gas cloud surrounding it can accelerate particles likely over $10^{20}$ eV.

### 2.1.3 Cascade development

The Earth’s atmosphere is quasi opaque for particle radiation at the very high energy levels considered in this thesis cannot reach the surface of the Earth. The incoming cosmic rays do not penetrate but instead interact with the nuclei in the air. Various possible interactions can occur depending on the incident particle type, energy, and interaction partner. In most cases, the original particle is destroyed and several new particles are created. These new particles inherit different fractions of the original energy.

In the beginning, most of the resulting particles still have energies much higher than any cross-section thresholds and again undergo similar interactions with the surrounding matter, creating a cascade-like effect. This descends further down the atmosphere until all particles reach a sufficiently low energy or are stopped, for example, in the ground. Based on this cascading process, the resulting interplay is called an atmospheric cascade or **E**xtensive **A**ir **S**hower (EAS). Depending on the exact kind of incoming cosmic ray, the emerging cascade shows significantly different behaviour. The two dominant classes in which the cascade can be divided are the so-called Electromagnetic and Hadronic cascades. A sample of selected representative showers for different initial particle and energies is displayed in Figure 2.3 and give an overview of characteristic properties. For example, it is visible that the extent or width depends less on the energy and more on the mass of the incoming particle. The individual development of the classes is explained in the following sections.

### Electromagnetic Cascades

Electromagnetic cascades result from high energetic photons, called Gamma ($\gamma$) rays or similar high energetic electron ($e^+$) respectively positron ($e^-$) radiation, hitting particles inside the atmosphere. Because of the fundamental physical rules, like Lepton conservation, the result of each interaction is mainly limited to the same three fundamental particles that can initiate the interaction: Photon, electron, and positron. Due to the low mass of all particles involved in this cascade type, the new particles generated are strongly boosted in the forward direction, showing only slight deviation in their path. This behaviour makes the shower much narrower than hadronic showers of comparable energy. The dominant physical effects with the resulting new particles are schematically displayed in Figure 2.4. Missing are rare occurrences like photohadronic interaction, where a large part of the photon energy transfers to the nucleus, which can result in the generation of hadronic cascades.

The dominant physical process for gamma rays, over a few MeV kinetic energy, is the so-called pair production in which it generates an electron-positron pair close to a nucleus. The base energy of the initial photon is distributed between the resulting two. Charged electrons and positrons are slowed down inside a medium and emit via bremsstrahlung additional high-energy photons along their path.
By reducing the initial centre-of-mass energy by dividing it among all fission products along the cascade, lower energy levels are achieved that allow other processes to be favoured. Here, Compton scattering and the photoelectric effect move into the foreground to produce new electromagnetic particles from photons. For electrons' respective positrons, continuous influences like ionisation dominate energy reduction.

### Hadronic Cascades

The evolution of hadronic cascades in the atmosphere differs substantially from the electromagnetic cascade described above. Foremost, the hadronic interactions are, within themselves, much more complex and can produce a vast range of particles within the energy limit given by the mass and initial kinetic energy.
Due to the higher mass of particles propagating through the air, the energy loss by ionisation or radiative losses is relatively low compared to the EM-Cascade. Therefore, the energy mainly divides into the creation of new particles in each interaction. The creation of particles with a high mass difference in a single interaction and the often resulting high transverse momentum arising from symmetry constraints pushes the lighter part in a more orthogonal direction of propagation. Overall, this results in a much weaker forward momentum, resulting in a larger transversal width of the shower. In addition, the energy distribution is much less homogeneous than in electromagnetic cascades, which, combined with the higher transverse momenta, leads to much greater inhomogeneity in the spatial distribution of the shower itself. This effect is further enhanced by the production of neutral pions, which decay into high-energy photons to form electromagnetic sub-cascades.

A combination of the shorter mean free path of the heavier hadronic particles, which causes the cascades to form much sooner and the high transversal momenta of the shower results in a much larger area of impact on the observation plane. The schematic representation of some process possible in a hadronic cascade are displayed in Figure 2.5. The physical simulation of hadronic cascade with a lower-mass (proton) and with a much higher mass (iron) initial particle is displayed in Figure 2.3f and 2.3i. A comparison shows the significant increase in width and the earlier development of these exemplary showers.

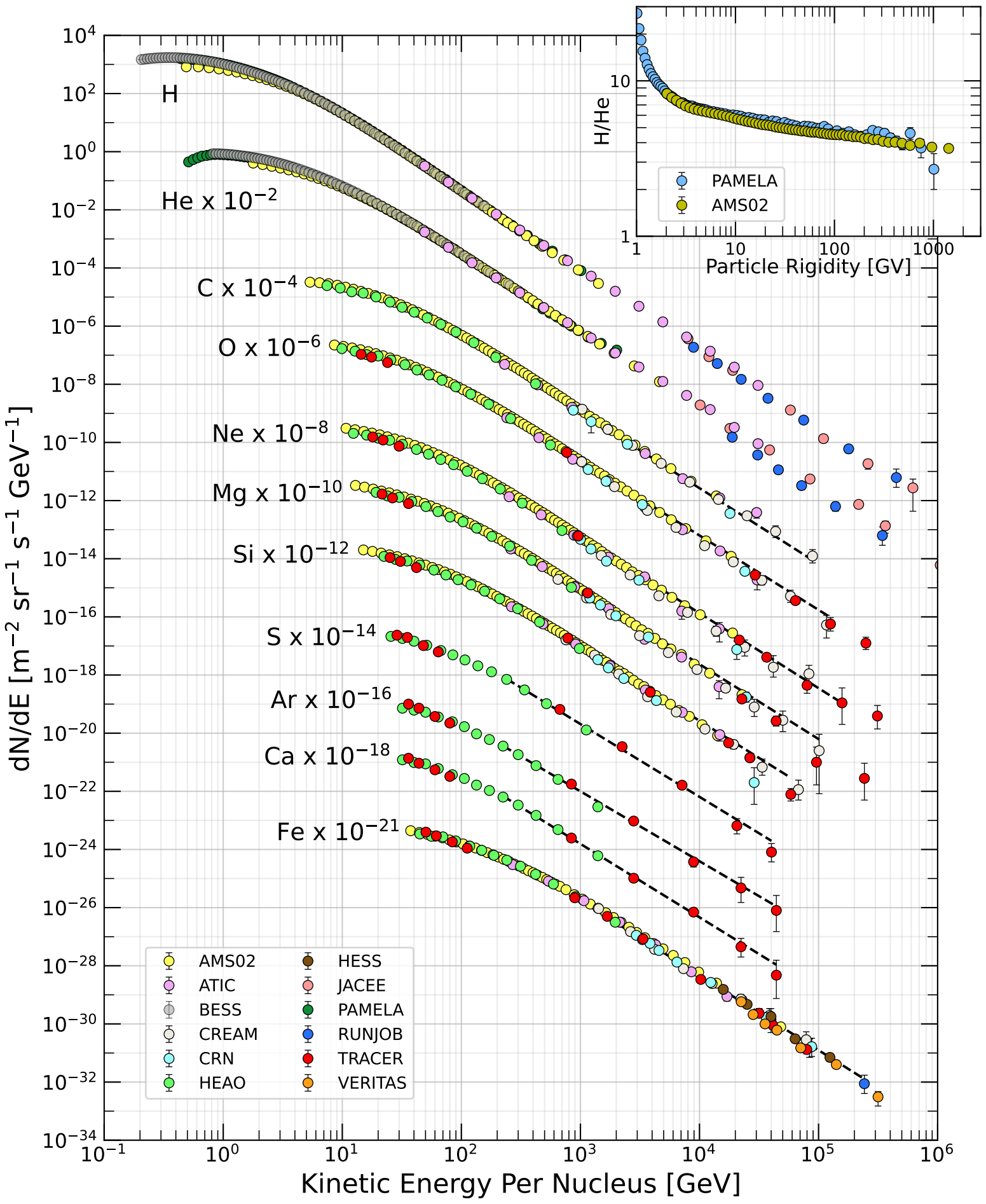

**Figure 2.1:** Displayed is the composition of cosmic rays derived from different Experiments.

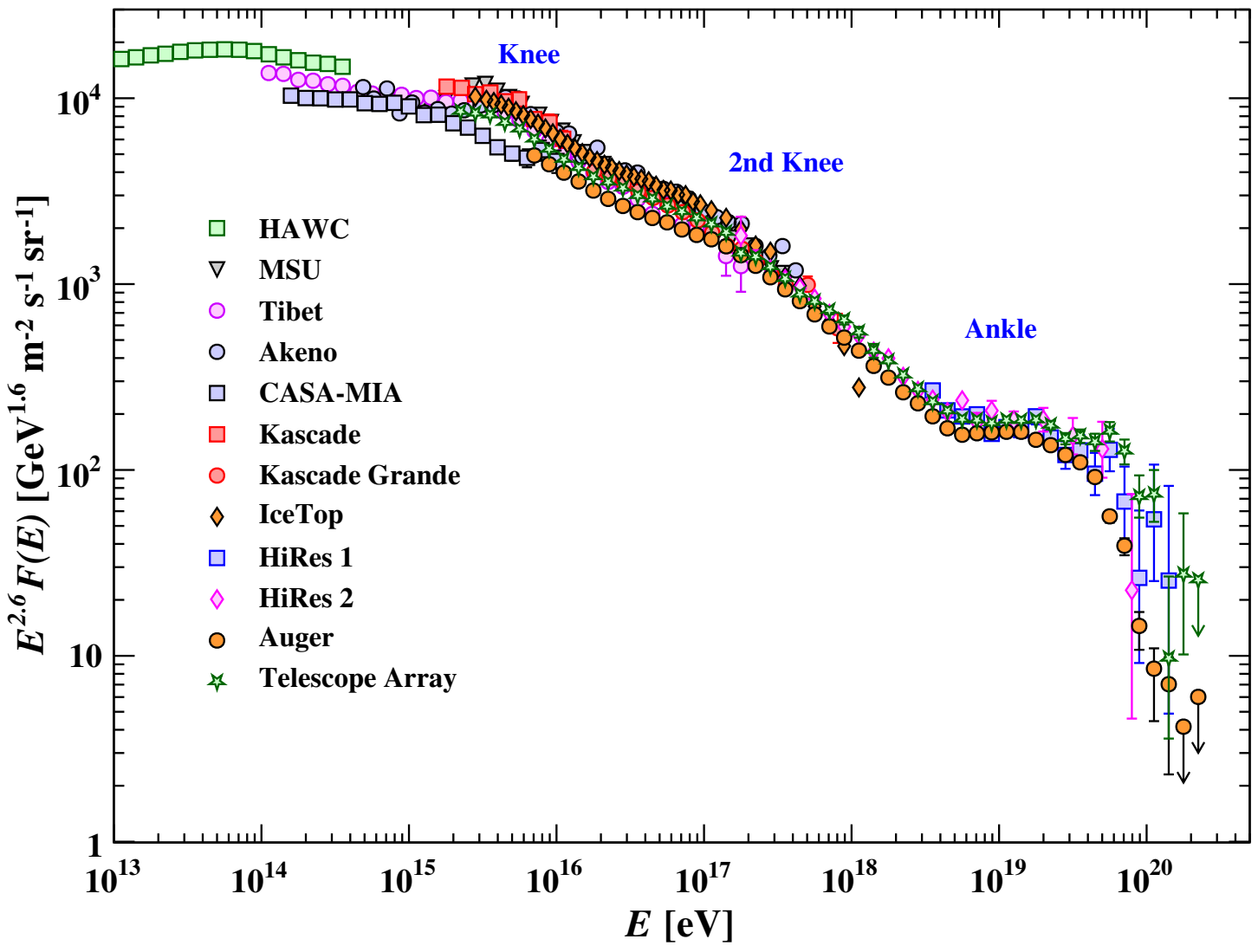

**Figure 2.2:** Shown is the flux of cosmic rays arriving at Earth before interacting with the atmosphere as a function of energy. Following the "leg" shape of the curve, the low-energy bend can be called the knee, and the high-energy bend the ankle. [37]

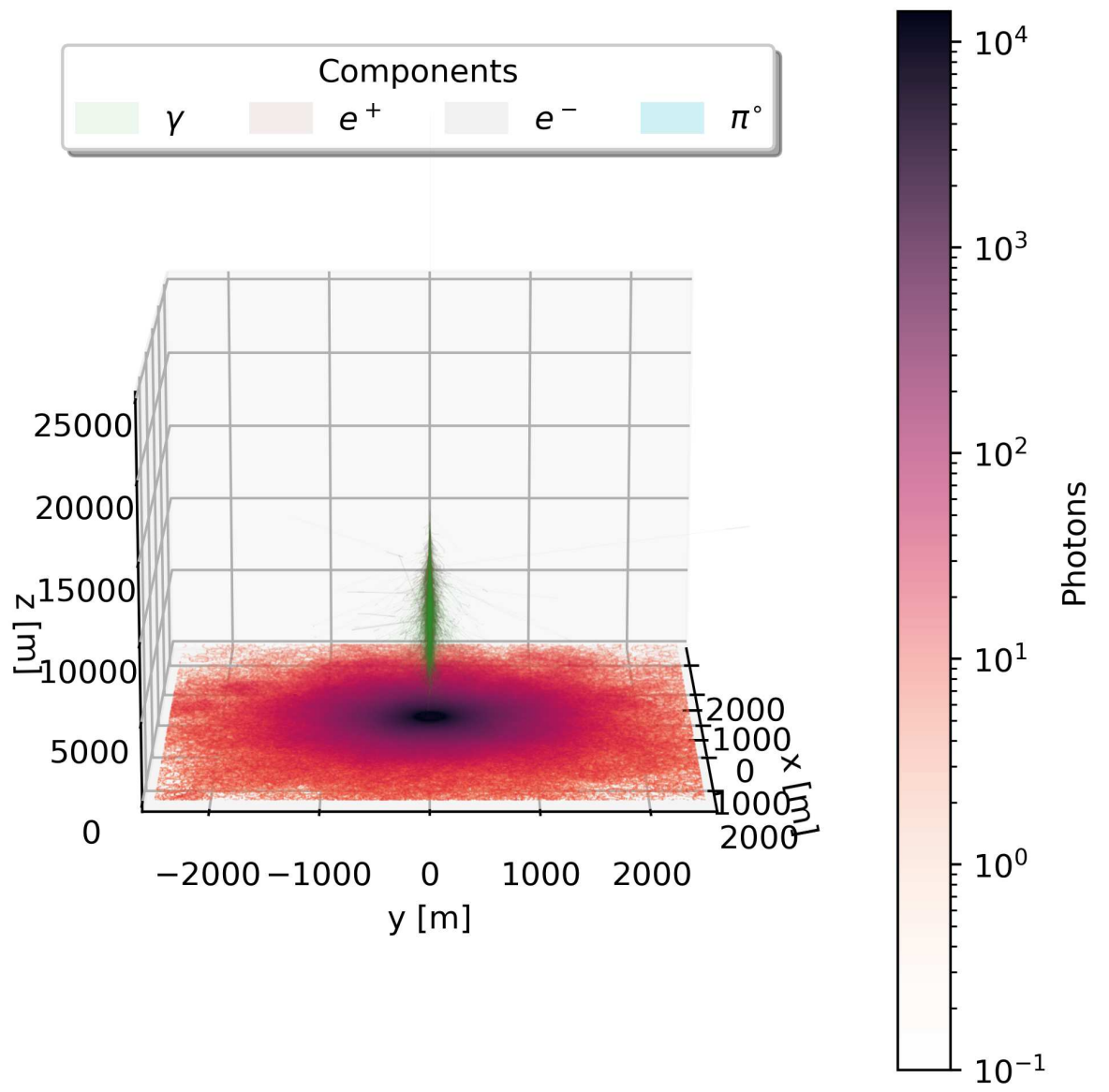

**a:** 500 GeV

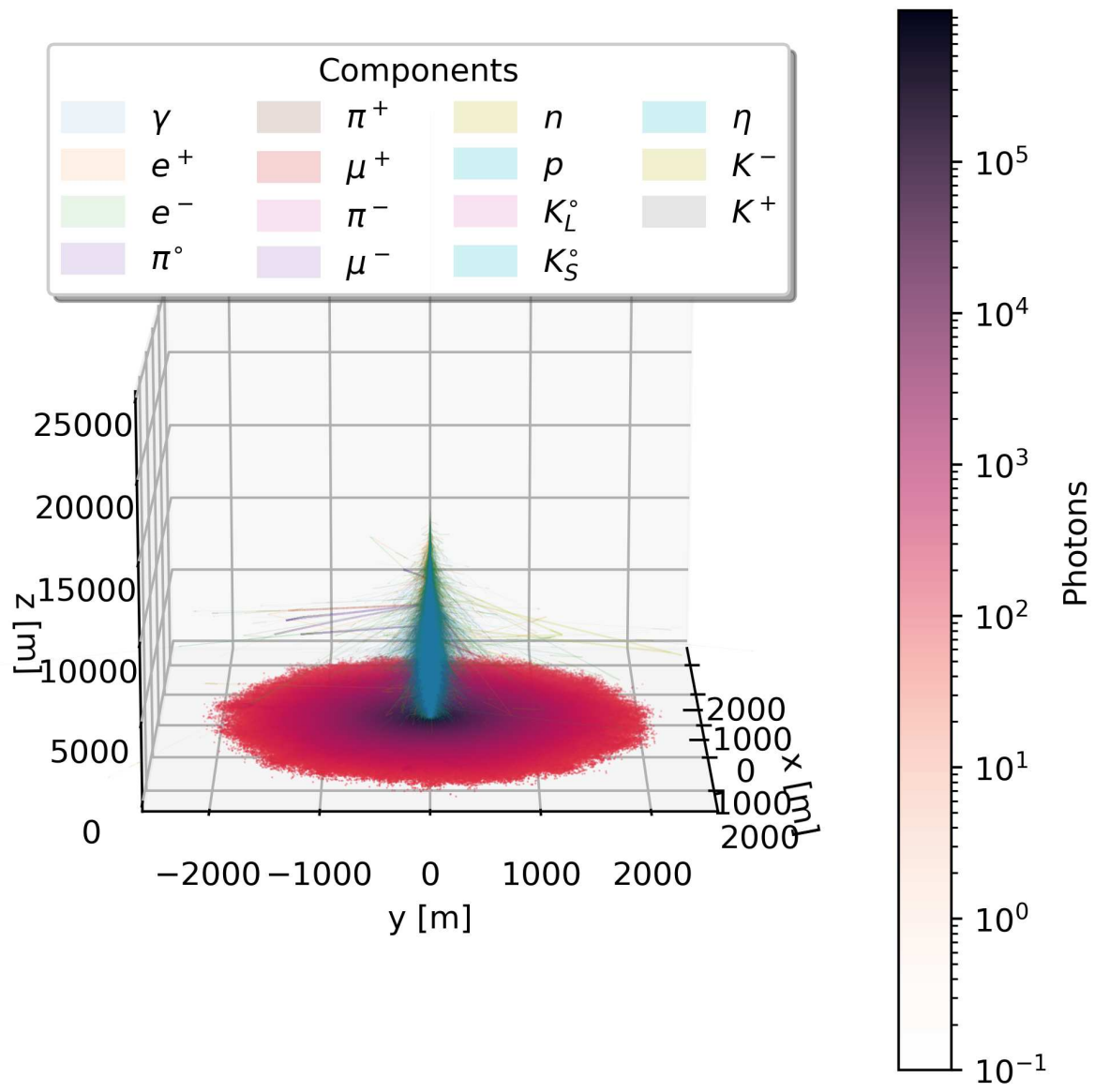

**b:** 10 TeV

**c:** Photon

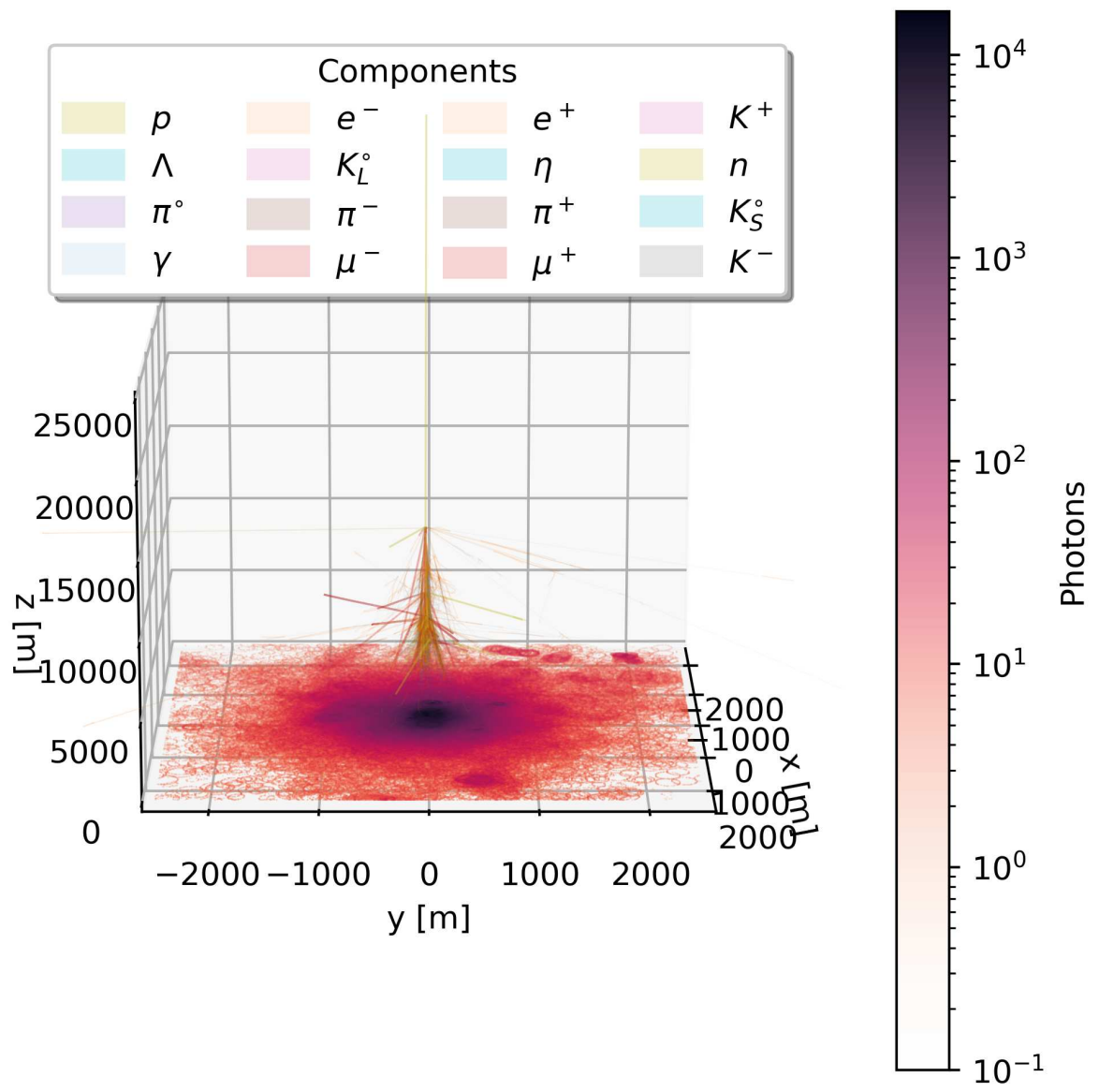

**d:** 500 GeV

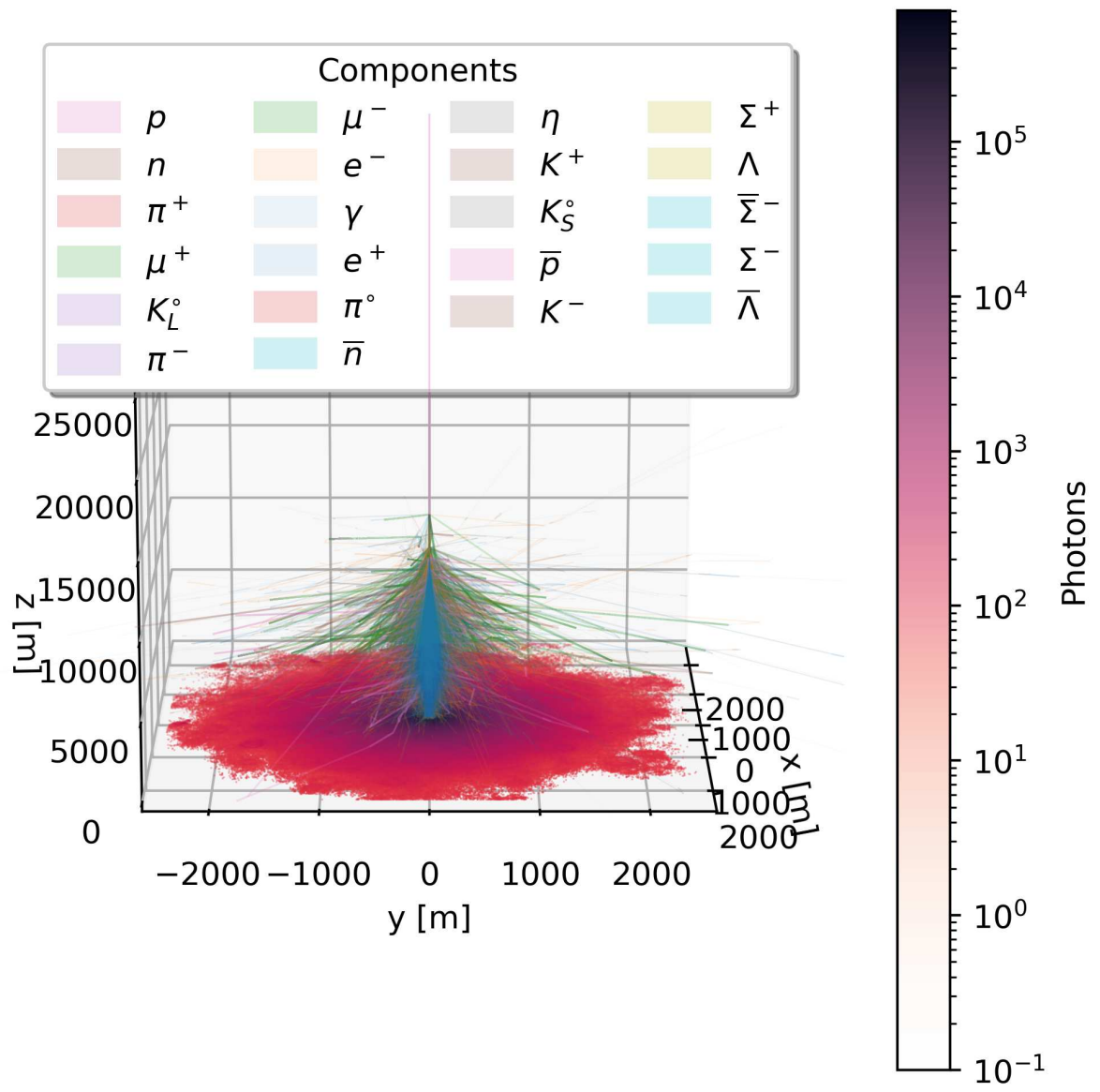

**e:** 10 TeV

**f:** Proton

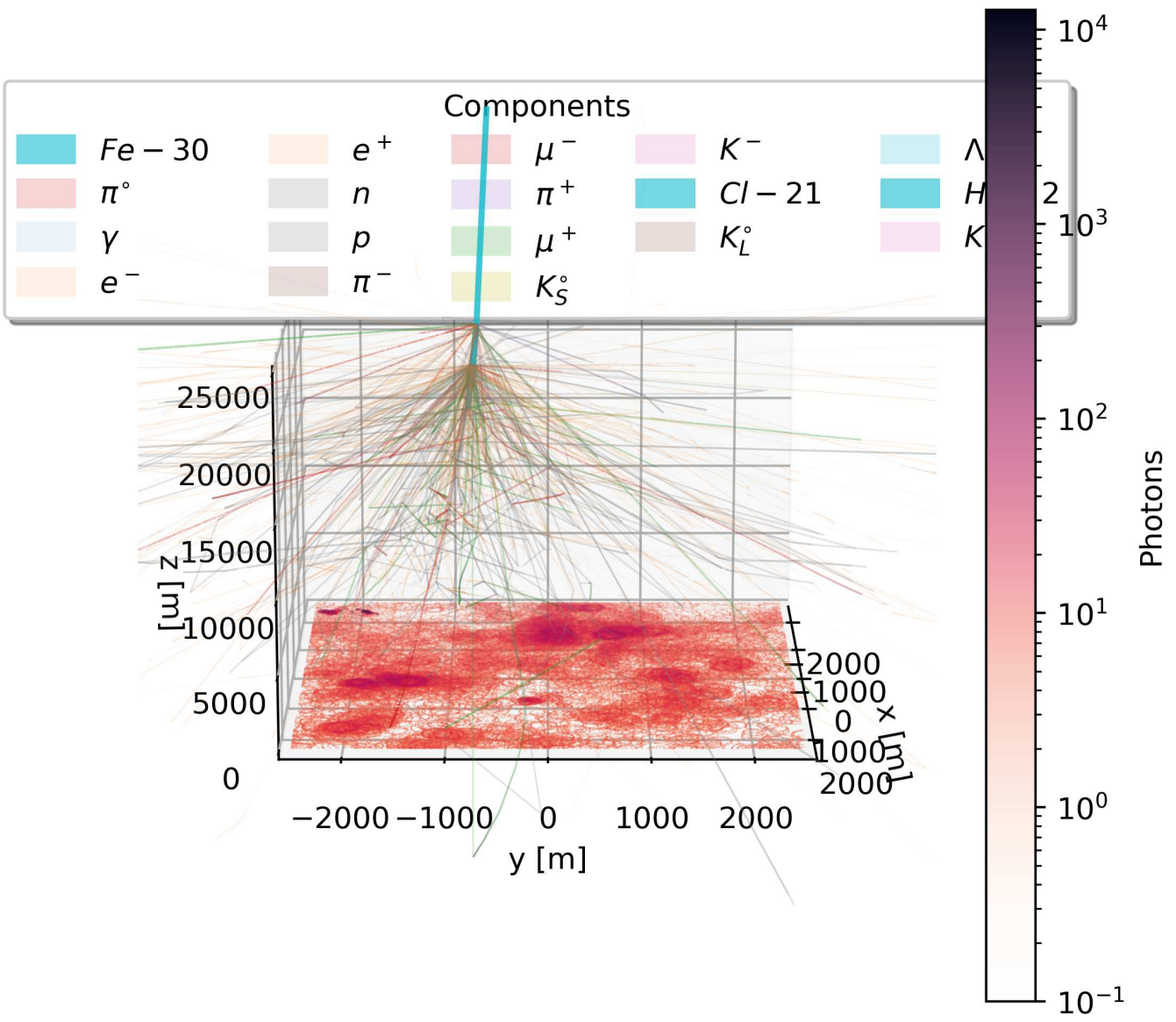

**g:** 500 GeV

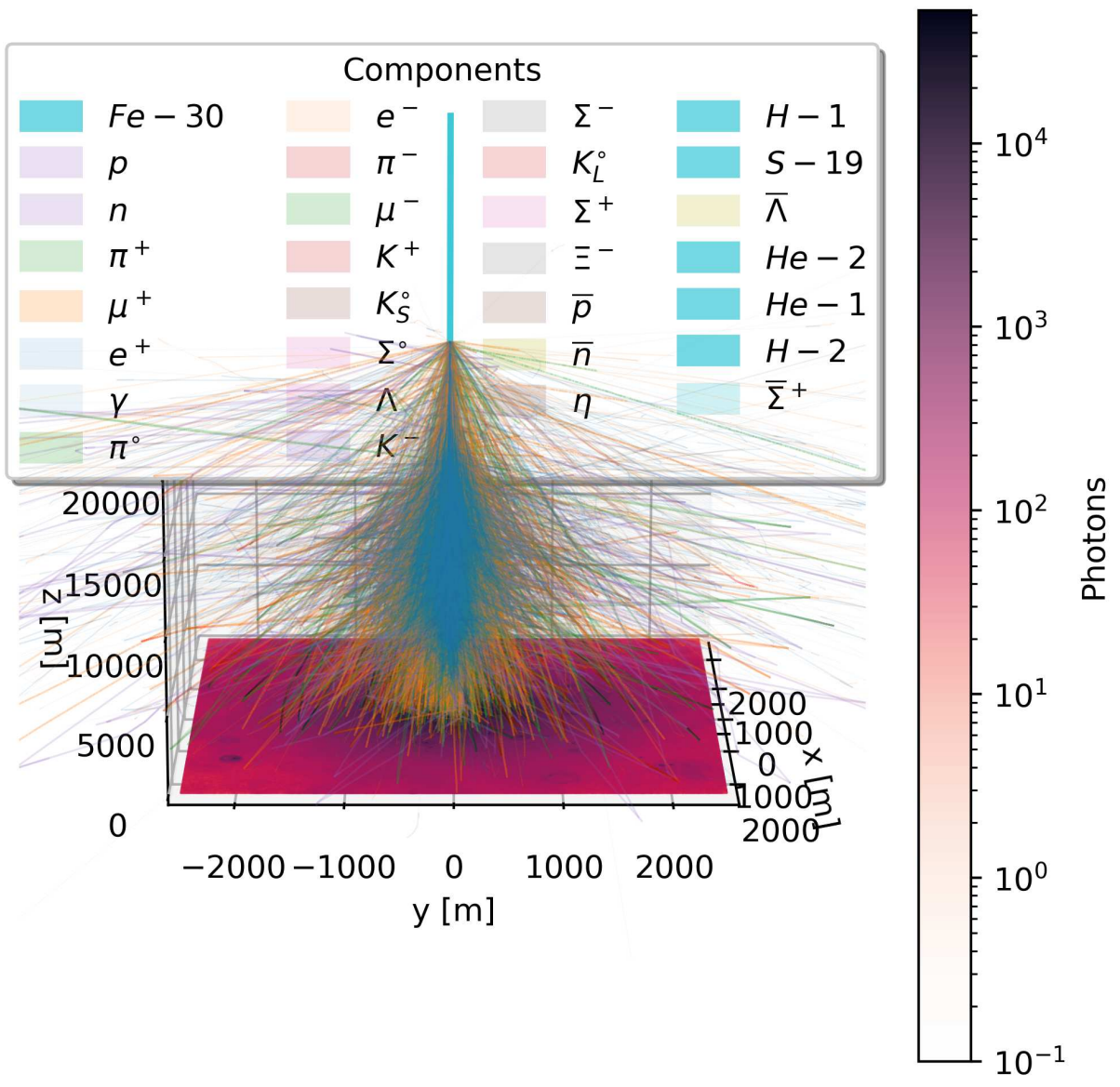

**h:** 10 TeV

**i:** Iron

**Figure 2.3:** Displayed are cascades for different particles for two representative energy levels. The size can be estimated accordingly for particles with masses and energies in between.

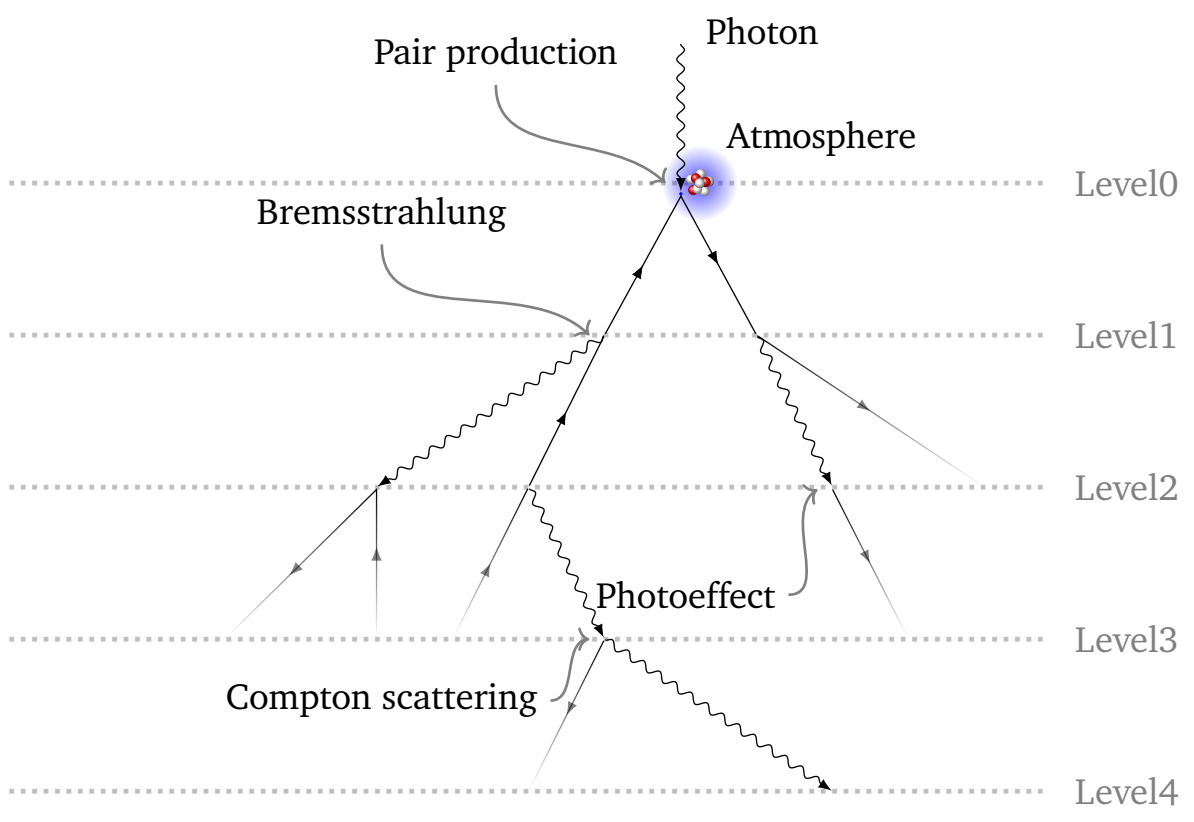

**Figure 2.4:** Displayed is the schematic representation of an electromagnetic cascade inside the atmosphere. The four most significant physical effects for the formation of the cascade are shown schematically for illustrative purposes only.

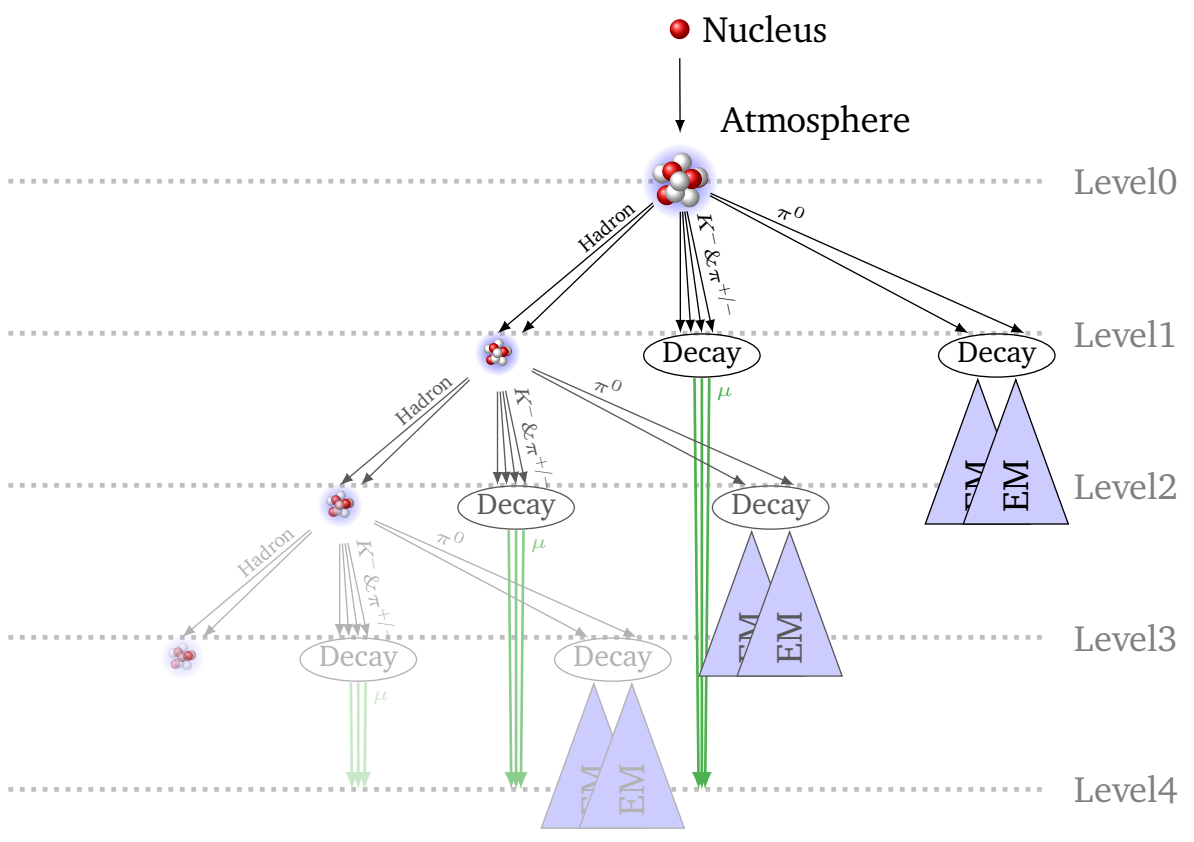

**Figure 2.5:** Displayed is the schematic representation of a hadronic cascade inside the atmosphere. Separate electromagnetic cascades are generated by the production of neutral Pions during the cascade. Due to the high variability, only the most significant products are specifically marked.

## 2.2 Measurement of Cosmic Rays

In the same way, as there are many manifestations and types of cosmic rays, there are equally various experiments to measure them, mostly depending on the targeted detectable emission, often multiple per experiment.
The "easiest" way is the direct evaluation of the incoming cosmic rays outside the atmosphere before they can interact at all. This method allows the direct reconstruction of energy and direction from the determined parameter. Currently, only satellite-based experiments allow this kind of measurement. The resulting disadvantages, such as the very limited instrumented area and high costs, limit its use to lower energy ranges with corresponding high radiation flux, which is displayed in figure 2.2. Some examples of this type of experiment are the AMS [50], CALET [68] and DAMPE [32].

The detection area must be significantly extended to obtain adequate data in the mid-energy range. Ground-based experiments that can be sufficiently scaled allow this coverage but necessitate the indirect observation of the particle cascade generated within the atmosphere using different observable parameters. As the atmosphere effectively becomes part of the detector, the experiment observes a much larger volume than would otherwise be possible. As a result, the instrument's sensitivity is significantly increased since the initial particle does not have to pass through the detector itself but can be detected, depending on the observable, up to a distance several orders of magnitude larger than the detector size. This allows the experiment to detect radiation of lower intensity in relevant quantities. Possible observable parameters include secondary emissions such as radio radiation caused by charge separation in the Earth's magnetic field [43] and the Askaryan effect [14], as well as Cherenkov and fluorescent light caused by various physical effects. In addition, the particle tracks of the cascade can be detected directly using appropriate sensor technology. The latter method, in particular, can be scaled up to several square kilometres of the instrumented area to provide meaningful statistics even in the highest energy ranges.

Neutrino radiation is a particular case in that it easily penetrates the atmosphere, detectors, or even the whole Earth because it couples exclusively via weak interactions with the surrounding matter. Large quantities of a dense, instrumentable medium, such as ice or water, are required for meaningful measurements. There, the neutrino has a chance to interact and produces, among other things, light emission that can be detected. IceCube, with a volume of $1\,km^3$, can be mentioned here as an example.

This work is concerned with the acceleration of simulation software for light-based experiments. More precisely, it is exclusively for experiments in a thin medium like the atmosphere; therefore, neutrino detectors are excluded due to the very particular medium of water and the complications that follow. Classical optical telescopes and cosmic ray Cherenkov telescopes work in fundamentally identical ways, the main difference being the detection method. The light production in atmospheric cascades is decidedly short-lived, covering only a few ns. Correspondingly, the installed sensors must have appropriate time resolution. Due to other external circumstances, such as individual pixel size, minimal mandatory light yield, and targeted energy range, an appropriate aperture angle optimising all constraints must be chosen for the telescope. As an example here, MAGIC Cherenkov telescopes use an angle of $3.5°$ [30], fluorescence detectors have a larger field of view up to $30°$ [12].

To meet the growing demand for observations, as well as the significantly broader diversification in cutting-edge science, many experiments now need to rely on a combination of various physical detection methods, including optical light emission, as opposed to relying only on a single observable parameter like the tracks of the cascade. This multi-observable method helps minimise systematic and absolute measurement errors. Overall optical light measurements from cascades are present at multiple locations, most prominent is the large-scale Cherenkov light experiment "**C**herenkov **T**elescope **A**rray" (CTA) which consists of over 50 telescopes with diameters ranging from approximately $4\,\mathrm{m}$ to $12\,\mathrm{m}$.

### 2.2.1 Cherenkov Radiation

The already mentioned Cherenkov effect [26] named after its discoverer Pavel Cherenkov, later mathematically described by Ilya Frank and Igor Tamm [27], describes the emission of optical photons from relativistic particles in a media. Due to the very high energy of the cosmic radiation impacting the atmosphere, a large part of the cascades secondary particles has a phase velocity that is far above the speed of light inside the medium, $c_n = {}^{c_0}/_{n}$ which is lower than the vacuum light speed $c_0$. A simple analogy would be the classical sonic boom during supersonic movement. However, instead of a pressure wave, the particles release electromagnetic waves in a forward direction in the form of photons. One particularity is that the particle moves faster than the emitted light, resulting in later emitted photons arriving earlier. The wavelength distributed covers a larger range with an approximate intensity distributed with $\approx {}^{1}/_{\lambda^2}$ and thus can be seen as blue light. The number of photons released per meter of particle track can be calculated sufficiently accurately by Frank-Tamm formula, despite simplification in its derivation [62]. Therefore, it will be used for this thesis:

$$\frac{\partial^2 E}{\partial x \partial \omega} = \frac{q^2}{4\pi}\mu(\omega)\omega\left(1 - \frac{c_0^2}{v^2 n^2(\omega)}\right) \tag{2.1}$$

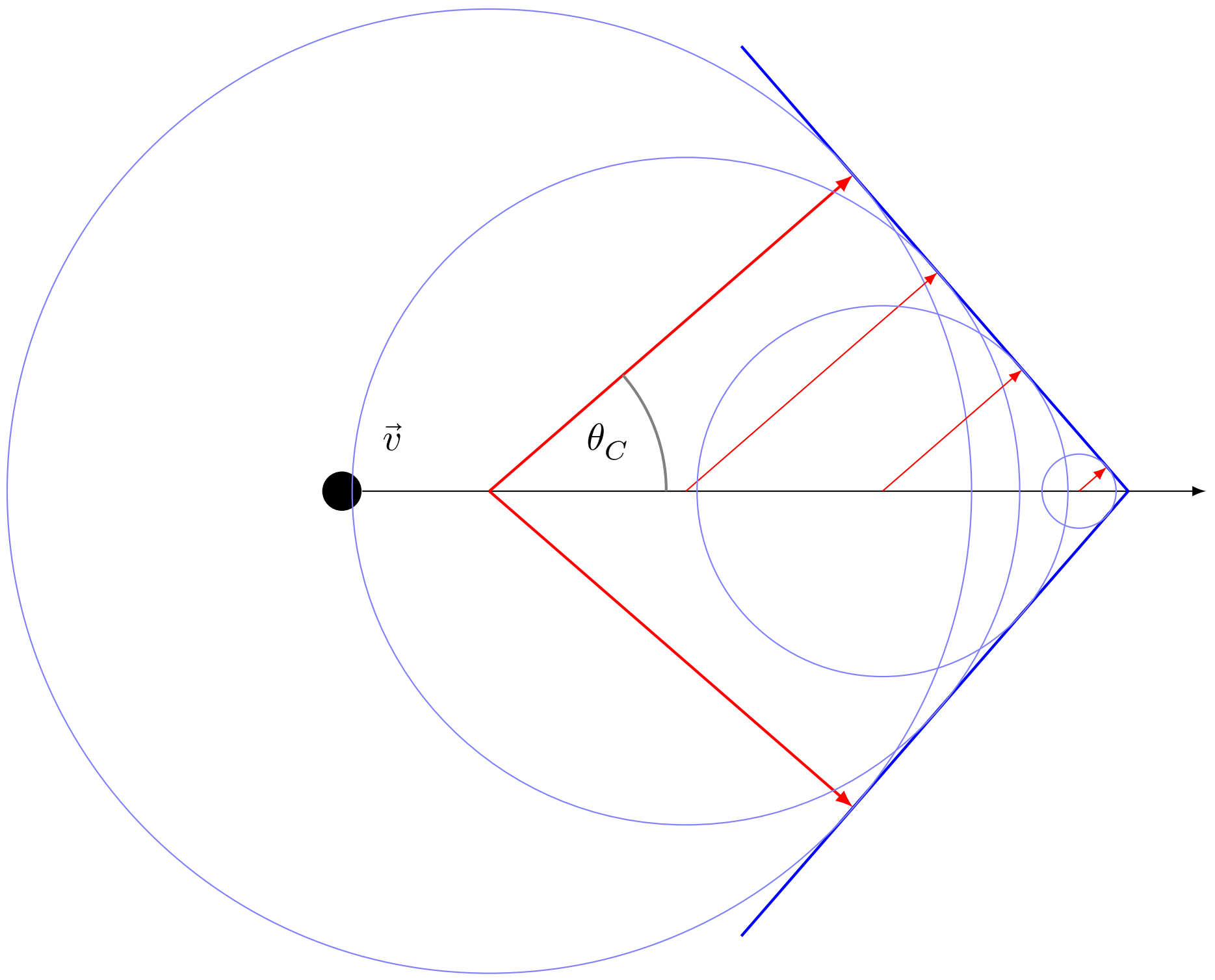

**Figure 2.6:** The Cherenkov light (blue) and the emission direction (red) of charged particles with a velocity $v$ above the speed of light in the surrounding medium $nc$ are shown. The emitted photons create a wavefront (dark blue) containing the Cherenkov light. For small angles $\theta_c$, i.e. a velocity close to the vacuum speed of light, all the photons arrive almost simultaneously. In media with a higher refractive index, the photons emitted later will arrive earlier at the detector.

With the charge $q$, the frequency-dependent permeability $\mu(\omega)$ and the refractive index $n(\omega)$. Because the air's permeability is quasi-constant over a wide frequency range [35], it can be replaced by $\mu_r \mu_0$. To translate this from energy to the actual number of photons, the identity $\omega = 2\pi\nu$, $\partial\omega = 2\pi\partial\nu$, and the photon Energy

$\partial E = \partial N h\nu$ can be used.

$$\frac{\partial^2 E}{\partial x \partial \nu} = \frac{q^2}{4\pi}(\mu_r \mu_0)(2\pi\nu)\left(1 - \frac{c_0^2}{v^2 n^2(\nu)}\right)(2\pi) \quad | \quad \text{(simplify)}$$
$$\frac{\partial^2 E}{\partial x \partial \nu} = \frac{\pi q^2}{1}\mu_r \mu_0 \nu \left(1 - \frac{c_0^2}{v^2 n^2(\nu)}\right) \quad | \quad (\text{E} \rightarrow \text{N})$$
$$\frac{\partial^2 N}{\partial x \partial \nu} = \frac{\pi q^2}{h\nu}\mu_r \mu_0 \nu \left(1 - \frac{c_0^2}{v^2 n^2(\nu)}\right)$$

For easier use, the frequency $\nu$ is converted to the more commonly used wavelength $\lambda$. For this $\nu = \frac{c_0}{\lambda}$ and $\partial \nu = -\frac{c_0}{\lambda^2}\partial\lambda$ is used.

$$\frac{\partial^2 N}{\partial x \partial \lambda} = \frac{\pi q^2}{h\left(\frac{c_0}{\lambda}\right)}\mu_r \mu_0 (\frac{c_0}{\lambda})\left(1 - \frac{c_0^2}{v^2 n^2(\lambda)}\right)\left(\frac{c_0}{\lambda^2}\right)$$
$$\frac{\partial^2 N}{\partial x \partial \lambda} = \frac{c_0 \pi q^2}{h\lambda^2}\mu_r \mu_0 \left(1 - \frac{c_0^2}{v^2 n^2(\lambda)}\right) \quad | \quad (\mu_0 = \frac{1}{\varepsilon_0 c^2})$$
$$\frac{\partial^2 N}{\partial x \partial \lambda} = \frac{\pi q^2}{h\lambda^2 \varepsilon_0 c}\mu_r \left(1 - \frac{c_0^2}{v^2 n^2(\lambda)}\right) \quad | \quad (v/c = \beta,\, q = ze)$$
$$\frac{\partial^2 N}{\partial x \partial \lambda} = \mu_r \pi z^2 \frac{e^2}{h\varepsilon_0 c_0}\frac{1}{\lambda^2}\left(1 - \frac{1}{\beta^2 n^2(\lambda)}\right) \tag{2.2}$$

This can be simplified further for human use to the well known form below.

$$\alpha = \frac{e^2}{2h\varepsilon_0 c_0}$$
$$\cos\vartheta = \frac{1}{\beta n} \tag{2.3}$$

$$\frac{\partial^2 N}{\partial x \partial \lambda} = 2\mu_r \pi z^2 \alpha \frac{1}{\lambda^2}\sin^2(\theta) \tag{2.4}$$

For real applications, formula 2.2 can be integrated over the wavelength of experimental observable photons to reduce the overall propagation workload by not including non-measurable photons. The actual simulation introduces an additional level of complexity originating from the energy loss of the particle along the track, wavelength dependence of the refractive index, called dispersion, and free-form atmospheric models, which can include discontinuities and non-monotonic behavior of the refractive index. As a result, boundary conditions are introduced into the integral that the software needs to verify along the entire track. For example, only for a part of the light spectrum, the condition $\beta > 1/n(\lambda)$, necessary for Cherenkov light, may be fulfilled. Furthermore, this may change throughout the path since both $\beta$ and the refractive index depend on the position. Therefore, the integral must be changed to account for all relevant dependencies: The location dependency is expressed as a Heaviside function $\Theta$:

$$N = \mu_r \pi z^2 \frac{e^2}{h\varepsilon_0 c_0} \int_{\lambda_1, \beta(l)>1/n(\lambda,l)}^{\lambda_2} \int_{l_1}^{l_2} \frac{1}{\lambda^2} \left(1 - \frac{1}{\beta^2 n^2(\lambda,l)}\right) \Theta\left(\beta\left(l\right) - 1/n\left(\lambda,l\right)\right) \mathrm{d}\lambda\,\mathrm{d}l$$

In addition formula 2.3 which is required for the calculation of the Cherenkov angle $\vartheta$ is also dependent on the location $l$ and wavelength $\lambda$ which results in:

$$\vartheta = \arccos\left(\frac{1}{\beta\left(l\right) n\left(\lambda,l\right)}\right) \tag{2.5}$$

For some of the most important particles, the energy limit for Cherenkov light is displayed in figure 2.7. The plot covers a wide refractive index range from air up to water. The marked values correspond to conditions at sea level. Experiments located at higher altitudes, e.g. Magic Telescopes at $2240$ m AMSL (Above Mean Sea Level) or HAWC $4100$ m AMSL, experiencing an overall decreased maximum air pressure, resulting in a reduced maximum refractive index.

Each experiment can choose between several strategies for the basic experiment-specific tuning of the particle simulation part of CORSIKA. One of these is the removal of particles below a specific energy. Therefore, the marked values in figure 2.7 can be used as default energy cuts for Cherenkov light production with the potential to be increased, resulting in a higher removal rate, for specific experiments. In addition, secondary effects must be considered, like the decay from a muon into an electron or the production of new particles by hadronic interaction. Therefore, cutting limits must be set carefully in the final simulation to ensure a good balance between simulation speed and accuracy.

### 2.2.2 Fluorescence Emission

Besides the above-described Cherenkov emission, fluorescence is one of the main contributors to optical light emission inside the cascade and creates another possible observation channel usable for experiments, with its own advantages and disadvantages. Both emission types are not mutually exclusive because they share similar wavelength domains and overlay each other in the experiment sensor.

One of the most advantageous features of fluorescence light is that the emitted light is highly proportional to the initial energy and can be used as a calorimeter for precise energy measurements, provided a sufficient understanding of the atmosphere is given. Several experiments utilise fluorescence emission, such as the Pierre Auger Observatory [5] and the Telescope Array [6], for improved energy measurement quality. With high enough angular resolution, it is possible to observe the longitudinal distribution of the shower and improve the initial mass reconstruction of the original particle.

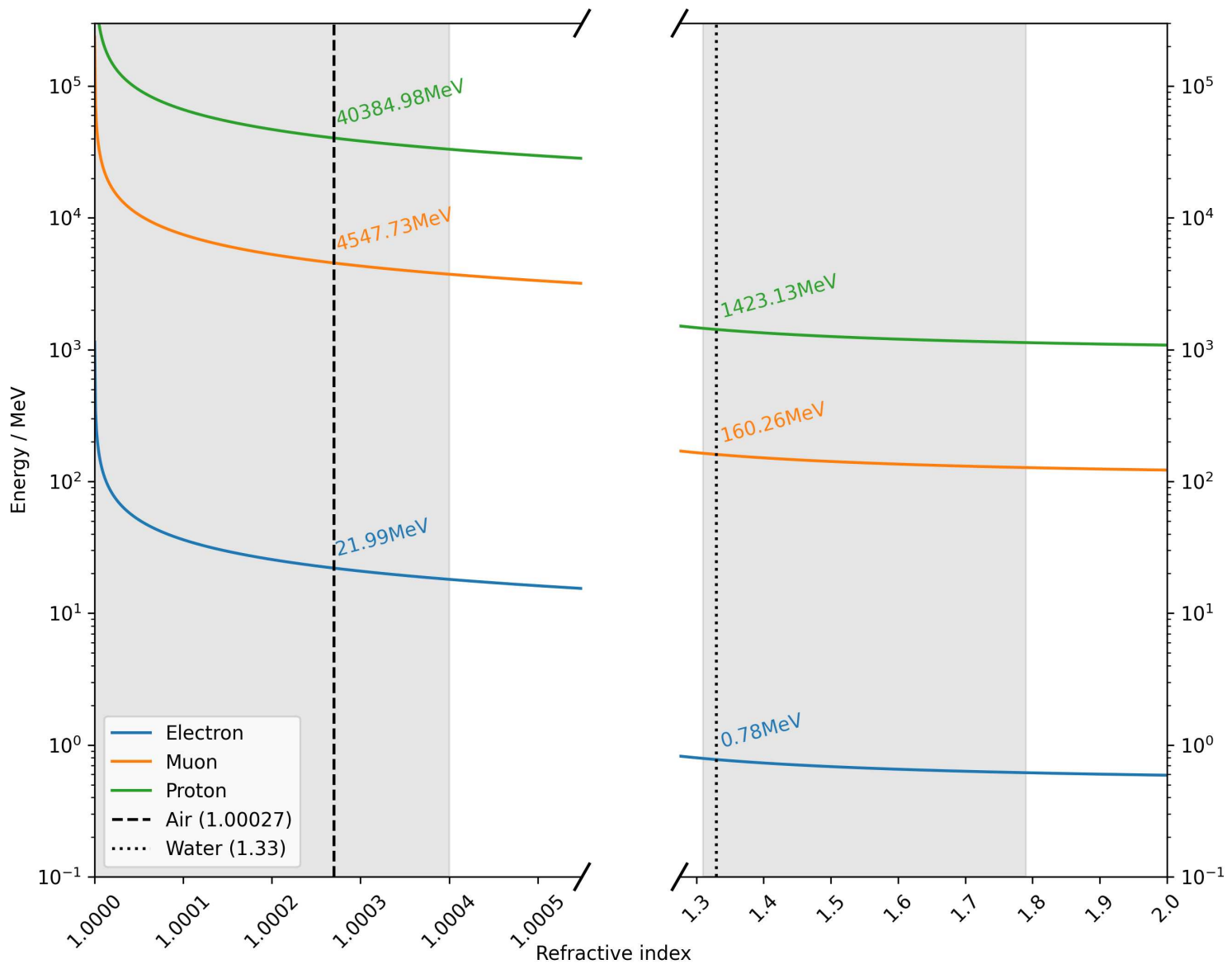

**Figure 2.7:** Energy limit for Cherenkov light emission for the most important particles with various refractive indices. The default values for air and water with their corresponding energy limits are highlighted. The shaded section marks possible areas that can be naturally reached by appropriate environmental conditions found on Earth.

The main contributor to fluorescence-driven light generation is the de-excitation of atmospheric nitrogen. The contribution of other molecules is far less pronounced or is found in wavelength bands with high background emission from other sources and is hence not usable for experimental observation. The actual excitation comes from the ionisation of atmospheric particles by cosmic rays and the resulting release of electrons. After a statistically driven time frame, with an exponential distribution, the nitrogen can de-excite over one of several possible channels in which near UV-Photons are emitted isotropically. To calculate the actual photon yield number, the following formalism used by the AirFly collaboration [15], among others, can be used:

$$Y_\lambda = \varepsilon_\lambda\left(p, T, e\right) \cdot \frac{\lambda}{hc_0} \cdot \frac{\mathrm{d}E}{\mathrm{d}x} \cdot \rho_{\text{air}} \left[\frac{\text{photons}}{\text{m}}\right] \tag{2.6}$$

The “Photon Yield” parameter $\varepsilon_\lambda\left(p, T\right)$ hides most of the underlying complexity by including all dependencies on atmospheric pressure $(p)$ and temperature $(T)$ as well as the water vapour pressure $(e)$. In addition, each parameter is independent for each wavelength $(\lambda)$. All the mentioned variables significantly influence the amount of light produced by particle traversal through the atmosphere. Therefore, a precise knowledge of the atmospheric conditions is necessary to calculate absolute values.

## 2.3 Atmosphere

The atmosphere, as the primary medium of cosmic ray airshower simulation, plays a crucial role in all processes that allow the detection of these showers. Therefore, the atmospheric models must be extensive and accurate enough to serve as a basis for all physical methods with the desired error limit. Because of the focus of this work on optical propagation, only the relevant atmospheric properties will be referred to in the following subsections. Those are the refractive index, fluorescence yield originating from the composition and pressure, scattering, and absorption coefficients. The last two are treated only approximately, so an exact differential model is not required.

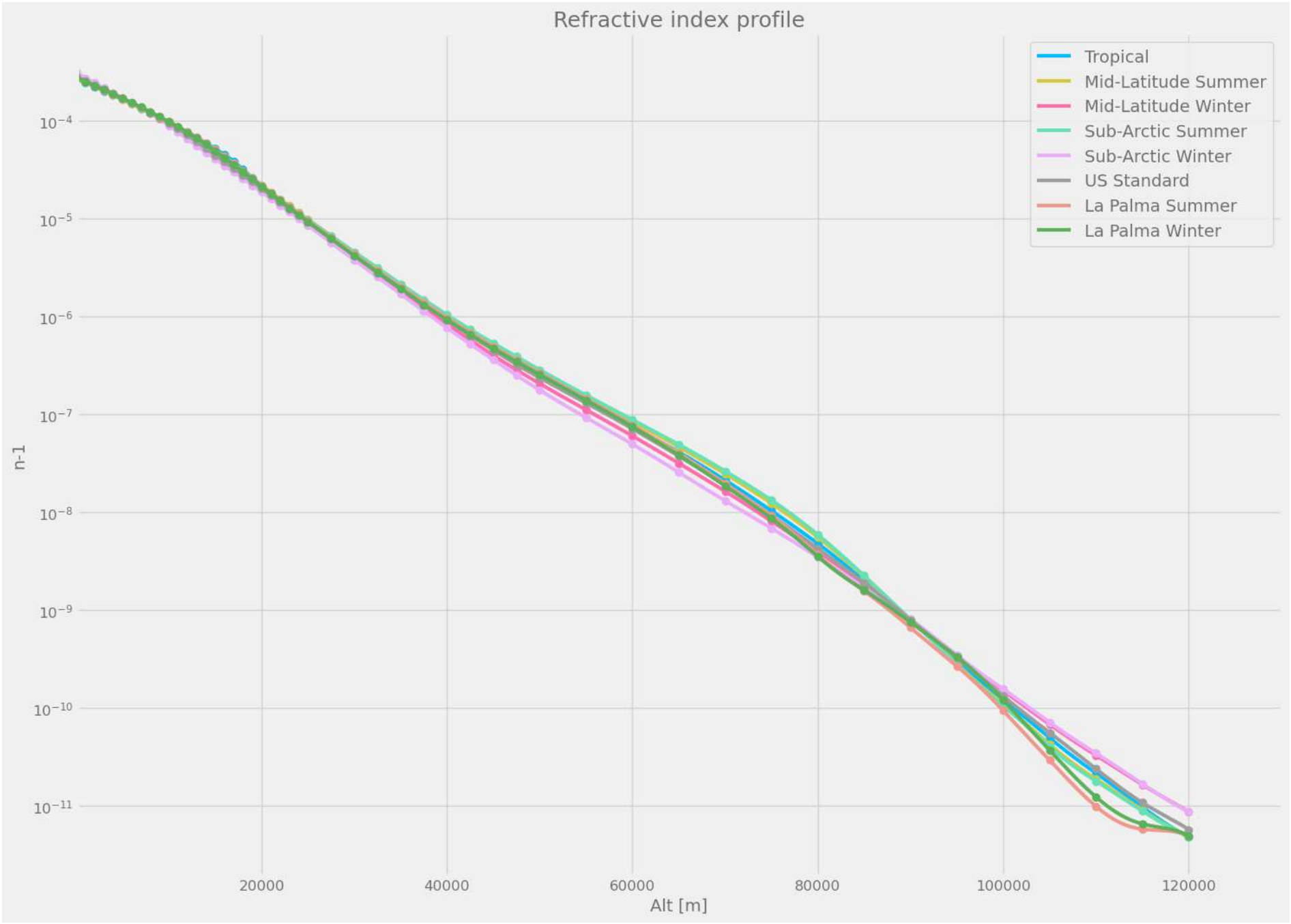

**Figure 2.8:** Refractive index profile of several standard atmospheric models commonly used for CORSIKA 7.

### 2.3.1 Refractive Index

The refractive index as an atmospheric property can only be determined by direct measurement with great effort since, for numerous altitudes, the value in the medium itself must be quantified by an optical gas-refractometer. For this purpose, almost only measurements by a weather balloon are suitable. If it is considered that the refractive index is dependent on the overall weather, which can change in comparison to the measurements in a short timeframe, such direct measurements with the required resolution are nearly impossible. A more suitable method of measurement has been found by using the empirical correlation between nominal quantities such as pressure, humidity and temperature, as described in the publication [47], further refined in [23]. These quantities can be determined much more quickly using weather models and, nowadays, for real-time measurements using lidar technology [74], directly from the ground to an altitude of a few 10 km. This information allows experiment operators to log changes over observation time and generate specifically tailored simulations.

The refractive index is needed to calculate the Cherenkov yield and the curvature of the photon track. Depending on the tracking algorithm, the refractive index is called only once per particle or, in the worst case, several times per photon. Accordingly, an appropriately optimised refractive index calculation algorithm for arbitrary heights above mean sea level must be used to avoid a possible bottleneck. The basic atmospheric data are primarily available in arbitrary tabular form and, for runtime reasons, cannot be used directly for the simulation. Using various algorithms, see section 4.1 for details, intermediate points between the tabulated values can be calculated and provide a continuous approximation for the refractive index. This curve resembles an exponential function directly resulting from the underlying atmospheric density distribution. Figure 2.8 shows the refractive index curve.

### 2.3.2 Absorption and Scattering

Two influences of minor importance in the overall context of the simulation are the absorption [42] and the scattering of photons by constituents of the non-idealised atmosphere. Here, water molecules in the form of humidity and droplets (clouds) and microscopic and macroscopic dust particulate absorb or scatter photons away. The amount of relative intensity decrease of arriving photons at the site of the experiment, which can usually be generated from atmospheric forecast data or local measurements using appropriate radiative transfer models, e.g. Modtran [20] can be applied after CORSIKA in the experiment specific code as is standard in many applications today or in CORSIKA itself which reduces the number of photons to be generated and tracked. The latter is the preferred method, as it reduces the calculation time.

In addition, other effects that typically do not require extensive computation and reduce the absolute number of photons measured, such as mirror reflectivity, sensor acceptance and other experimental optical phenomena, can be exploited. Overall, the combined influences can reduce the number of photons that need to be handled by one or two orders of magnitude, which can significantly speed up the simulation. A concrete example of the optimisation potential for a Silicon Photomultiplier-based Cherenkov-Telescope is given in section 4.3.1.

# 3 Simulation Fundamentals

The last chapter (Chapter 2) described the differences and independent goals of the two established fields of experimental particle physics, specializing in elementary particle and astroparticle physics, respectively. This chapter focuses on the underlying simulation technique, the Monte Carlo simulation, which is widely used in physics. The aim is to describe the basic principles and to provide a detailed insight into the specifically derived methods relevant to the later sections.

Since both fields of particle physics use very similar methods, the problems and hurdles that arise are mainly identical. One of these obstacles is that the information measured in the detectors differs significantly from the fundamental parameters of the original event. However, these parameters are the only ones that can be measured and must be used to obtain scientific results.

For example, in a straightforward case, the measured intensity of the radiation emitted by the particle can be used to infer its original energy. The situation is further complicated because a large proportion of all particle-physical processes and interactions are inherently statistical, i.e., random. For this reason, events originating from identical initial situations can appear completely different in the detector or almost identical for unequal initial conditions. Since there are many such random processes between the input and output variables of the process being observed, a closed-form formalism for describing the transfer function is extremely difficult and, in many cases, impossible. Another common problem is additional signals in the detector due to so-called background events. By producing the same observable properties (e.g. light), physically irrelevant processes can superimpose on the desired signal. Often, the background dominates the measurement and is orders of magnitude more intense than the desired signal.

In order to infer the measured values with the accuracy required for many experiments today, simulations are a fundamental tool for analysis. The simulation can easily retain the original physical quantities, such as energy, particle type, or direction, in conjunction with the observables produced. The resulting data are essential for the development and use of essential methods like signal-cleaning or estimators to determine the original quantities. Without it, analysis of the observations would be almost impossible. As a typical example, labelled data sets can be generated to train machine learning methods for signal-background separation. In addition, basic detector characteristics like acceptance, efficiency, and resolution can be estimated.

The ultimate goal of the simulation is a complete description of the experimental setup, based on all measurements and theories known to date, to enable an exact representation of physical reality and, consequently, an almost complete reproduction of the measurement results. This last point is essential for the method's functioning: the observable parameters of simulation and measurement must be almost indistinguishable in large parts; otherwise, no conclusions can be drawn. However, due to the limitations of available computing power and the complexity of the system in question, losses in precision often have to be accepted.

## 3.1 Monte-Carlo-Simulation and Random Numbers

The goal of the simulation in the context of cosmic ray physics is the propagation of individual particles, starting at the top of the atmosphere, depending on the definition about $120\,\mathrm{km}$ height above sea level, down to the so-called observation level, the height where the experiment is physically placed. This simulation needs to include the reproduction of all physical processes that occur during the travel of the particles, such as the interaction with the atmosphere and the decay of the particles. The problem here is that most particle physics processes are probability-based, an example being the decay of the neutral pion $\pi^0$, which leads with over $98\,\%$ branching ratio to $\gamma + \gamma$, but may lead to $e^+ + e^- + \gamma$ or several other more exotic variations [76]. As a result, the classical analytical approach is not possible. Combining individual probability functions numerically and, in a final step, sampling from them to obtain statistical data at the observation level [61] is still feasible. However, this approach nonetheless lacks the ability to reproduce three-dimensional distributions and secondary effects such as Cherenkov or fluorescent light. To overcome the problems of the analytical methods, Monte Carlo methods can be used, which are based on the principle of repeated random sampling of the physical processes. The software tracks each particle through the atmosphere and applies all appropriate processes; if a statistical process is involved, a random number taken from corresponding theoretically defined distributions is used to calculate the result. Because of the random nature of the process, this type of simulation is called Monte-Carlo simulation, named after the Monte-Carlo Casino in Monaco [58]. With the large number of particles and interactions, the amount of random numbers required is equally high. To give an overview, the figures 2.3 shown in the previous chapter require the amount of individual random numbers listed in table 3.1.

| | 0° Zenith | |
|---|---|---|
| | 500 GeV | 10 TeV |
| Photon | $4.4 \times 10^7$ | $1.2 \times 10^8$ |
| Proton | $2.1 \times 10^7$ | $1.2 \times 10^8$ |
| Iron | $1.1 \times 10^7$ | $1.9 \times 10^8$ |

**Table 3.1:** Total amount of random number calls required for the simulation of Cascades displayed in Figure 2.3.

Due to the large number of random numbers required, as seen in the table 3.1, it is clear that the computational cost of generating these numbers can have a substantial influence on the overall runtime. A more detailed analysis of the calls shows that a large part, $>90\,\%$, comes from the Cherenkov functions, which will become relevant later. However, speed is not the only criterion; the quality of the random numbers is of equal or greater importance. The independence of each generated random number is one noteworthy factor of quality to avoid artefacts for multidimensional data, or else problematic behaviour similar to those visible in the spectral test [55] can occur in the data. Reproducibility also needs to be considered, as some simulations are terabytes in size, and available data centres are rarely able to store this for long periods of time. For the method of random number generation, abbreviated as RNG - **R**andom **N**umber **G**enerator, a whole set of necessary quality criteria can be formed from the requirements of a physically correct simulation:

- **Periodicity** — Most random generators have the property that the sequence of numbers generated will be repeated in whole or in part after a certain number of invocations, thus reaching a period. The length may be too small for very simple generators compared to the amount of required values.

- **Coverage** — The coverage of the complete phase space, within the limits of the available computer architecture, is essential. Otherwise, due to rigid coordinate systems and fixed conditions in the code, non-physical voids would occur in the results.

- **Statistical Independence** — The statistical independence of the following number generated from all its predecessors in the sequence is already given by the requirement for random numbers. Without this property, the basic algorithm fails. Several test programs allow the investigation of this property, such as the TestU01 suite [52] or the Dieharder framework [24], implementing various statistical tests.

- **Computational Cost** — The runtime is usually not of the highest relevance, but it has some importance due to the high number of calls. This includes not only the generation of each number but also the initialization of the generator at startup and fast-forwarding when loading from checkpoints.
- **Reproducibility** — For simulations of particular large events that cannot be stored permanently due to memory limitations, it must be possible to reproduce them at any time. The RNG must be able to reproduce the random sequence using storable values.

Regarding statistical independence, the highest quality random numbers are available through dedicated physical hardware, which provides an actual random data stream based on physical measurements. Due to the need to make physical measurements of an entropy source, the overall speed is comparatively low ($\approx$1 ms) and therefore not feasible for large scales. In addition, the entire random stream must be stored for playback, as there is no way to regenerate it. Currently, the best solution for simulation is the use of deterministic pseudo-random number generators based on a combination of mathematical functions. The algorithm uses a set of pre-defined numbers, called the seed and mutates these on each call to produce a new number based on the seed and the last number generated. This seed must be provided during initialization and allows for reproducibility.

Over the last few decades, several algorithms have been developed to provide high-quality random numbers, the most commonly used today being the Mersenne Twister [57], the Xorshift [56] and the Philox [66] algorithms. No absolute recommendation can be made as each algorithm has advantages and disadvantages.

CORSIKA 8 utilizes a Philox variant based on an extended “square” algorithm [10]. It is one of the so-called counter-based RNGs, which means that the seed is a simple number that is incremented with each call. This behaviour makes initialization and fast-forwarding trivial. Performance is excellent, and, as an added benefit, it can be easily parallelized if necessary without losing the deterministic sequence through parallel computing effects like race conditions. A “Race Condition” is an effect created by variation in the machine-codes execution order between threads caused by unequal execution speed (e.g. outside influence like OS or cache timings).

## 3.2 CORSIKA

CORSIKA [38] is one of the most established and de facto standard simulation tools for cosmic ray-induced atmospheric cascades. With its origins in late 1989 [45], its codebase now comprises over 78,000 lines of Fortran77 code with several large extensions, mostly written in C or C++. Unlike many other simulations, CORSIKA does not implement all physical models but uses several partly interchangeable interaction models or libraries specialized for different cascade parts.

A complete rewrite of the old code base in C++ is currently underway to provide a more modern and much more flexible infrastructure for the future, called CORSIKA 8 [11]. The fundamentally modular approach will be retained to allow each simulation to be tailored to the experiment's needs, thus reducing the computational complexity to the required level. For example, the activation of advanced secondary production such as fluorescence light, Cherenkov light, or radio emission slows down the simulation by a factor of about 10 for optical light and much more for radio emission (which is not part of this work), mainly due to the number of optical photons produced, which is several orders of magnitude higher than the number of particles in the standard particle simulation.

### 3.2.1 Basic Structure

The core algorithm of all previous versions of CORSIKA is based on an iterative approach, starting with the first particle that starts the cascade. The first particle is placed in a temporary storage buffer called the stack. For processing, the first particle is removed from the stack and propagated through the atmosphere until it reaches a certain distance, cut level, or interaction point. The particle gets removed if one of its attributes is below a cut level. If no interaction occurs, the simulation places the particle back at the end of the stack with updated values. In case of an interaction, the resulting new particles are placed at the end. The code now loops until there are no particles left. The actual physics in CORSIKA 8 is applied during the propagation step by individual interaction modules. All parts of the optical photon generation are implemented as so-called continuous loss functions, which, as the name implies, are used to apply non-discrete energy losses. In the case of Cherenkov light emission, which is generated along the entire particle path, the resulting energy losses are discarded because they are nonessential in the overall context. In contrast, fluorescence emission is generated directly by ionization losses, which are handled by other modules.

For both emission mechanisms, the exact path of the particle is critical because the emission is only generated along it. Path modifications such as magnetic field and multiple scattering must be considered throughout the propagation of an individual particle. With the usual endpoint formalism, where all changes are applied only at the end of the path, the maximum step length must be significantly shortened compared to the path length usually chosen. The reason for this is the available angular resolution of the detector, where the maximum path length must be in a single pixel.

With the planned CTA telescopes achieving an individual resolution of $0.067°$ per pixel [7], a conservative estimate for the lower limit can be determined with a maximum angular resolution of $0.05°$. The required spatial resolution of the simulation is displayed in figure 3.1 depending on the distances to the telescope's centre.

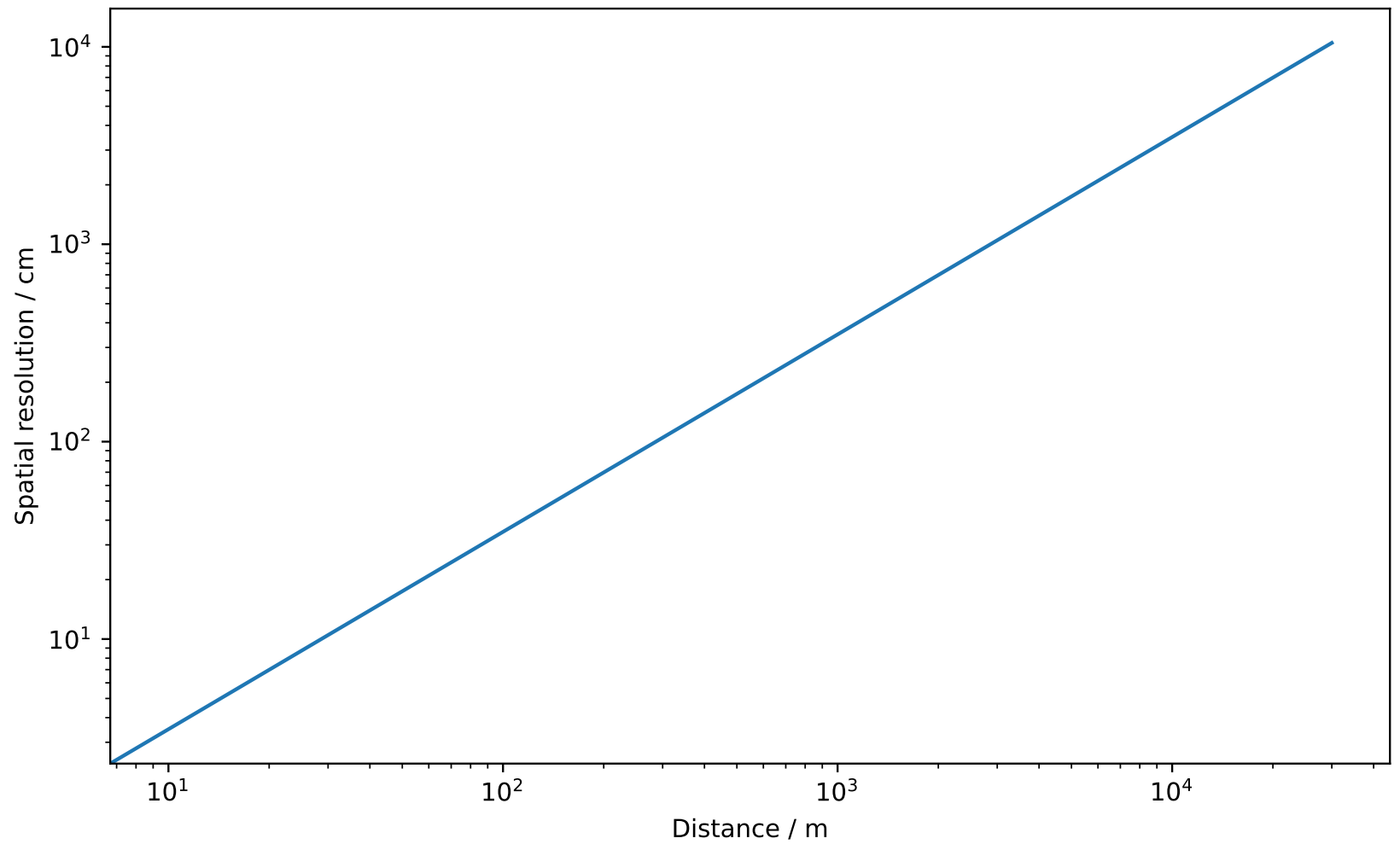

Figure 3.1: The blue line shows the maximum allowed deviation orthogonal to the telescope viewing direction. The simulated segmented particle trajectory must have a deviation from the real trajectory less than this value to be below the resolution limit of a telescope. The lines correspond to a maximum angular resolution of $0.05°$.

## 3.3 Atmosphere Model

To be able to run simulation code in an acceptable time, several simplifications of the realistic world are necessary. One of these is using a modelled atmosphere based on actual measurements. The naturally occurring atmosphere shows an incredibly high complexity where local weather phenomena, air pressure variation due to the topographic layout, and time variation are only a few effects that otherwise would need to be represented. Several distinct atmospheric models have been developed for CORSIKA to cover a wide assortment of experiments. Each uses different degrees of complexity reduction and enables the fast calculation of the available parameters. These models also create the basis for the emission and propagation of Cherenkov light. For Cherenkov, in particular, their role is to provide the location-dependent refractive index for determining the number of photons emitted and the deflection of the photon along its path. It also provides density and composition data for the fluorescence and interaction methods that process the cascade.

With various processes requiring a variety of data with different levels of accuracy, there is no single solution that fits all criteria. With a focus on optical light propagation, only the methods for accessing the appropriate subset of environmental information required to describe the effects mentioned above are presented here. Four modules (e.g. Homogeneous, Planar, Cylindrical, Spherical) to access atmospheric data were developed and tested. Fundamentally, these four use the CORSIKA 8 environment framework, allowing users to combine and define volumes where their respective model applies. The simplest example would be the combination of two spheres, one displaying the earth and the second surrounding it with the atmospheric informations. Analysis and implementation details can be found in section 4.1.

Each model can only use a single, in itself flexible, tabulated refractive index curve as its core. In the case of large experimental sites where the atmospheric conditions vary and several refractive index profiles are available for different spatial positions, the average must be taken. At present, no current or planned experiment requires this level of detail, but it is a possible future extension. The downside of allowing this degree of atmospheric variation is the loss of all symmetric characteristics, which makes most of the currently implemented algorithm-based optimizations impossible.

### 3.3.1 Homogeneous Atmsphere

The simplest atmosphere model available is the homogeneous atmosphere model, which assumes a constant refractive index for all positions and altitudes. It discards physical reality for the default use case of CORSIKA, atmospheric cascades spanning several kilometres in size. However, it is still useable for test studies or the injection of particles close to the detection site. In addition, the simulation of showers in near homogeneous media such as water is another use case, even more so through the possibility of stacking multiple different media and models. Outside the typical simulation, it is used as a medium for unit testing because of its predictable nature. The calculation of refractive index equations is only required for boundary changes where the photon transfers from one medium to the next.

### 3.3.2 Planar Atmosphere Model

The first structurally slightly more complex model, the planar or layered Atmosphere model, is shown in Figure 3.2a. This model assumes that the atmosphere is homogeneous in the horizontal direction and only varies in height. The assumption is good considering the dimension of the critical cascade part for optical emission is roughly $\approx$50 km in length, compared to the earth radius of 6370 km. Even at 100 km distance, the vertical error corresponds to less than 5 % (1.57 km) of the cascade size.

The planar model is the preferred method for cascades close to vertical but works up to $45°$ with only minor deviations. Since the refractive index gradient always runs in the same direction, as a general principle parallel to the vertical axis, the deflection calculation can occur almost entirely in 2D space. In addition, several optimizations are possible because the photon never leaves the plane spanned by the initial direction and the gradient of the refractive index.

### 3.3.3 Cylindrical Atmosphere Model

The Cylindrical methods take another degree of freedom into account in addition to the height dependency of the planar model, the horizontal distance $d$ from the coordinate systems origin (usually the centre of the experiment) to the emission point, with $d^2 = x^2 + y^2$, is added. Afterwards, the refractive index gradient only changes in its Zenith angle $\theta$ where the rotation angle $\phi$ is kept constant, as displayed in figure 3.2b This still results in some level of optimization potential; but not as much as the planar model. The result is a highly accurate tracking up to the highest inclinations possible, whereas the planar model would not give accurate results. The limiting factor is its loss in accuracy for photon impact positions far from the centre of the coordinate system, where the gradient will have a substantial rotation angle that cannot be presented in this model. Fortunately, this angle scales with the distance from the earth's centre, which results in significant impact only for experiments with an instrumented area of several tenths of kilometres in size. Very large-scale experiments like the Pierre Auger Observatory [9] or the planned Telescope Array extension [67] will need to do tests if the introduced error is inside their resolution limits and can be accepted.

### 3.3.4 Spherical Atmosphere Model

The spherical model differs from the Cylindrical model in how many degrees of freedom the refractive index gradient has per photon. Compared to only one angle per photon in the Cylindrical model, the spherical model requires an additional change of the refractive index gradient rotation angle in every step. Figure 3.2c displays the atmosphere profile.

The spherical model has the additional advantage that, by treating all three spatial coordinates separately, even large detectors, several kilometres in size, where distances of more than 10 km from the coordinate origin are possible, can be treated accurately. This is where all other approximations fail. As the number of degrees of freedom increases, so does the resulting computational complexity.

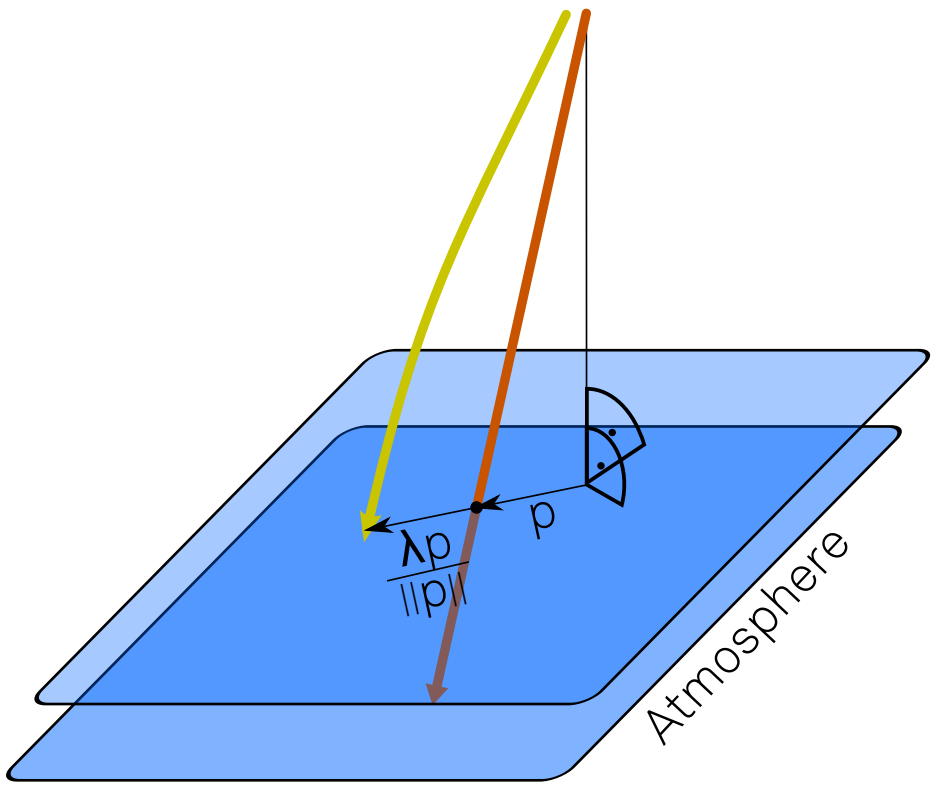

**a:** Planar

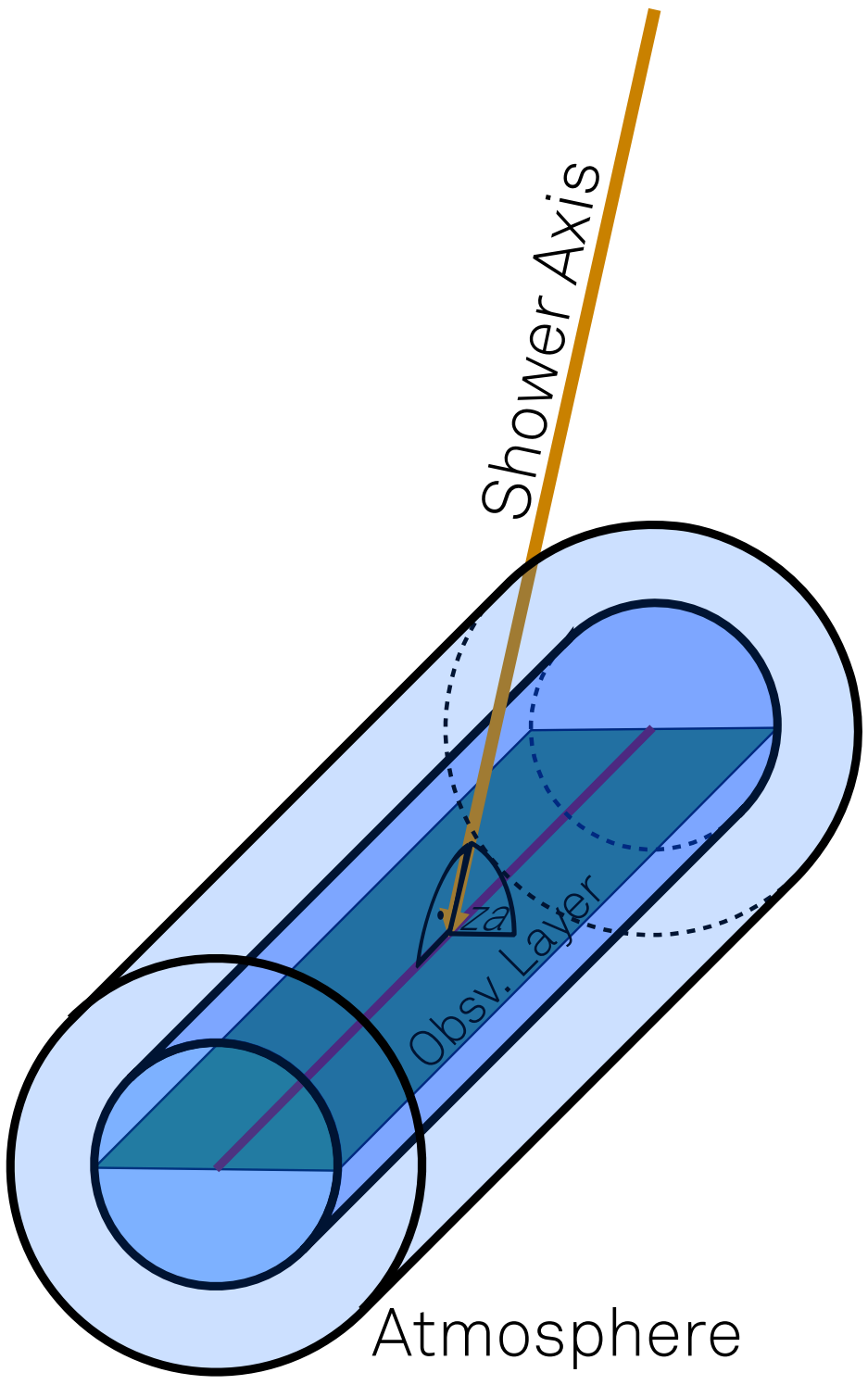

**b:** Cylindrical

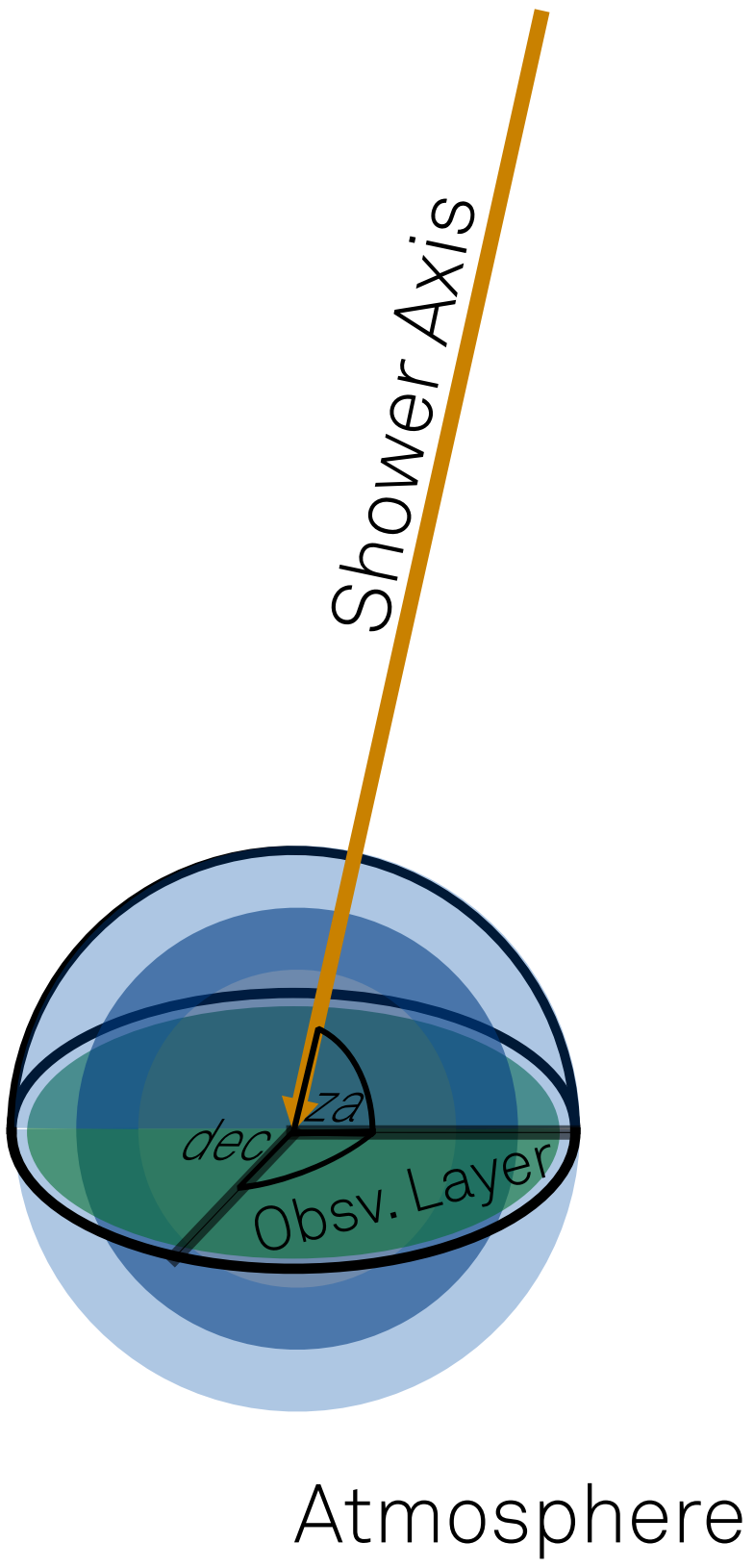

**c:** Spherical

**Figure 3.2:** Schematic representation of different atmospheric models with varying degrees of complexity.

## 3.4 Atmosphere Tracking Methods

Photons generated by CORSIKA are not stored at the point of emission but are treated as particles and propagated through the atmosphere to avoid re-implementation of this shared necessity. The code is flexible enough to suit a wide range of experiments, from classical ground-based telescopes to balloon or satellite experiments. Once the photons have reached an area close to the experimental site, defined as an observation plane (or other geometric bodies), they can be stored or passed on to experiment-specific code, where high-precision ray tracing can be used through the detector hardware. There are several different approaches to this atmospheric propagation, but the following section focuses on three different implementations (3.4.1, 3.4.2, 3.4.3), each with individual advantages and disadvantages. They are the necessary precursors of a fourth method used later, combining many of their positive features and explained in the later section 4.4.1.

### 3.4.1 Straight Line

By far the fastest approach, by several orders of magnitude in terms of computational time and complexity, is simple straight-line propagation, where all atmospheric effects, such as density and refractive index changes, are entirely neglected. This approach sacrifices physical accuracy for all angled photons above a few degrees of inclination, as shown in figure 3.3, where the linear propagation is plotted against the physically correct one.

The input to the tracking algorithm consists only of a starting point $\vec{A}$ and an emission direction $\vec{u}$ of the photon, calculated according to 4.2. The intersection point $\vec{I}$ with an observation plane ($\vec{N} \cdot \vec{x} = \vec{p} \cdot \vec{n}$) is generally calculated by the following set of equations:

$$\vec{A} = (A_1, A_2, A_3)^\top, \qquad \vec{u} = (u_1, u_2, u_3)^\top$$
$$\vec{N} = (N_1, N_2, N_3)^\top, \qquad \vec{p} = (p_1, p_2, p_3)^\top$$

$$l = n \cdot u \tag{3.1}$$

$$\vec{I} = \begin{cases} \frac{\vec{N}\cdot(\vec{A}-\vec{p})}{l} \cdot \vec{u} + A & |l| \leq \epsilon \\ (\infty, \infty, \infty)^\top & \text{else} \end{cases} \tag{3.2}$$

with $\epsilon \to 0$ limited through computer numerical resolution

Due to the very frequent use of the straight line approximation of at least once (e.g. 4.4.1) per photon, with millions of photons per cascade, the reduction of each arithmetic operation can lead to improvements in the overall throughput. Therefore, additional constraints, in this case for ground-based experiments, can be applied:

- The normal of the observation plane is parallel to the Z-axis of the orthonormal basis. $\vec{n} = (0, 0, \pm 1)^\top$.
- Any horizontal photons have been removed beforehand, resulting in $\forall u : u_3 \neq 0$.

This leads to the following equation, which removes most of the complexity:

$$\vec{I} = \frac{A_3}{u_3} \cdot \vec{u} + A \tag{3.3}$$

### 3.4.2 Numerical Approach

A physically much better approach, which considers the refractive index gradient, approximates the natural atmosphere by discretising the continuous curve and dividing it into homogeneous planes. Only using Fermat's principle at the plane boundaries avoids the complex consideration of bending the light curve through a continuously varying medium. This approach can treat nearly all possible atmosphere configurations and is a secure fallback solution if other possibilities do not work. The algorithm uses the classical solution for discrete transitions on the layer boundaries: Snellius' law of refraction.

$$n_1 \cdot \sin(\alpha_1) = n_2 \cdot \sin(\alpha_2) \tag{3.4}$$

$$\tag{3.5}$$

In the limit case of infinitesimally thin layers $\lim_{d \to \infty}$, the propagation should thus approximate a continuous method despite this straightforward approximation. However, for the actual calculation on standard computer hardware, the classical step-by-step implementation via rotation matrices quickly becomes computationally intensive and, even more problematically, numerically unstable since the refractive index difference approaches zero for too small a layer thickness. Therefore, the best approach is to reshape this formula to directly calculate the photon pointing vector instead of rotating it via $\Delta\alpha$ around a potentially continuously changing axis. An implementation is

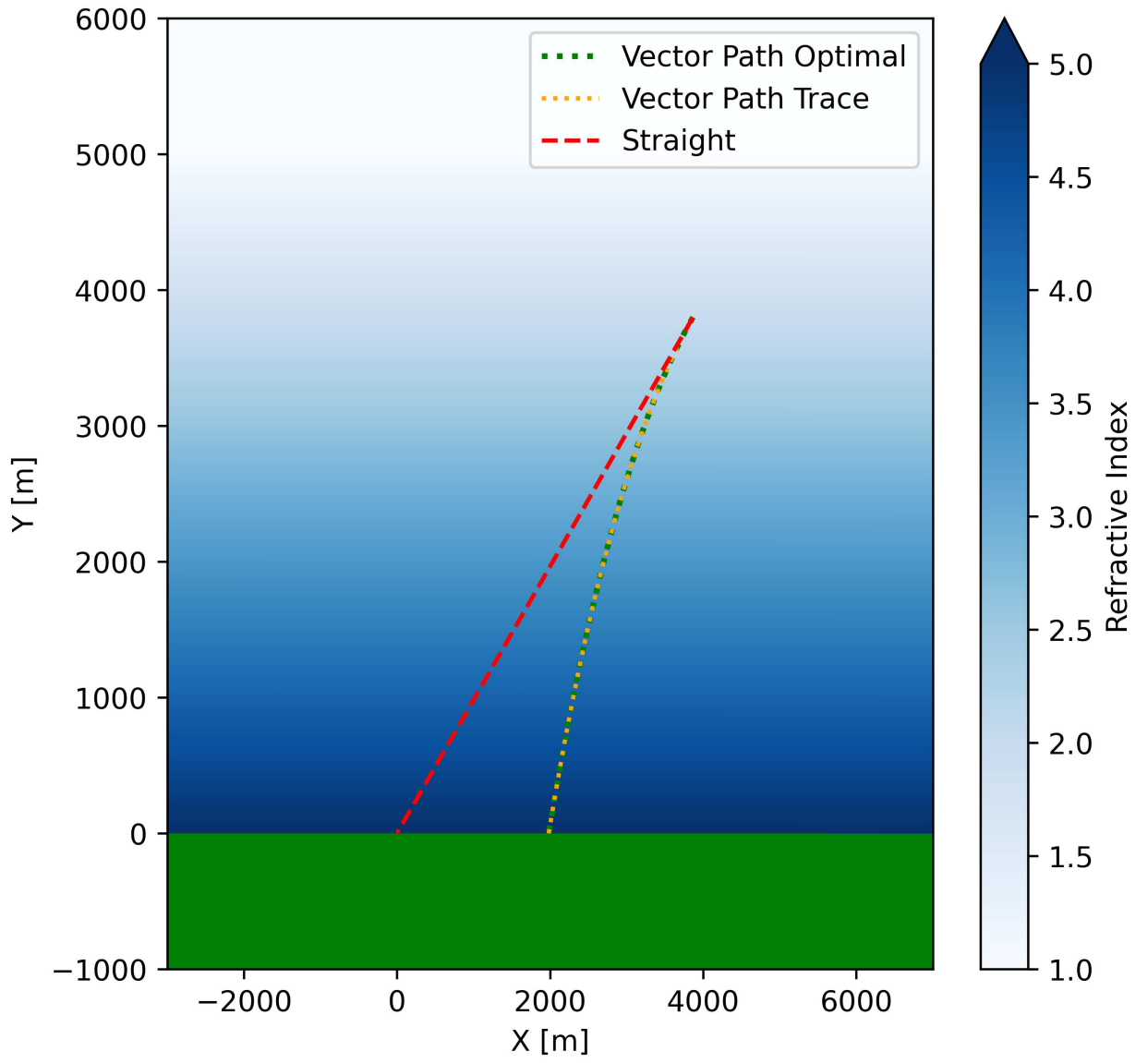

**a:** Comparison of the propagation with the physical method compared to the straight line method for photons with a high inclination angle. As a reference, the earth's surface is displayed.

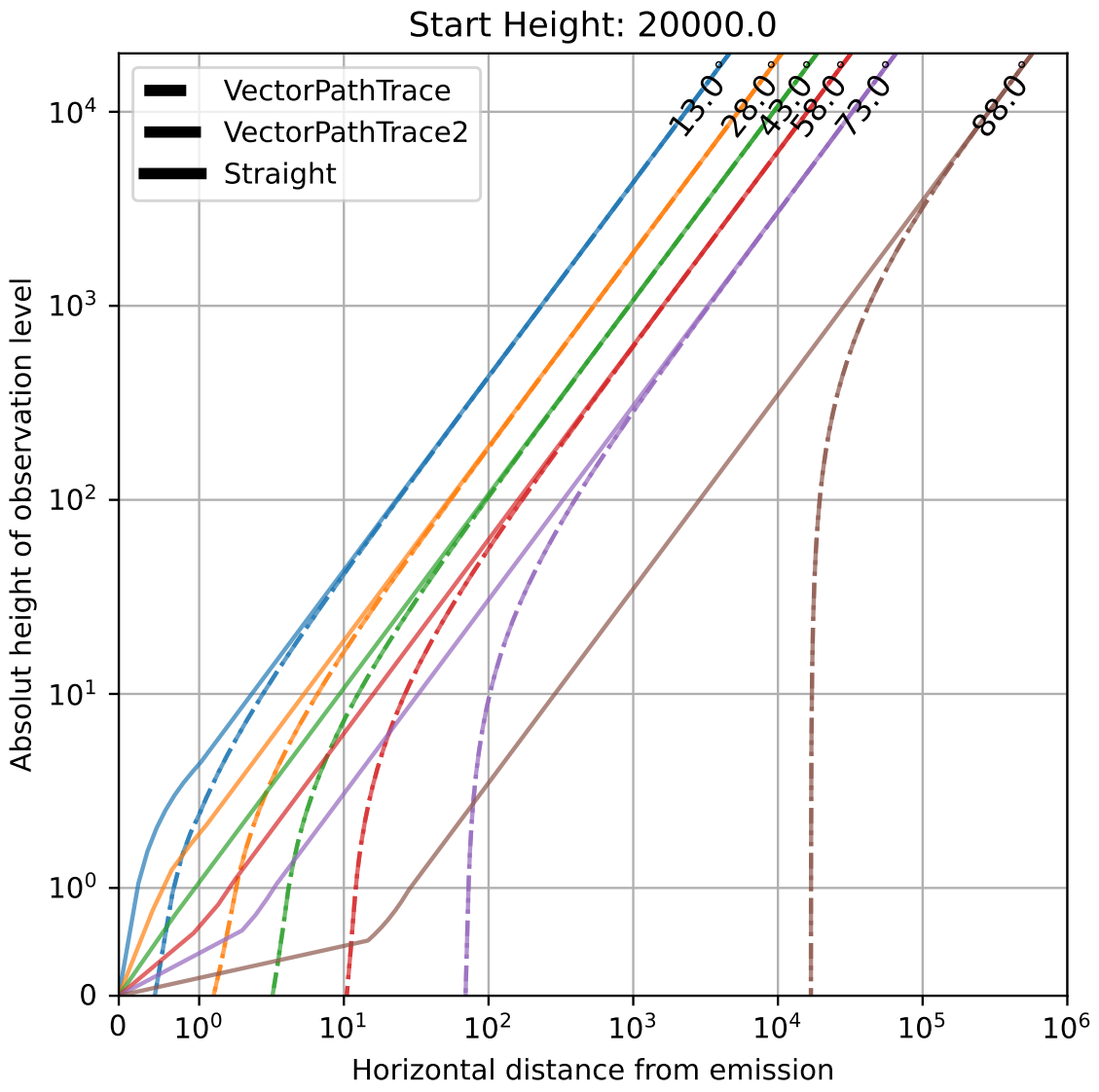

**b:** Deviation of the linear propagation from physics-based methods.

**Figure 3.3:** Displayed is the simple linear propagation and the result of the physical correct propagation through the atmosphere. The logarithmic axes are changed to a linear one in the range from 0 to 1 to include the origin.

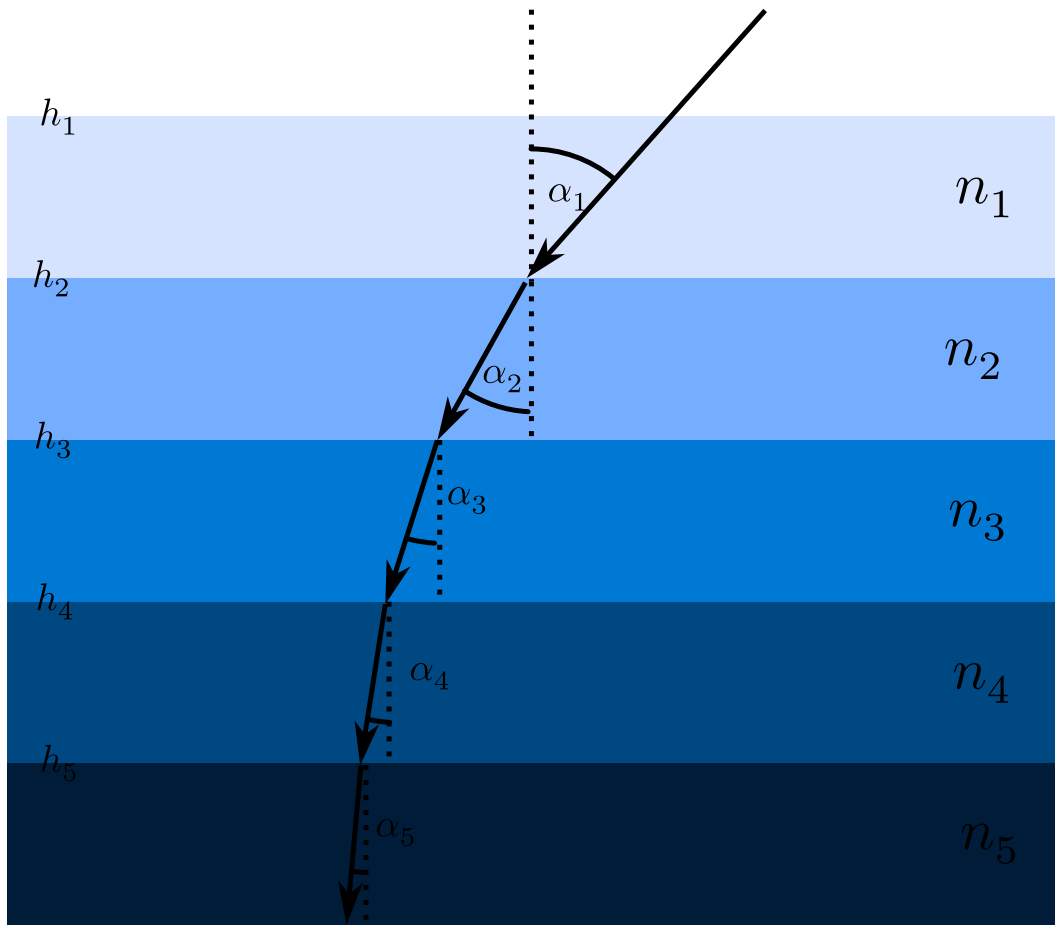

**Figure 3.4:** Schematic representation of layered atmosphere model calculated with individual application of Snell's law.

possible using Fresnel's law in vector notation [75]:

$$n_1\left(\vec{A}_1 \times \vec{N}\right) = n_2\left(\vec{A}_2 \times \vec{N}\right) \tag{3.6}$$

$$\text{with } |r| = |n| = 1$$

$$\text{with optical length } \vec{A}'_i = n_i \vec{A}_i$$

$$\vec{A}'_1 \times \vec{N} = \vec{A}'_2 \times \vec{N}$$

$$\left(\vec{A}'_1 - \vec{A}'_2\right) \times \vec{N} = 0 \tag{3.7}$$

This shows that $\vec{A}'_1 - \vec{A}'_2$ needs to be parallel to $\vec{N}$, therefore:

$$\vec{A}'_1 - \vec{A}'_2 = c\vec{N} \tag{3.8}$$

$$\text{multiply with } \vec{N} \text{ on both sides}$$

$$\left(\vec{A}'_1 - \vec{A}'_2\right) \cdot \vec{N} = c\vec{N} \cdot \vec{N} = c * |\vec{N}|^2 = c$$

$$\vec{A}'_1 \cdot \vec{N} - \vec{A}'_2 \cdot \vec{N} = c \tag{3.9}$$

$$\text{use the geometric dot-product definition}$$

$$|\vec{A}'_1||\vec{N}|\cos(\alpha_1) - |\vec{A}'_2||\vec{N}|\cos(\alpha_2) = c$$

$$n_1\cos(\alpha_1) - n_2\cos(\alpha_2) = c \tag{3.10}$$

$$\tag{3.11}$$

With $n\cos(\alpha) = n\sqrt{1-\sin(\alpha)^2}$, $n\sin(\alpha) = n\sqrt{1-\cos(\alpha)^2}$ and the "original" Fresnel Law from above the second part of the formula with the unknown $\alpha_2$ can be

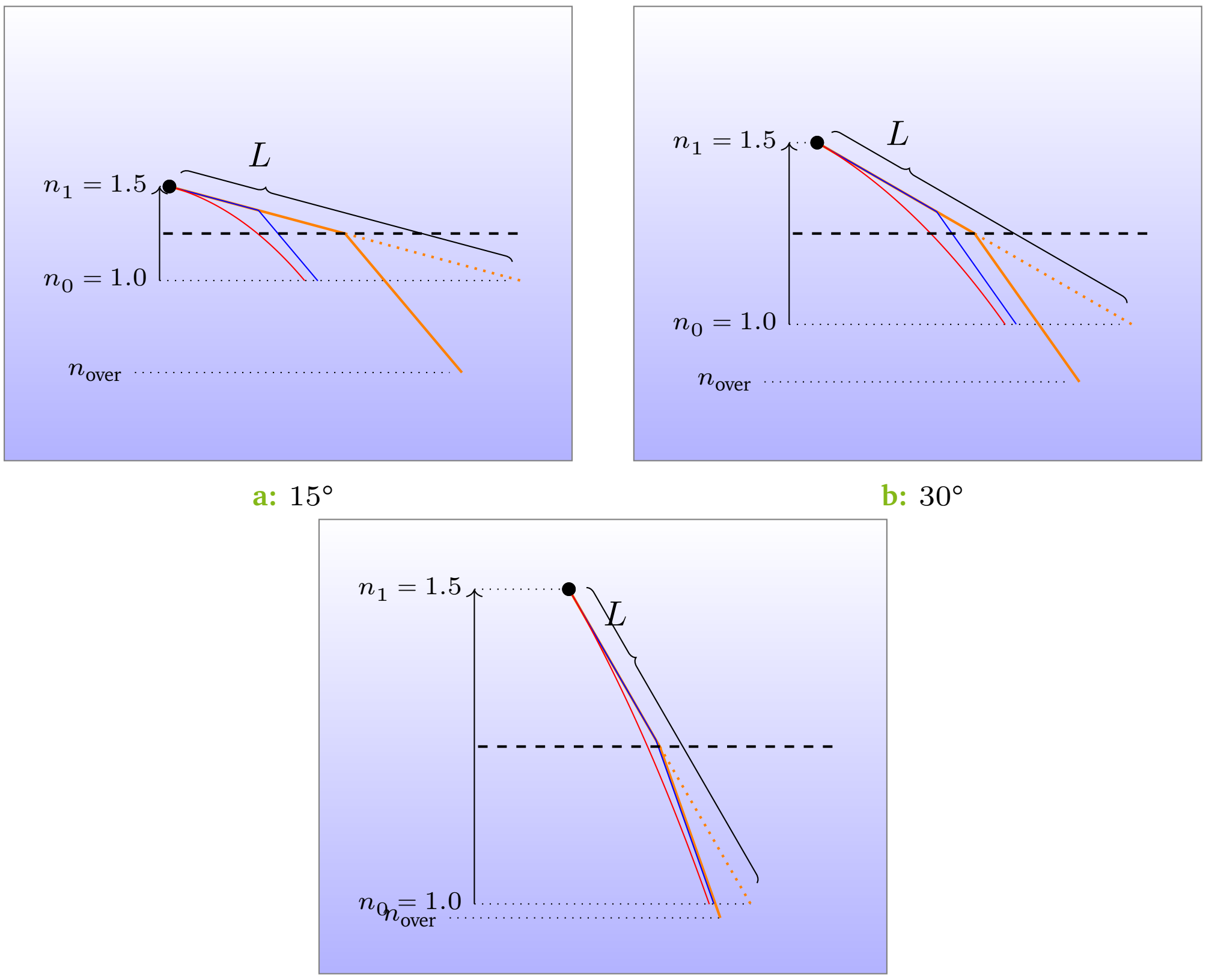

**a:** 15°

**b:** 30°

**c:** 60°

**Figure 3.5:** Displayed is the microscopic view of the numerical approach for propagating photons through the atmosphere. The steplength $L$ is drawn as the straight dashed line without refraction applied. The orange solid line is calculated according to Snell's law with the layer boundary in the centre; the blue line is based on the same calculation but rescaled to end at the correct atmospheric depth. The red curve is the actual photon path from analytical calculations.

eliminated:

$$n_1 \cos(\alpha_1) - \sqrt{n_2^2 - n_2^2 \sin(\alpha_2)^2} = c \tag{3.12}$$

$$n_1 \cos(\alpha_1) - \sqrt{n_2^2 - (n_2 \sin(\alpha_2))^2} = c$$

$$n_1 \cos(\alpha_1) - \sqrt{n_2^2 - (n_1 \sin(\alpha_1))^2} = c$$

$$n_1 \cos(\alpha_1) - \sqrt{n_2^2 - \left(n_1 \sqrt{1 - \cos(\alpha_1)^2}\right)^2} = c$$

$$n_1 \cos(\alpha_1) - \sqrt{n_2^2 - \left(\sqrt{n_1^2 - n_1^2 - \cos(\alpha_1)^2}\right)^2} = c$$

$$n_1 \cos(\alpha_1) - \sqrt{n_2^2 - n_1^2 + n_1^2 \cos(\alpha_1)^2} = c \tag{3.13}$$

$$\tag{3.14}$$

This results in a simple formula for $A_2'$:

$$A_2' = A_1' + cN$$

$$A_2' = A_1' + -\left(n_1 \cos(\alpha_1) - \sqrt{n_2^2 - n_1^2 + n_1^2 \cos(\alpha_1)^2}\right) N$$

$$A_2' = A_1' + \left(\sqrt{n_2^2 - n_1^2 + n_1^2 \cos(\alpha_1)^2} - n_1 \cos(\alpha_1)\right) N \tag{3.15}$$

Which can be utilized much faster and with less numerical error than recalculating rotation matrices for every step. In figure 3.5, a single step of this formula is displayed for a linear atmosphere. The figure shows that the method underestimates the actual path deviation, at least for the linear atmosphere where an analytic solution exists.

### 3.4.3 Analytic Approach

A closed formula is indispensable to verify the accuracy of the numerical approximation; several different approaches can be utilized to obtain the analytical solution of the photon path in varying media. The original technique from which Snells' law was derived is based on Fermat's principle of minimising the optical path length.:

$$\int_{p_{\text{start}}}^{p_{\text{end}}} = n\left(x, y, z\right) \mathrm{d}s\,. \tag{3.16}$$

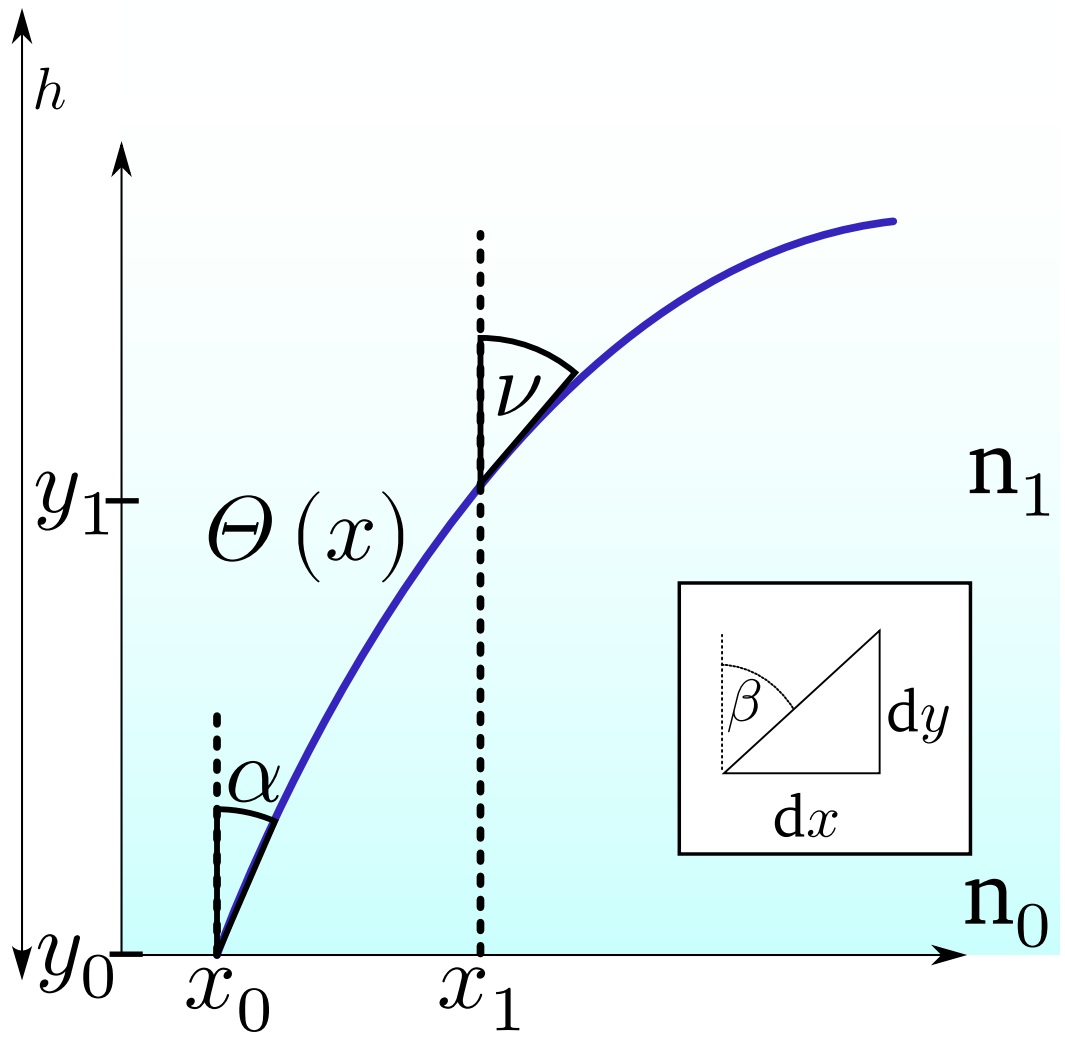

**Figure 3.6:** Schematic display of all variables needed for the analytic refraction model in different atmospheres.

After transforming the formula into a second-order differential equation, it is possible to solve it.

Another approach pursued here reverses the derivation by starting with Snells' law, similar to the numerical algorithm described in the previous section 3.4.2. Instead of converging the layer height to 0, the steplength in flight direction is minimized to an infinitesimally small length. The resulting differential form of Snell's law can then be integrated over the flight path to calculate individual positions.

Determining a well-defined formula with arbitrary atmospheric profiles is not easily possible, resulting in picking only specific profiles that lead to a solution. The derivation uses the classical formula of the angular deflection according to Snellius at the position $x$ and a point infinitesimally $x + \mathrm{d}x$ far away.

$$n(\vec{x}) \cdot \sin(\Theta(\vec{x})) = \text{const.} \tag{3.17}$$

reduce to 2D problem

$$n(x) \cdot \sin(\Theta(x)) = n(x + \mathrm{d}x) \cdot \sin(\Theta(x + \mathrm{d}x))$$

$$\text{with} f(x+y) = f(x) + \int_x^y \frac{\mathrm{d}f(z)}{\mathrm{d}z}\,\mathrm{d}z$$

$$= \left(n + \frac{\mathrm{d}n}{\mathrm{d}x}\,\mathrm{d}x\right)\left(\sin(\Theta) + \cos(\Theta)\frac{\mathrm{d}\Theta}{\mathrm{d}x}\,\mathrm{d}x\right)\mathrm{d}x$$

$$= n\sin(\Theta) + \left(\frac{\mathrm{d}n}{\mathrm{d}x}\sin(\Theta) + n\cos(\Theta)\frac{\mathrm{d}\Theta}{\mathrm{d}x}\right)$$

$$0 = \left(\frac{\mathrm{d}n}{\mathrm{d}x}\sin(\Theta) + n\cos(\Theta)\frac{\mathrm{d}\Theta}{\mathrm{d}x}\right)\mathrm{d}x$$

$$\Rightarrow \frac{\cos(\Theta)}{\sin(\Theta)}\,\mathrm{d}\Theta = \cot(\Theta)\,\mathrm{d}\Theta = -\frac{1}{n}\,\mathrm{d}n \tag{3.18}$$

### Linear Refractive Index Profile

With a linearly interpolated atmosphere profile between $(h_1, n_1)$ and $(h_2, n_2)$: $n(h) = n_2 - \frac{y-h_1}{h_2-h_1} \cdot (n_2 - n_1)$:

$$\frac{\mathrm{d}n}{n} = \frac{\mathrm{d}n}{n_2 - \frac{(y-h_1)}{h_2-h_1} \cdot (n_2 - n_1)}$$

$$= \frac{\mathrm{d}n}{n_2 - \frac{y}{h_2-h_1} \cdot (n_2 - n_1)}$$

$$= \frac{\mathrm{d}n}{n_2 - \frac{n_2-n_1}{h_2-h_1} \cdot y}$$

$$\text{with } \frac{n_2 - n_1}{h_2 - h_1} = c_1$$

$$= \frac{\mathrm{d}n}{n_2 - c_1 y}$$

$$\text{with } \frac{\mathrm{d}n\,\mathrm{d}y}{\mathrm{d}y} = -c_1\,\mathrm{d}y$$

$$= \frac{1}{n_2 - c_1 y}(-c_1)\,\mathrm{d}y$$

$$= -\frac{\mathrm{d}y}{\frac{n_2}{c_1} - y} \tag{3.19}$$

Now formula 3.19 and 3.18 can be combined and integrated:

$$\int_{\alpha}^{\nu} \cot(\Theta)\, \mathrm{d}\Theta = \int_{0}^{y} \frac{\mathrm{d}y}{\frac{n_2}{c_1} - y}$$

$$\text{with } c = \frac{n_2}{c_1} = \frac{h_2 n_2 - h_1 n_2}{n_2 - n_1}$$

$$\Big[\ln(\sin(\Theta))\Big|_{\alpha}^{\nu} = -\Big[\ln(c-y)\Big|_{0}^{y}$$

$$\ln(\sin(\nu)) - \ln(\sin(\alpha)) = -\left(\ln(c-y) - \ln(c)\right)$$

$$\text{with } \ln(a) - \ln(b) = \ln\left(\frac{a}{b}\right)$$

$$\frac{\sin(\nu)}{\sin(\alpha)} = \frac{c}{c-y}$$

$$\Rightarrow y = c \cdot \left(1 - \frac{\sin(\alpha)}{\sin(\nu)}\right) \text{ or } \Rightarrow \sin(\nu) = \frac{c \cdot \sin(\alpha)}{c-y} \tag{3.20}$$

To transform this in a $x/y$-coordinate system the trigonometric identity $\cot(\beta) = \mathrm{d}y/\mathrm{d}x$ in combination with $\sin(\beta) = 1/\sqrt{1+\cot(\beta)^2}$ can be used.

$$\sin(\nu) = \frac{1}{\sqrt{1+\cot(\nu)^2}}$$

$$= \frac{1}{\sqrt{1+\left(\frac{\mathrm{d}y}{\mathrm{d}x}\right)^2}}$$

$$\Rightarrow \frac{1}{\sqrt{1+\left(\frac{\mathrm{d}y}{\mathrm{d}x}\right)^2}} = \frac{c \cdot \sin(\alpha)}{c-y}$$

$$\Rightarrow \left(\frac{dy}{dx}\right)^2 = \frac{c^2 - 2cy + y^2 - \sin(\alpha)^2 c^2}{\sin(\alpha)^2 c^2}$$

$$= \frac{(c-y)^2 - \sin(\alpha)^2 c^2}{\sin(\alpha) c}$$

$$\text{with } \sin(\alpha) \cdot c = f$$

$$\left(\frac{dy}{dx}\right)^2 = \frac{(c-y)^2 - f^2}{f^2} \tag{3.21}$$

$$\tag{3.22}$$

This first-order nonlinear differential equation 3.21 can be solved by the method of separation of variables and integration. See Appendix A.1.1 for more information.

$$
\begin{aligned}
\int \frac{1}{f}\mathrm{d}x &= \int \frac{\mathrm{d}y}{\sqrt{(c-y)^2 - f^2}} \\
\frac{x}{f} &= -\log\left(\frac{\sqrt{(c-y)^2 - f^2} + c - y}{f}\right) + \text{const.} \\
&\text{Inertial condition x(0) = 0} \\
&= -\log\left(\frac{\sqrt{(c-y)^2 - f^2} + c - y}{f}\right) + \log\left(\frac{\sqrt{c^2 - f^2} - c}{f}\right)
\end{aligned}
\tag{3.23}
$$

# 4 Methods and Implementation

Quantitative values of runtime measurements given in this chapter were obtained on the DGX A100 computer system, which has the following components:

| Component | Specification |
|---|---|
| CPU | 2x AMD EPYC 7742 64-Core Processor |
| GPU | 8x NVIDIA A100-SXM4-40GB ([28]) |
| RAM | 1TB DDR4 |

Table 4.1: Hardware specifications of the DGX A100 system

The measurements were performed on an otherwise idle machine on a single CPU core, unless otherwise stated, and are generally comparable. The software used was built with hardware-based optimization, using SIMD (Single Instruction Multiple Data) standards from SSE up to AVX512 [29] to cover a wide range of currently existing hardware without having to expect massive differences in performance through hardware changes.

## 4.1 Atmosphere

Most of the Atmosphere data available does not come as a continuous function as required for a simulation where photons can be placed at arbitrary positions. For use in one of the four atmospheric models described in 3.3, the refractive index curve must be calculated from the environmental data collected at the experiment. As the most general implementation, the simulation expert provides the height-dependent refractive index data as a human-readable tabulate file with arbitrary binning. If this format is unavailable, Sellmeier's formula [71] can be used to deduce the refractive index from pressure, temperature, and humidity.

The step of calculating intermediate data points between given supports, commonly called interpolation, is a wide field with several algorithms present. As a prerequisite, some assumptions can be made:

- With its natural origin, the function will be continuous.
- The function will be smooth, i.e., the first and second derivatives will be continuous.
- The refractive index will be monotonically decreasing with increasing height.
- The basic function will be close to an exponential with deviations from diverse causes (e.g. atmospheric layers).

Besides simply choosing the nearest support point, one of the most common techniques is the linear interpolation between neighboring supports. A more complex method is the spline interpolation [70], which fits several support points. It consists of a piece-wise defined polynomial function of degrees one (slinear), two (quadratic), or three (cubic). Based on the knowledge of the theoretical course of the refractive index curve, the use of an exponential interpolation between the supports also seems reasonable, as confirmed by the later analysis.

$$n_{\text{linear}}(h) = n_i + \frac{n_{i+1} - n_i}{h_{i+1} - h_i} \cdot (h - h_i) \tag{4.1}$$

$$n_{\text{exponential}}(h) = n_i \cdot \exp\left(\frac{\ln(n_{i+1}) - \ln(n_i)}{h_{i+1} - h_i} \cdot (h - h_i)\right) \tag{4.2}$$

The CORSIKA 7 refractive index tables were interpolated and plotted to test the different interpolation methods, as seen in figure B.1. As an additional check, because the original values between data points are unknown, the data sparsity was increased artificially by removing every second element and checking the interpolated results against the data. With sparser data and inhomogeneous binning, the spline interpolation showed oscillating, often called ringing, which can create unphysical results depending on the displayed degree of deflection.

The relative difference between two consecutive data points was calculated and plotted in figure 4.1 as an additional check. This method is quite similar to the first derivatives of the resulting function. The target is to check visually for excessive changes between neighboring values, which could lead to artifacts in the final simulation. By comparing the overall quality, the exponential interpolation showed the best results in derivative smoothness and deviation from the data with fewer samples.

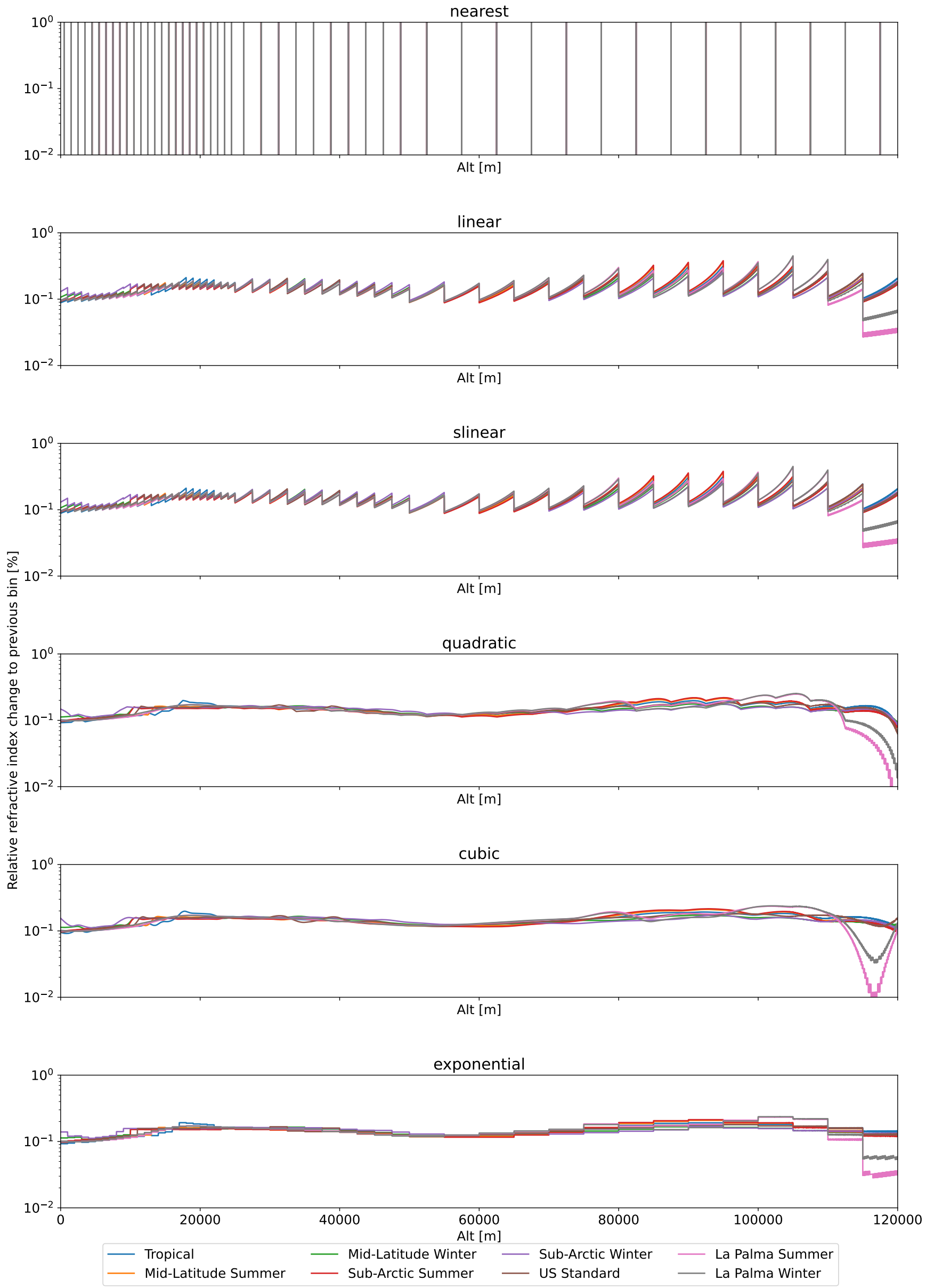

**Figure 4.1:** Displayed is the relative difference between successive data points, which is used to check for excessive changes between neighboring values. An unsmooth function can lead to artifacts in the final simulation. In combination with figure 3.3 this plot can be used to determine the best interpolation method.

To reduce the runtime CORSIKA uses a combination of interpolation methods. Before the simulation starts one of the more complex procedures is used to upscale the refractive index table to a resolution of 1 m. During the simulation, the fast linear interpolation provides accurate results with the high number of underlying support points. Adding more supports can increase the precision later if demanded, but with some diminishing returns (the increase in precision does not scale linearly with the number of supports). The interpolation can be done in constant time, regardless of the table size, by directly mapping the height to the table's indices. An example implementation is shown in listing 4.1. An essential precondition is that the program can not sample the refractive index outside the defined volume, and its boundaries are matched to the tabulated data; otherwise, the algorithm requires additional checks to avoid invalid memory access.

```
float interpolate(float h)
{
    const int idx = floor(h / step_size)
    const float frac = h / step_size - idx;
    return (1 - frac) * data[idx] + frac * data[idx + 1];
}
```

## 4.2 Photon Generation

The photon generation is strictly geometry-driven and has little optimization potential from an algorithm point of view. The three individual steps are represented schematically in figure 4.2, where subfigure 4.2a (first step) shows the generation of photons on a cone with the individual opening angle $\Theta_c$ in the coordinate frame's origin with downwards facing direction. The photon-generating particle processed contains, combined with the refractive index, the data (e.g. velocity) mandatory to calculate the direction as well as the number of photons as described in section 2.2.1. Afterwards, the photons are randomly distributed along the cone by sampling a number $c_{\text{rng}}$ from the interval $[0, 2\pi]$ and calculating the direction vector $\vec{p}$ via the following equation:

$$\vec{p} = \begin{pmatrix} \cos(c_{\text{rng}}) \cdot \sin(\Theta_c) \\ \sin(c_{\text{rng}}) \cdot \sin(\Theta_c) \\ -\cos(\Theta_c) \end{pmatrix}$$

The second step (subfigure 4.2b) rotates the individual photons via matrix multiplication into the particle's flight direction. Figure 4.3 displays the names of all required dimensions for a better understanding.

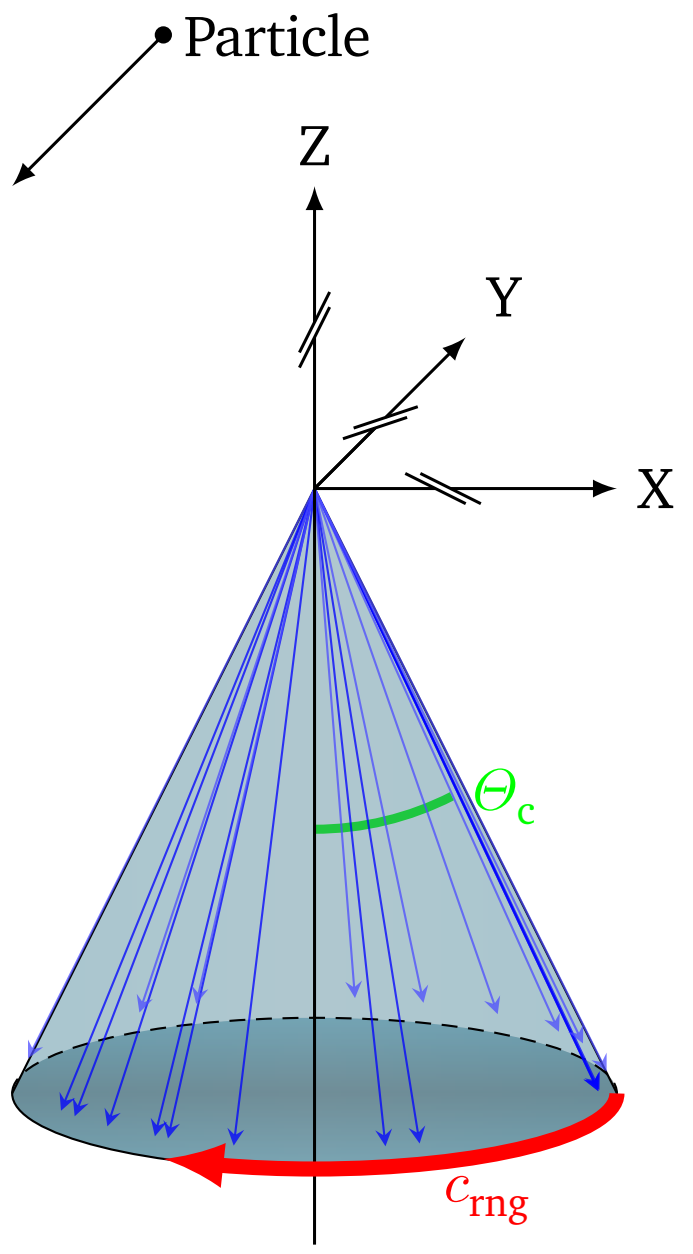

**a:** Photon generation on a cone

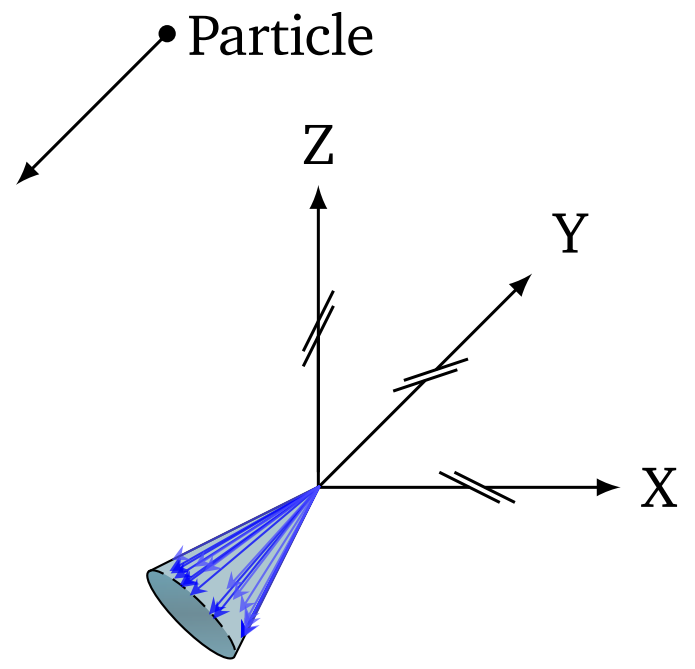

**b:** Rotate into particle flight direction.

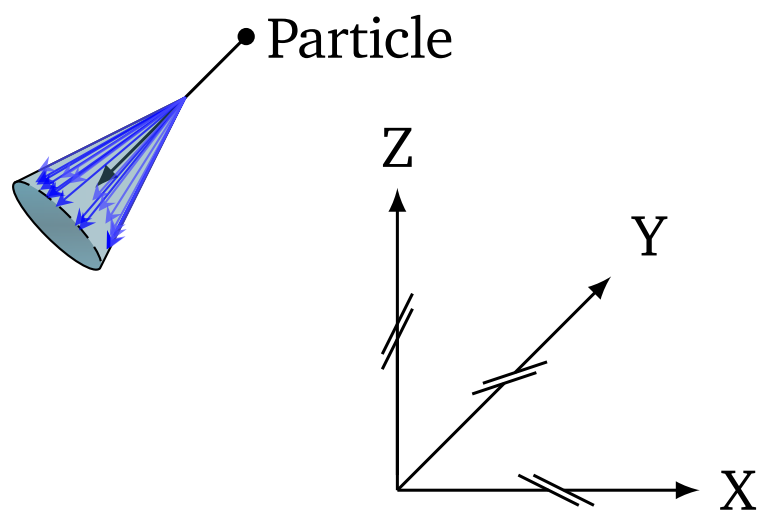

**c:** Translate to position along the particle track.

**Figure 4.2:** Schematic representation of the individual step for Cherenkov photon generation

The central axis of the photon cone, $D$, needs to be transformed to the particle flight direction $P$. The fact that the photon direction is randomly sampled and that the transformation does not need to keep the angle around the axis allows for several methods of rotation matrix generation that can be optimized for the computation speed of the transformation. The direct method is via the classic rotation matrix $M_r$ with the rotation axis $A = \frac{D \times P}{||D \times P||}$ and the angle $\alpha$ between the two vectors. The rotation matrix is then calculated via the following equation:

$$M_{\text{r}} = \begin{pmatrix} \cos\alpha + A_x^2(1-\cos\alpha) & A_xA_y(1-\cos\alpha) - A_z\sin\alpha & A_xA_z(1-\cos\alpha) + A_y\sin\alpha \\ A_xA_y(1-\cos\alpha) + A_z\sin\alpha & \cos\alpha + A_y^2(1-\cos\alpha) & A_yA_z(1-\cos\alpha) - A_x\sin\alpha \\ A_xA_z(1-\cos\alpha) - A_y\sin\alpha & A_yA_z(1-\cos\alpha) + A_x\sin\alpha & \cos\alpha + A_z^2(1-\cos\alpha) \end{pmatrix}$$

with $D = (0,0,-1) \rightarrow A = \left(P_y, -P_x, 0\right)$

$$= \begin{pmatrix} \cos\alpha + P_y^2(1-\cos\alpha) & P_y(-P_x)(1-\cos\alpha) & (-P_x)\sin\alpha \\ P_y(-P_x)(1-\cos\alpha) & \cos\alpha + (-P_x)^2(1-\cos\alpha) & -P_y\sin\alpha \\ -(-P_x)\sin\alpha & +P_y\sin\alpha & \cos\alpha \end{pmatrix}$$

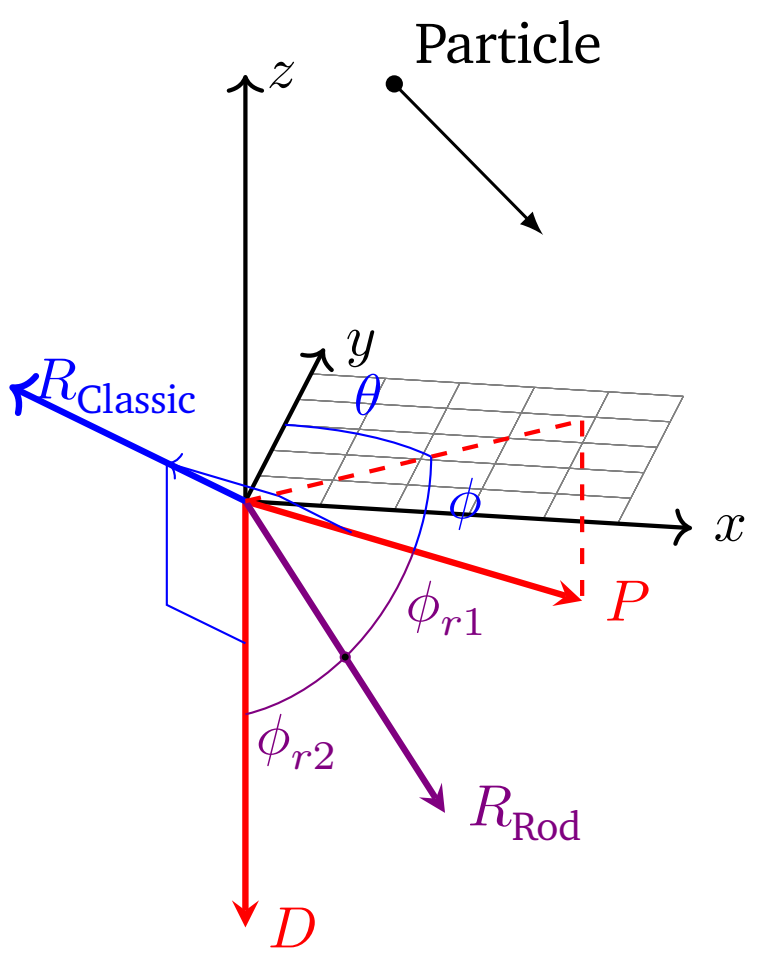

**Figure 4.3:** Schematic representation of two distinct rotation methods.

For each particle track segment, the matrix needs to be calculated once.

Another way would be using the Euler-Rodrigues rotation formula [31] and rotating $\beta = 180°$ around the axis $A = |P|D + |D|P$ bisecting the two vectors $D$ and $P$ as displayed in figure 4.3. With the general formula

$$R'_{\text{Rod}} = R_{\text{Rod}}\cos\beta + (R_{\text{Rod}} \times \vec{p}) + R_{\text{Rod}}\left(R_{\text{Rod}} \cdot \vec{p}\right)(1-\cos\beta)$$

or in Matrix form:

$$M_{\text{Rod}} = \mathbb{1} + \sin\beta \cdot R_{\text{Rod}} + (1-\cos\beta)R_{\text{Rod}}^2$$

Inserting the rotation angle ($\beta = 180°$) leads to the following simplified matrix:

$$M_{\text{Rod}} = \mathbb{1} + 2 \cdot R_{\text{Rod}}^2$$
$$= \begin{pmatrix} -2A_y^2 - 2A_z^2 + 1 & 2A_xA_y & 2A_xA_z \\ 2A_xA_y & -2A_x^2 - 2A_z^2 + 1 & 2A_yA_z \\ 2A_xA_z & 2A_yA_z & -2A_x^2 - 2A_y^2 + 1 \end{pmatrix}$$
$$\text{with } D = (0,0,-1) \rightarrow A = |1|\,(0,0,-1)^T + |-1|\,\left(P_x, P_y, P_z\right)^T$$

Which is computationally less expensive than the large number of trigonometric function calls and better to calculate in parallel through vectorization.

The third and last step (subfigure 4.2c) is the translation of the photons to the center of the track segment or, for higher physical accuracy, to a random position along the track via a simple addition of the displacement vector.

### 4.2.1 Particle path length limitations - Substepping

For the efficient calculation of the cascade, the individual particle tracks are interpolated linearly between a predetermined start and endpoint provided to the tracking algorithm. The distance between those two points and the resulting steplength is determined by the modules employed for the simulation. This limit can be caused by angular deviation limits given by the magnetic field or an impending stochastic interaction. The simplified treatments of tracks as linear with default length are sufficiently precise enough for classic particle cascades. However, because the photon generation uses this linear trace for each photon's point of origin calculation, problems arise if there are significant deviations from the actual path.

With the high angular resolution of the **C**herenkov **t**elescope **a**rray (CTA), compared to particle detectors, approaching $0.02°$ [54] and an approximate time resolution of $500$ ps the spatial resolution used in the simulation must match. This results in a linearly decreasing resolution with increasing distance to the telescope of $\approx 34.9$ cm (perpendicular to the pointing direction) for every $1$ km of distance. Timing is not affected and results in constant $\approx 15$ cm of depth resolution.

The simple solution of forcibly reducing the step length to match the resolution requirements, as was done in CORSIKA 7 and for the first results in CORSIKA 8, leads to a massive growth in runtime. This increase is due to the fact that all modules are called for each step, even if the regular interaction length is much larger. Each steplength reduction boosts the overall number of model calls and the runtime, at least linearly.

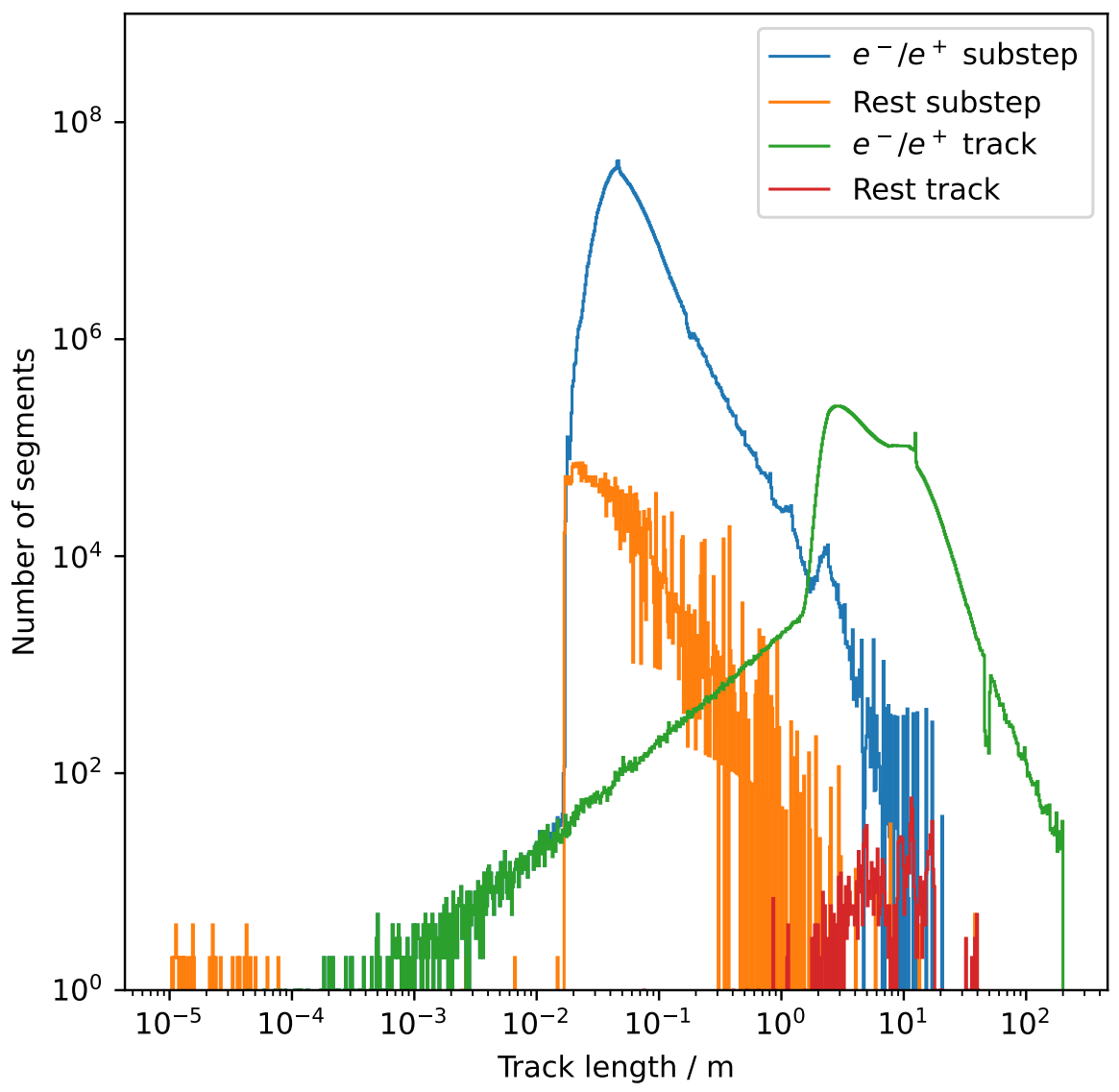

**a:** CORSIKA 7

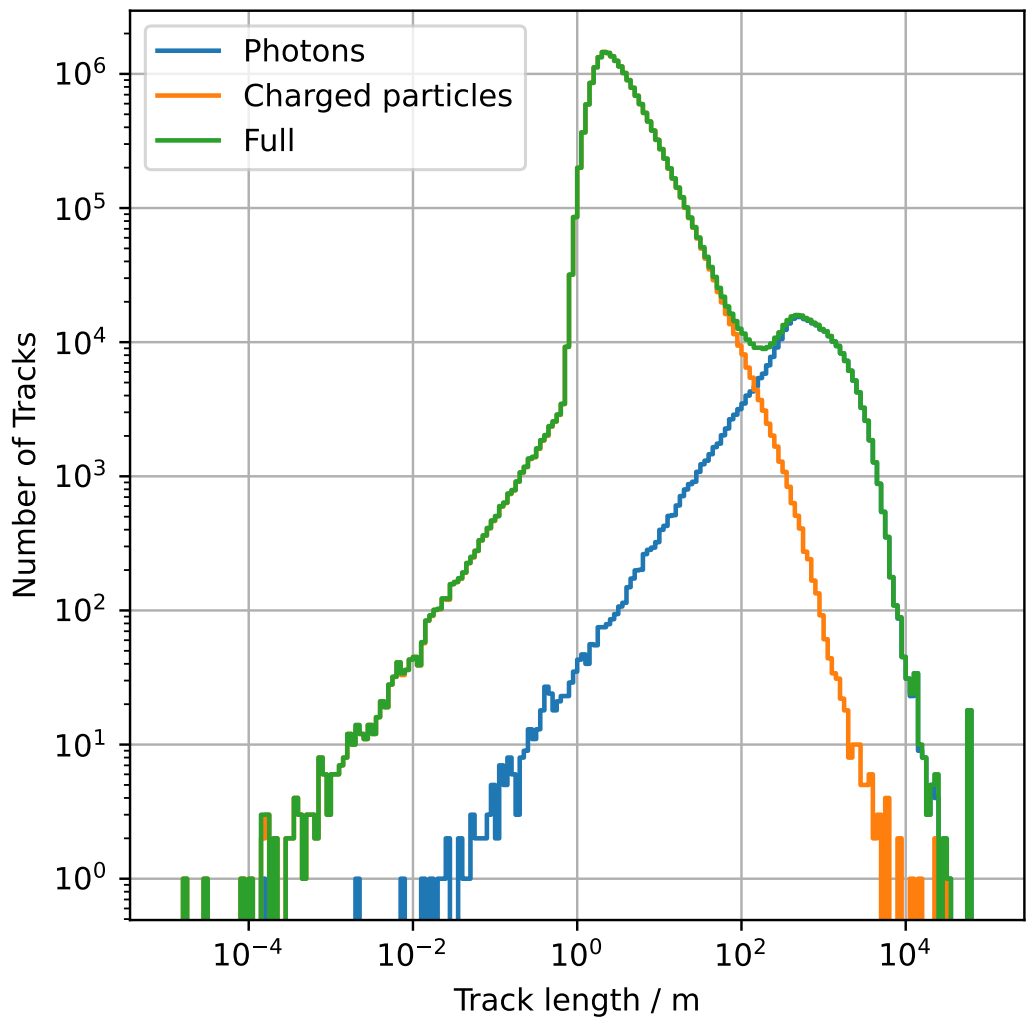

**b:** CORSIKA 8

**Figure 4.4:** Displayed are the resulting track length distributions of simulations with CORSIKA 8 default configuration and CORSIKA 7 with and without IACT extension. It is visible that the simulation without Cherenkov light utilizes much lower steplength compared to the basic simulation.

To avoid slowing down the entire simulation due to the track length requirements of the Cherenkov light generation module alone, a different approach was taken, and a secondary loop was introduced to step through the physically calculated particle path to generate a sequence of linear tracks with sufficient accuracy. Currently, only two effects are implemented that change the path of the particle in a meaningful way: magnetic deflection and multiple scattering. Effects originating from magnetic fields are easily calculated through the Lorentz Force via solvers like the Leap-Frog Method used in CORSIKA 8. Those already allow the querying of arbitrary position between start and end of the path $f_{\text{mag}} : [0, 1] \to \mathbb{R}^3$ with $f_{\text{mag}}(0) = p_{\text{start}}, f_{\text{mag}}(1) = p_{\text{end}}$ resulting in a correctly curved trajectory.

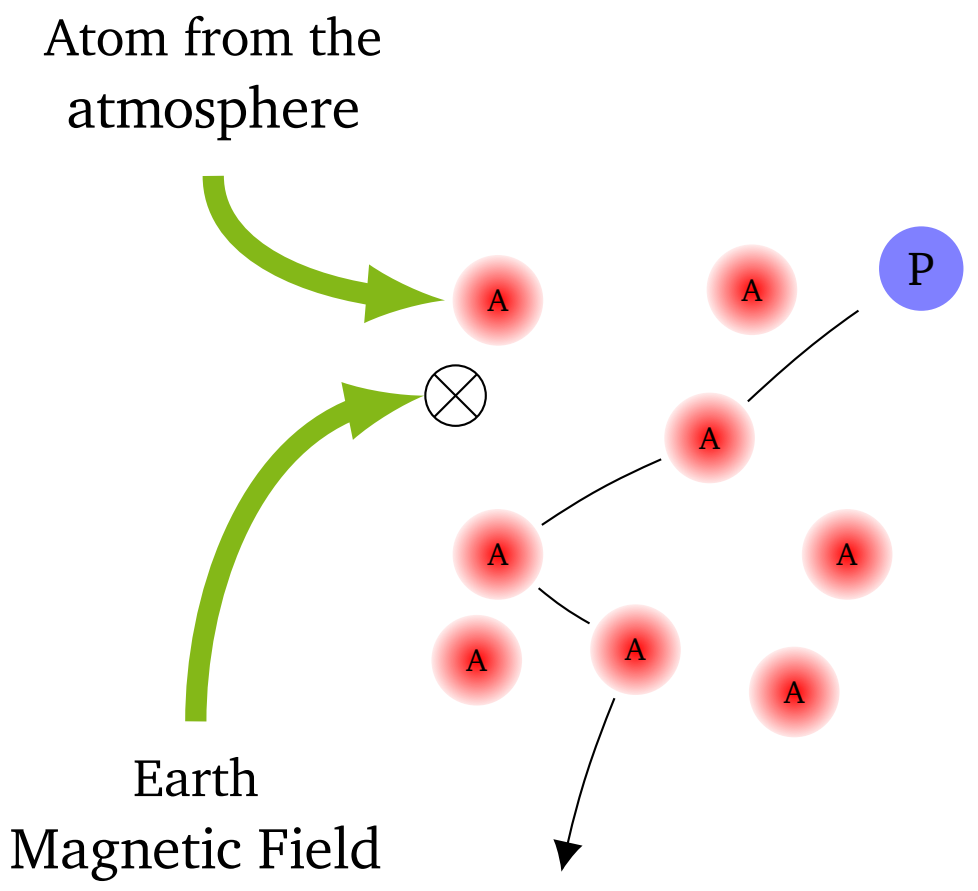

**Figure 4.5:** Schematic representation of a multiple scattering path through the medium, with $A$ being the Atmospheric constituent and $P$ the propagated particle.

Multiple scattering calculated by the leptonic propagator PROPOSAL [34] is more challenging to divide into substeps. As the name suggests, multiple scattering is the cumulative effect of (up to) millions of individual scattering events, which cannot be treated stochastically due to runtime limitations. Therefore, an approximation is in use, which employs the initial energy and the particles traversed grammage to calculate a theoretical scattering angle distribution from which to sample. Two such approximations are Highland [40] and Molière [21], where the latter shows much better agreement with actual measurements for high scattering angles. The axis deviation for Molière scattering, used as default, is shown in figure 4.6.

What makes calculating intermediate positions difficult is the fact that the physical path itself is a quasi-random walk from the start to the end point given by the simulation, where each step follows the distribution with the correspondingly reduced grammage in figure 4.6. There are two possibilities: instead of calculating the multiple scattering path deviation to the next interaction in a single step, several substeps could be sampled according to the required precision and stored. Alternatively, a path with the highest probability could be used based on the theoretical distributions.

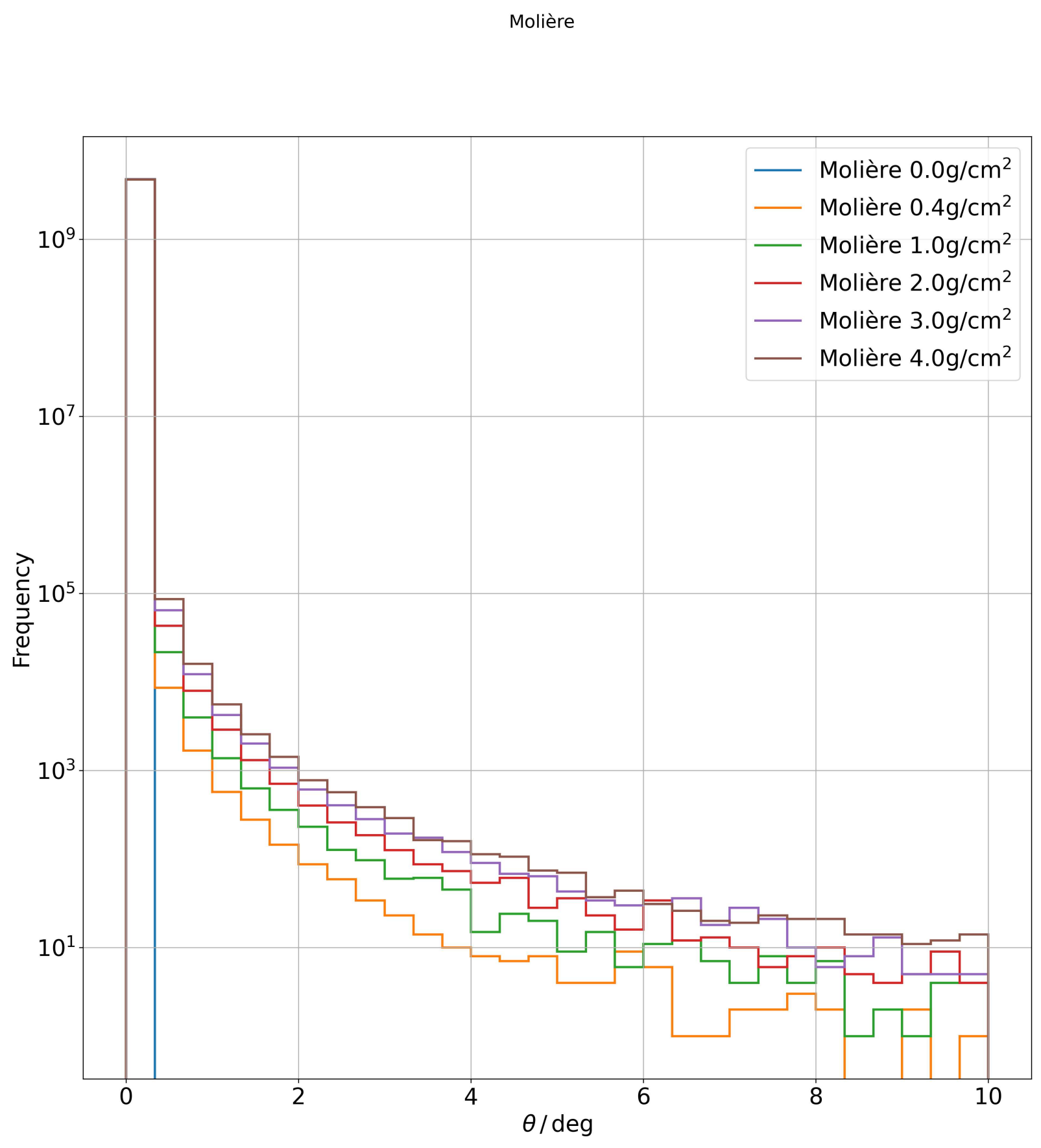

**Figure 4.6:** Displayed is the multiple scattering angular distribution of 100 GeV particles through different thickness of atmosphere.

For the time being, the first approach is used as it is the more straightforward method and also results in a more accurate calculation due to the additional displacement included, which emulates a random walk between each substep. With a constant run time of $\approx 7\,\mu s$ for the interpolated Moliere method and $\approx 15\,\mu s$ for the non-interpolated method, the induced simulation overhead is acceptable for development and can be improved later. Scaled to the vertical $500\,\mathrm{GeV}$ photon demonstration shower calculated in CORSIKA 7, with a combined electron and positron track length of $2.03 \times 10^6$ m and $33 \times 10^6$ substeps, this would add $231$ s of runtime per shower. Adaptive substepping and photon generation along the path can reduce the number of substeps necessary, thereby reducing the number of calls to the scattering implementation.

### 4.2.2 Fluorescence Photon generation

After calculating the number of photons from the ionization energy loss of the particle and the atmospheric light yield parametrization, the photons need to be physically generated and propagated. The emission in itself is isotropic and shows no preferred direction. The decay time distribution of the ionized state until the light production is measured to be of the order of several ns [19] and currently not considered. The simplest photon direction generation is done in spherical coordinates with a random azimuth angle $\phi$ and a polar angle $\theta$ sampled from a uniform distribution.

$$\begin{pmatrix} x \\ y \\ z \end{pmatrix} = \begin{pmatrix} \sin\phi\cos\theta \\ \sin\phi\sin\theta \\ \cos\phi \end{pmatrix}$$

Alternatively, $x, y, z$ can be sampled from Gaussian distributions and afterward normalized, which can be faster on some hardware architectures. This method requires a dedicated check for the case $x = y = z = 0$.

The photon direction can be modified from an isotropic emission to a pattern that favors the direction pointing to the experiment. With the number of photons scaled accordingly so that the density on the unit circle is kept constant, the overall number decreases. Here, only the reduction to a half sphere is implemented and can be selected for experiments where the telescopes are close together. More complex methods are required for experiments with larger distances between detectors, such as the Pierre Auger Observatory [5], where the experiment can contain the whole atmospheric cascade, which often requires more time than the actual photon generation.

## 4.3 Acceleration methods

With the increasing complexity of the simulation and the growing amount of data to be processed, the need for faster computing becomes increasingly apparent. In addition, reducing overall energy consumption and better use of existing hardware is an increasingly important topic.

To accelerate a computing program, three distinct possibilities exist:

1. Algorithmic changes that reduce the workload through better methods or simplifications.
2. Tuning of the already used algorithm to utilize hardware features better. A common example running on nearly all available hardware is using vectorization to perform the same operation on multiple data points simultaneously.
3. Dividing the workload of the program into parallel running tasks.

All methods will be utilized together to improve the performance of photon propagation in particular. With this, some methods are not exclusive to the photon propagation but can also be applied to other parts of the simulation. A primary focus is on the third point, parallelization, as it is the most promising long-term solution. With the slowdown in single-threaded computing performance over the last few years, as shown in figure 5.1, and the availability of parallel hardware in the form of CPUs or massively parallel hardware accelerators, commonly referred to as GPUs, the need for parallelization is obvious. Because of the broad field of parallel computing, the topic will be discussed in a separate chapter 5.

In general, algorithmic changes provide a stable improvement in performance; however, through the wildly different nature of accelerator cards and CPUs, not all improvements work equally well on all available architectures. A well-working method over all platforms efficiently reduces the workload by filtering out unnecessary work. The additional work for filtering must not exceed the benefit of the reduced workload.

### 4.3.1 Absorption, Scattering, and Efficiency

Most of the photons travel several kilometers through the atmosphere before arriving at the detector level. During this time, the photons are subject to absorption and scattering processes, which reduce their number on the side of the experiment significantly. The transmission curves, displayed in figure 4.7a, are generated with the partially free, but somewhat limited, MODTRAN web version [20] and are based

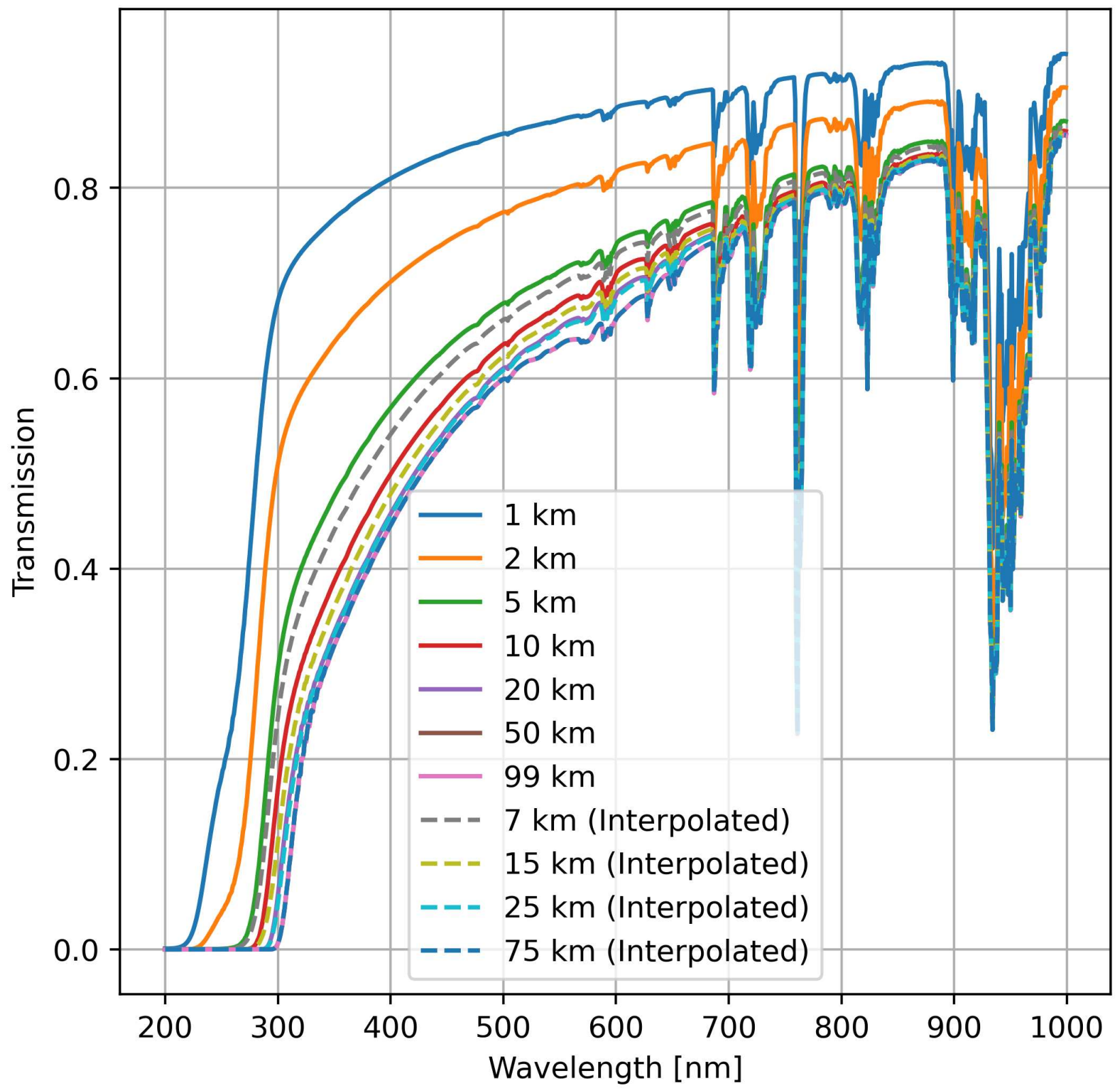

**a:** Relative extinction probability of photon for different wavelengths and emission heights with vertical flight direction.

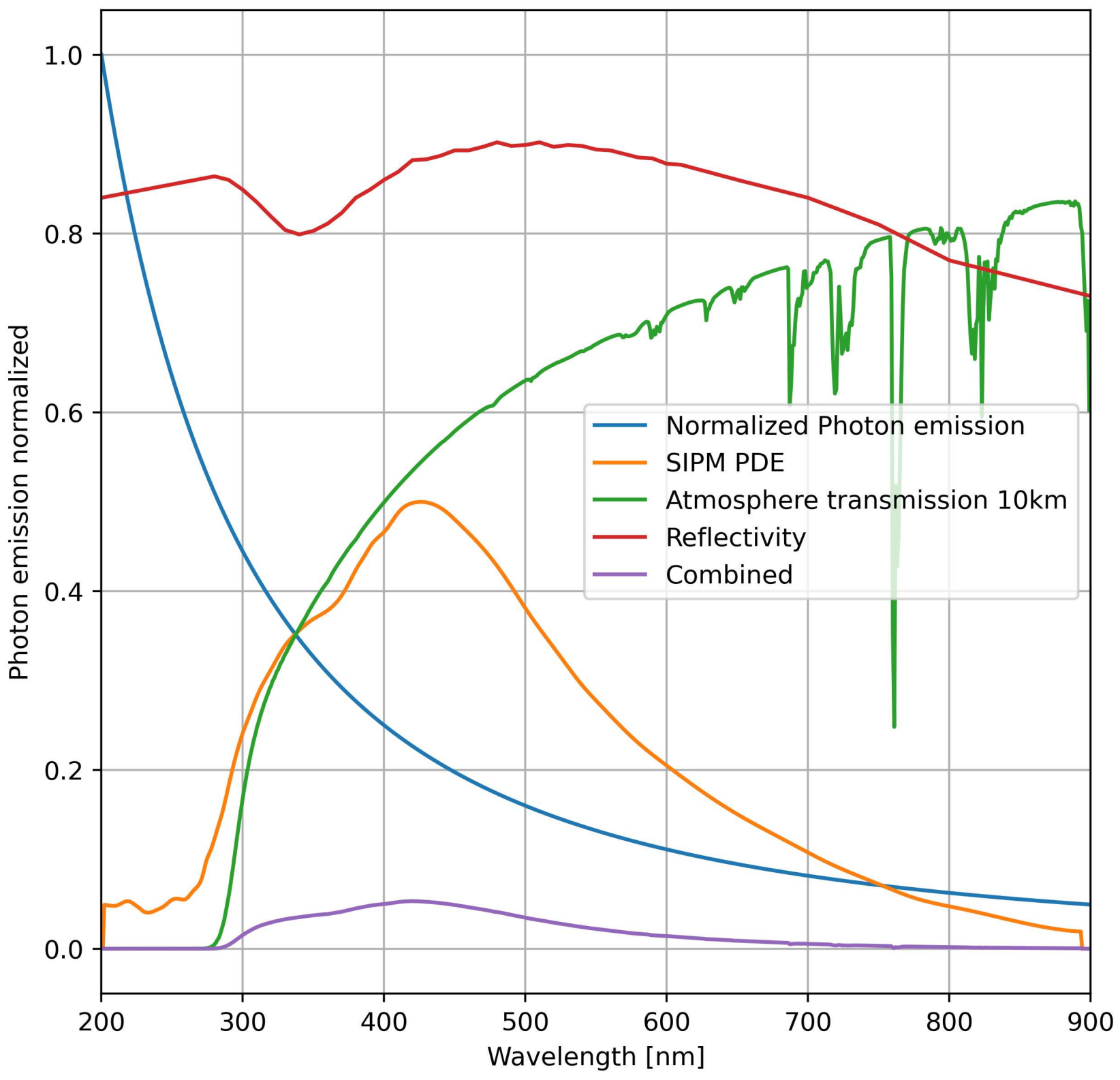

**b:** Influence of individual hardware components constituting a toy SiPM-based Cherenkov telescope with polished Aluminum mirrors.

**Figure 4.7:** Displayed are the influencing effects on the overall photon acceptance.

on the US standard atmosphere. Experiments require dedicated measurements of the transmission or appropriated licensing to access real-time information for the required data. Besides the atmosphere, photons interact with several objects like mirrors and sensors until they are converted to electronic signals. The combined effect of all those processes is called detector efficiency, is experiment-specific, and can be measured in dedicated calibration runs, for example, through muon ring intensity. For an artificial SiPM-based telescope, the wavelength-dependent detection probability of individual photons is displayed in figure 4.7b.

To utilize this reducing effect for accelerating the simulation, the number of photons to generate must be reduced by a factor given by the product of all those effects. The problem is that required values such as photon travel distance and angle are unknown beforehand and can only be estimated roughly by the current particle position relative to the telescope. Some values could be reused from the pre-staged filter calculations described in section 4.3.2 where the minimal distance to the telescopes is already calculated. However, the overall use of approximated values makes it impossible to reach the quality necessary for the entire simulation. The solution is to approximate the reduction factor intentionally low, for example, by using the height instead of the real distance. In addition, the minimal reduction factor (or highest detection probability) is used for angle-dependent effects such as sensor efficiency. The resulting probability is stored together with the photon information, and the experiment-specific code can scale the detailed removal probability, as it was done for CORSIKA 7, with the following method:

$$r_{\text{random}} < \frac{p_{\text{real}}}{p_{\text{approx}}} \text{ with } r_{\text{random}} \in [0, 1]$$

To avoid artificial biases with this method, it is of utmost importance that the applied reduction factor is never higher than the real one because a retrospective increase in photons is impossible.

The flight length and emission height distribution for a $500\,\text{GeV}$ photon shower production are folded with the transmission curve, SIPM detection, and mirror reflection probability to estimate the degree of reduction possible. The original distribution and the results are shown in figure 4.8.

### 4.3.2 Filter Level 1

The generation of photons from individual particles requires several refractive index calculations, which require memory loads because of the underlying table. This memory access will seldom be cached because of the height difference between photons,

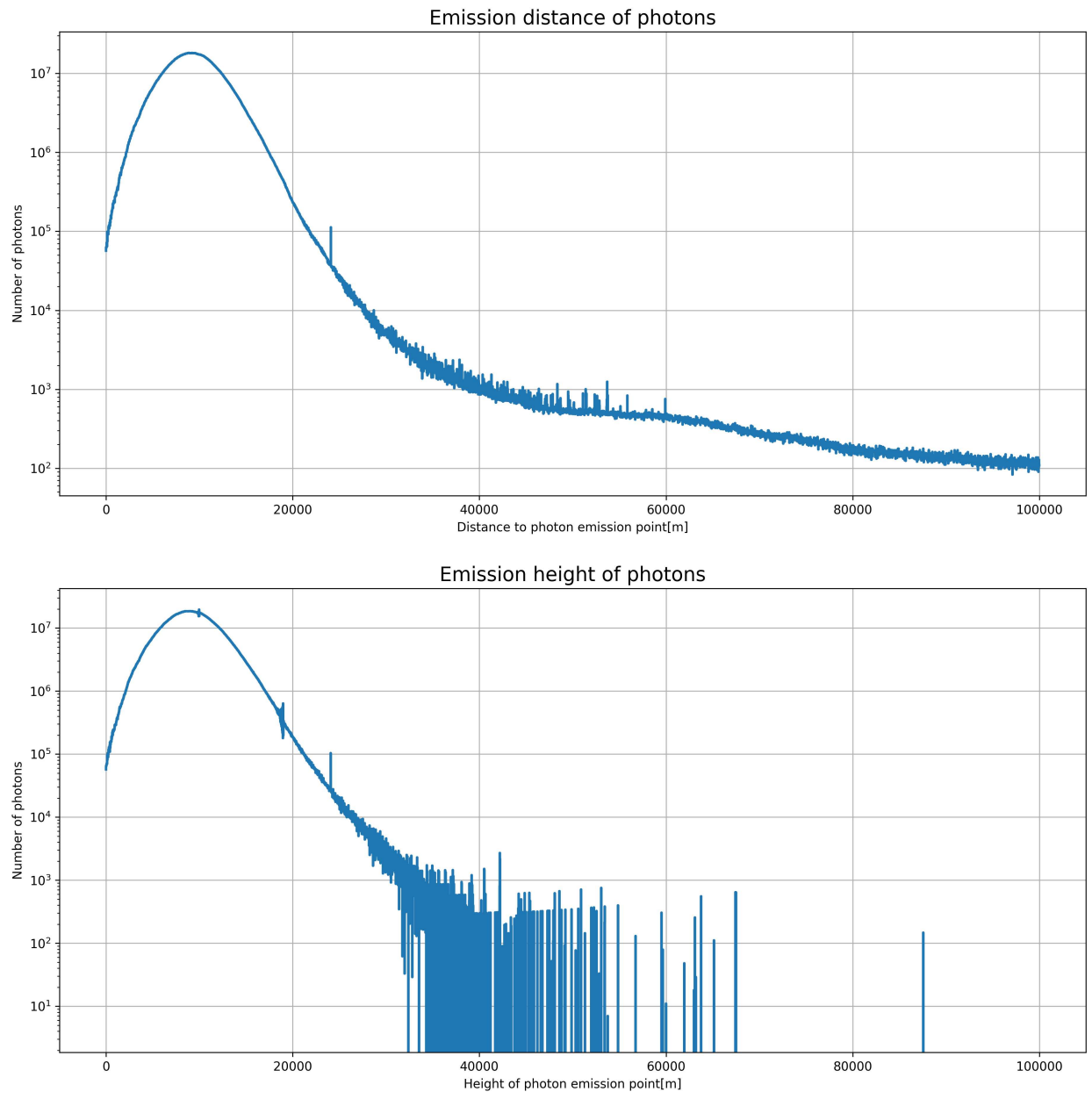

**a:** Displayed is the flight length distribution of photons for a 500 GeV Photon airshower production for 200 Showers in the $0°$ to $45°$ Zenith range.

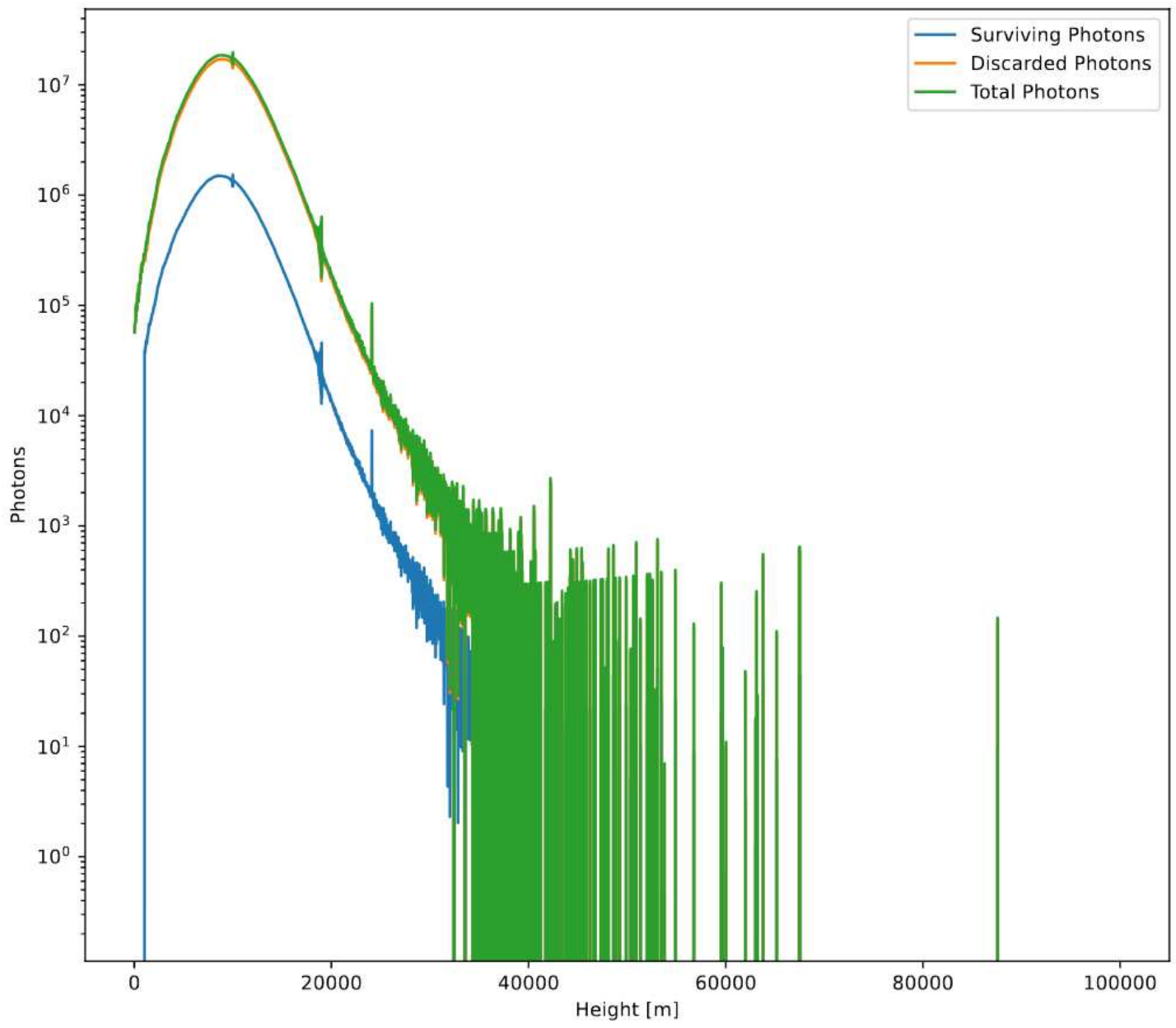

**b:** Removed photon distribution

**Figure 4.8:** Removal rate in a 500 GeV airshower production

even for the same track. In addition, the photon generation and transformation into the correct frame is quite costly through the trigonometric functions. Including the overhead introduced by calculating millions of photons, the early removal of particles that do not contribute to the measurement is a promising approach to reduce the overall workload.

The most fundamental prerequisite for this method is that the instrumented area of the Cherenkov experiments is smaller than the illuminated area, producing particles where the light pool does not contribute to the detector's signal. From this, it follows that the light emission needs to be directional like Cherenkov light; isotropic emission, like fluorescence, can nearly always contribute to the measurement and can not be confidently filtered.

For the Cherenkov Telescope Array (CTA) the figure 4.9 displays the light pool of a 500 GeV photon cascade against a possible telescope layout. With an increase in energy, Zenith angle, and even more so for heavier initial particles, the size of the cascade is increasing significantly. This fact is even more pronounced when real-world simulations include cascades that are shifted from the array centre and only partially clip the instrumented area. This behaviour will likely be relevant for the lifetime of the CORSIKA 8 simulation as experiments simultaneously recording several square kilometres of sensor area will not be possible through technological and financial limitations allowing direct implementation in the core photon propagation module. A simplified approach was already utilized in a previous work [18] and showed auspicious results of up to 90 % reduction in runtime for small individual telescopes.

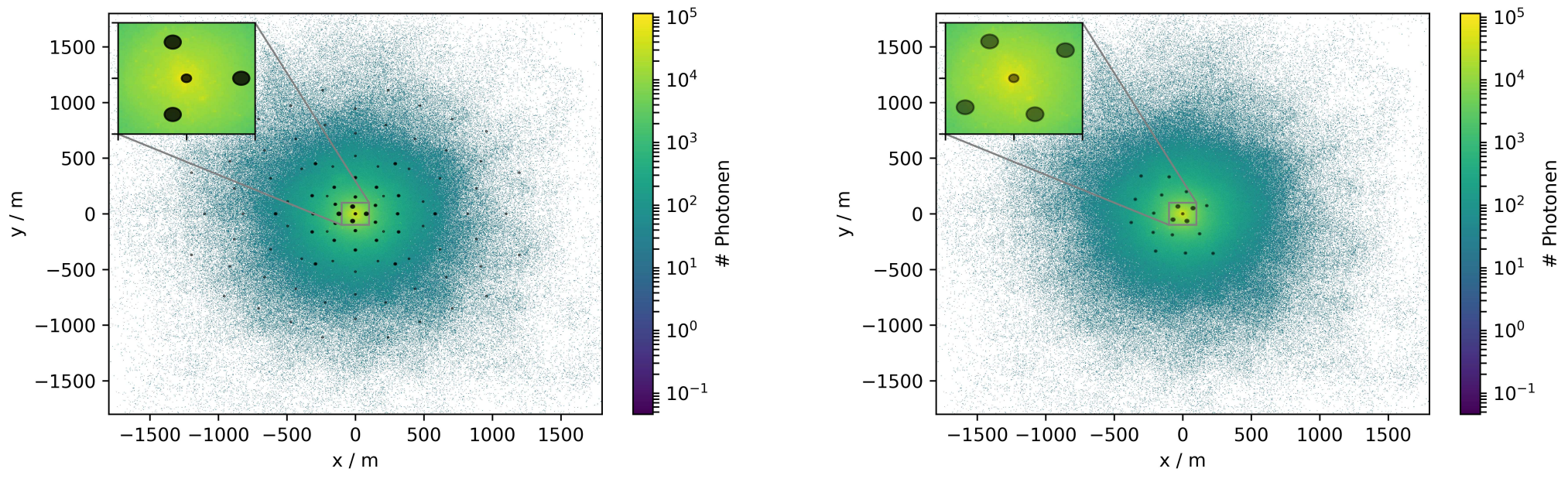

**a:** Southern Array    **b:** Northern Array

**Figure 4.9:** Displayed the Illuminated region of a 500GeV electromagnetic airshower with one possible layout (radius to scale) of the Cherenkov telescope array.

Therefore, the first filter stage (LV1 Filter) is applied to reduce the number of photon-emitting particles significantly, hence the number of subsequent calculations required, employing the “best” possible cuts available.

The theoretical description of Cherenkov radiation, as described in 2.2.1, was used to find a suitable cutting parameter. With the in-air highly collimated emission in the particle’s motion direction, with an opening angle of less than $1.4°$, simple geometric constraints are ideally suited to be used as cutting parameters. To be noted is that both the charged particles and, later, the photons are deflected during the propagation step through various effects. Therefore, the restrictions must provide enough leeway or otherwise consider the possible deflections.

### Cut optimization

The most straightforward cut parameter is the angle between particle flight direction and the experiment calculated by a simple dot product

$$\cos(\alpha_i) = \frac{\vec{p} \cdot \vec{d}}{\|\vec{p}\|\|\vec{d}_i\|} \tag{4.3}$$

$$\alpha_{\text{max}} = max(\alpha_0, ..., \alpha_n) \tag{4.4}$$

where $\vec{p}$ is the traveling vector of the particle and the direction to a singular point of interest $\vec{d}_i$, e.g. the instrumented area. Complexity is added by the fact that most experimental sites include multiple relevant instruments and that those instruments are usually not a point but extended over several $10\,\text{m}^2$. The following list of options represents some chosen possibilities for the point of interest listed with increasing computing complexity that was reviewed:

1. Center of array
2. The Closest distance to (circular) array border
3. Center of each telescope
4. The Closest distance to each telescope border 2D
5. The Closest distance to each telescope border 3D

Through the vector nature of the calculation, each one can be implemented on GPU or CPU while using hardware acceleration like vectoring.

### 4.3.3 Filter level 2

The second filter tested is based on the fact that the deflection of optical light from a straight line path is comparably tiny for short distances and can be approximated through this effortlessly. Using methods similar to those described in the previous section 4.3.2, photons can be removed before the physically correct propagation is applied. Ultimately, however, the usefulness of this method is shown to be somewhat limited. The number of photons is already high, and the reduction is comparably small, as shown in 6.2.2. The much faster propagation method described in section 4.4.1 is hardly more expensive regarding runtime than the straight-line calculation. This method would only be relevant if much more complex propagation methods had to be used, which should not be necessary for typical atmospheric conditions and ground-based experiments.

### 4.3.4 Filter level 3

The third filter applicable would be executed directly after the straight-line propagation required for the fast photon propagation algorithm but before its correction stage (see section 4.4.1 for details). The idea is to remove photons before possible expensive lookups in tabulated values are performed. Comparable to the enumeration for the first filter stage 4.3.2, different distance metrics with different performance impacts can be used.

Depending on the dimensionality and hardware used for the calculations, this leads to different improvements in throughput.

## 4.4 Photon Propagation

The physics-inspired approaches of photon propagation algorithms introduced in Section 3.4 range from simple straight-line calculations to capable iterative path-tracing algorithms. The choice of the algorithm depends, in general, on the desired accuracy and the available computing time. For thin media and the approximations of symmetrical model data (plane parallel, cylindrical, or spherical) as described in section 3.3, a hybrid approach was developed to combine the performance of the straight-line propagation with the accuracy of the iterative models without the massive overhead.

### 4.4.1 Interpolation Tracing

The iterative nature of all the photon tracing algorithms introduced is due to the support of various refractive index profiles, even dynamically changing during the runtime of the simulation. This added complexity, combined with the required number of photons to be traced to the observation plane, makes these iterative solutions much slower than acceptable for CORSIKA 8. To give an overview, the runtime of photon propagation through $10\,\mathrm{km}$ of atmosphere with a step length of $10\,\mathrm{m}$ are displayed in table 4.2. With optimization of the step length, e.g. depending on the refractive index gradient, an improvement of the values displayed is possible but not by several orders of magnitude required to be competitive.

The newly derived method instead exploits the fact that for most large-scale productions, the atmosphere is kept the same for a significant amount of individual events. Therefore, a correction table can be calculated in advance from the difference between the swift linear propagation and the slower but more accurate iterative propagation. The symmetries obtained by the atmospheric models shown above (section 3.3) keep the table small and, above all, low dimensional enough to make it efficiently usable. The table directly depends on the number of parameters underlying the atmospheric model used for the calculation. Figure 4.10 shows the schematic representation of the situation, including the lookup values needed to calculate the correction table for planar and cylindrical atmospheres. For the more straightforward case of a planar atmosphere, the offset $d$ can be removed; for the more complex case of a spherical atmosphere, the angle $\alpha$ becomes two-dimensional and gets an additional pointing angle. Distance $d$ for three dimensions equals the radius in the polar coordinate frame of reference. It follows that the basic tables are 2, 3, and 4-dimensional.

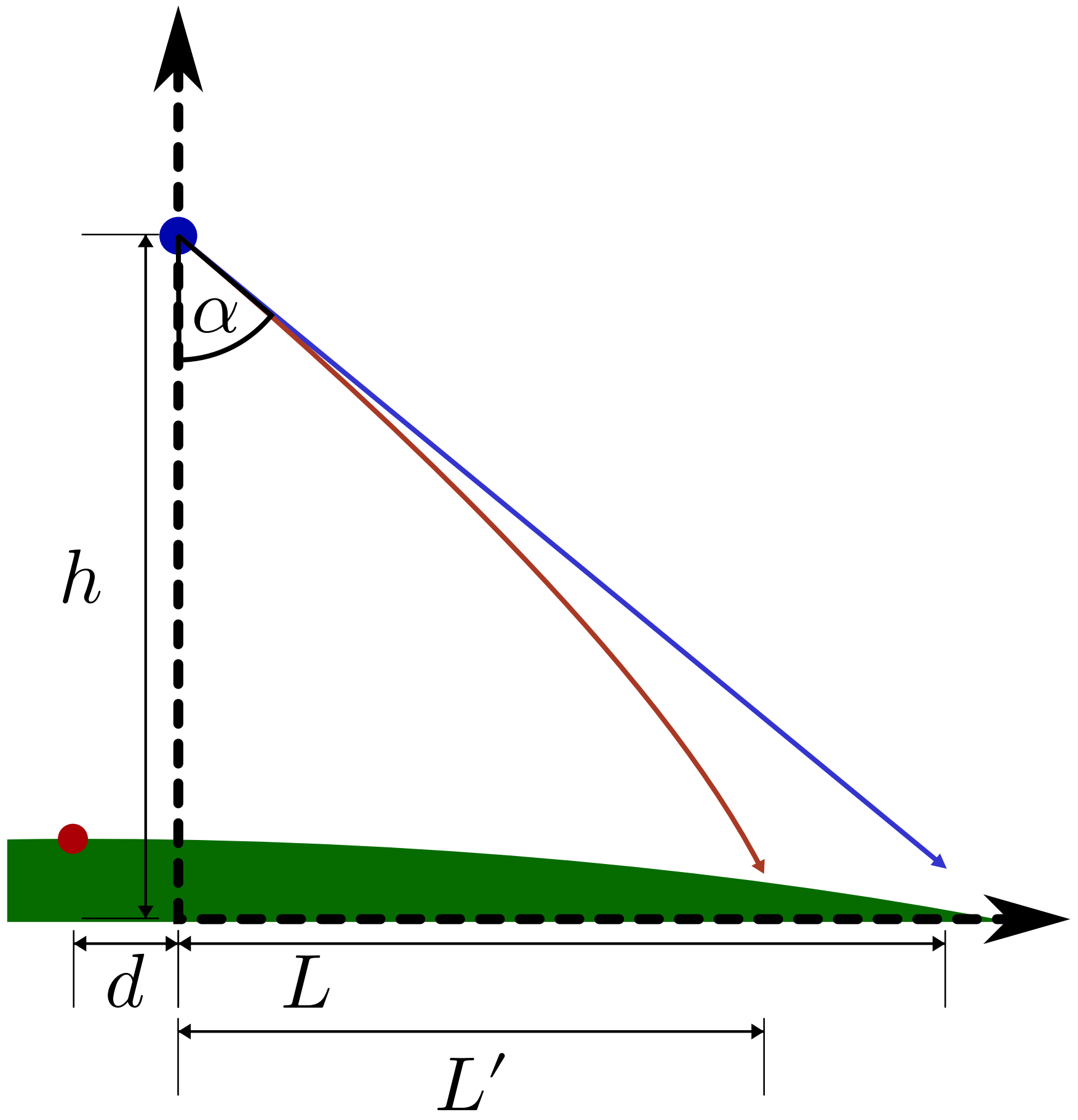

**Figure 4.10:** Schematic representation of the correction applied to the straight line propagation for calculating the realistic photon position. The red point marks the experiment's center, and the blue the photons' origin.

Table 4.2: Runtime comparison of various propagation methods for individual photons for two atmospheric models. The internal step length was kept constant and set to 10 m for physically accurate results. The emission point was set to 10 km in height with a nearly vertical starting angle.

| | 0° Inclination | | 60° Inclination | |
|---|---|---|---|---|
| **Method** \Runtime | Planar [ms] | Cylindrical [ms] | Planar [ms] | Cylindrical [ms] |
| Linear | $(3.1 \pm 0.06)10^{-4}$ | $(3.1 \pm 0.05)10^{-4}$ | $(3.0 \pm 0.01)10^{-4}$ | $(3.0 \pm 0.01)10^{-4}$ |
| Rotation matrix | $0.219 \pm 0.001$ | $0.279 \pm 0.001$ | $0.459 \pm 0.01$ | $0.580 \pm 0.01$ |
| Vector form | $0.157 \pm 0.001$ | $0.217 \pm 0.001$ | $0.290 \pm 0.01$ | $0.410 \pm 0.05$ |
| Vector form (Float32) | $0.146 \pm 0.002$ | $0.221 \pm 0.001$ | $0.290 \pm 0.01$ | $0.439 \pm 0.05$ |

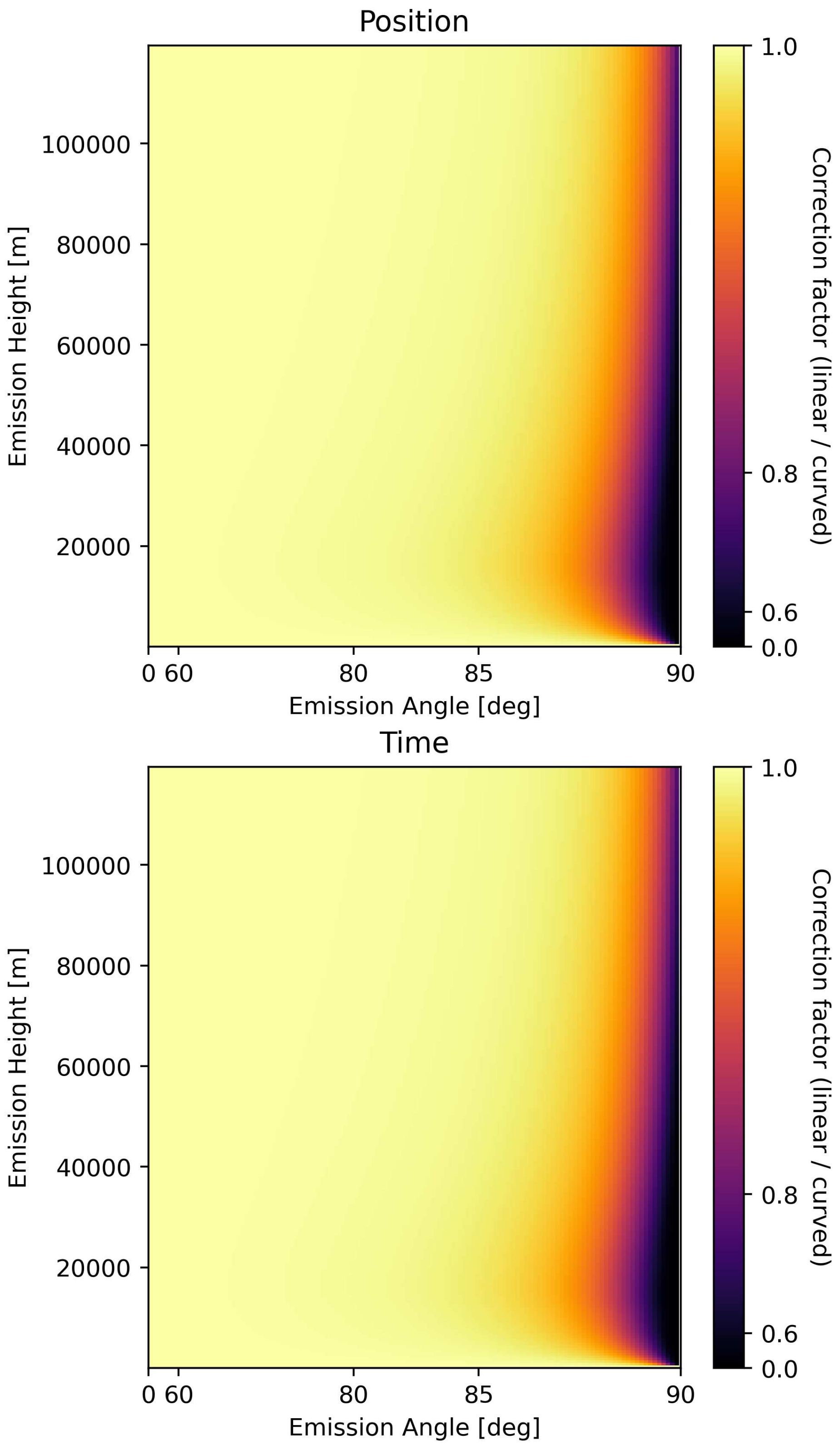

**Figure 4.11:** Interpolation data for planar atmosphere

# 5 Parallel Computing

With the growth of large-scale and increasingly precise measurements of physical properties, the requirements for underlying simulations are also increasing and often grow faster than the technology available for simulation, leading to considerable costs for additional infrastructure. This discrepancy leads to the fact that some measurements can only be performed with great effort because the needed simulations are not available or, as is often the case, statistically massively limited. Collaborations spent often up to several 100 000 h CPU Time on complex and extensive programs like the ones necessary for the simulation of particle interactions and propagation through the atmosphere (e.g. CORSIKA [38]) on individual production runs to be able to analyse their results. The runtime can even exceed this by orders of magnitude if rare physical processes need to be covered.

In addition, the underlying code base of CORSIKA has grown over many generations of physicists, spanning more than 30 years. This prolonged timeframe means modern principles and techniques are difficult to support, and the latest hardware improvements are only coincidentally exploited through compiler optimisation. This leads to the dilemma that the straightforward approach to reduce runtime by buying more and newer hardware does not scale well with the increasing cost.

The evolution of computing infrastructure in recent years, as shown in Figure 5.1, highlights the problem. Whereas a quasi-exponential increase in performance was measurable for a long time, nowadays, only a slight increase in performance can be detected for individual processes. Instead, hardware manufacturers push more and more performance via parallel architecture implementation, which classic simulations can only benefit from by multiple simultaneous executions.

The “physicist parallelisation”, the X-fold execution of the program, was a cheap way to harness the parallel processing power of the strictly sequential/single-threaded program execution. However, this does not scale optimally, partly due to multiple memory allocations from each application and the resulting increase in cache pressure. Especially concerning is the memory consumption on modern systems with 64 or more cores; only a few data centres can guarantee enough memory to utilise all cores completely with the simulation software. The problem becomes even more pronounced when looking beyond classic x86 CPUs to other architectures such as ARM or RISC, which are on the rise, for example, with Nvidia's Grace architectures, where the ratio of available cores to RAM can shift even further.

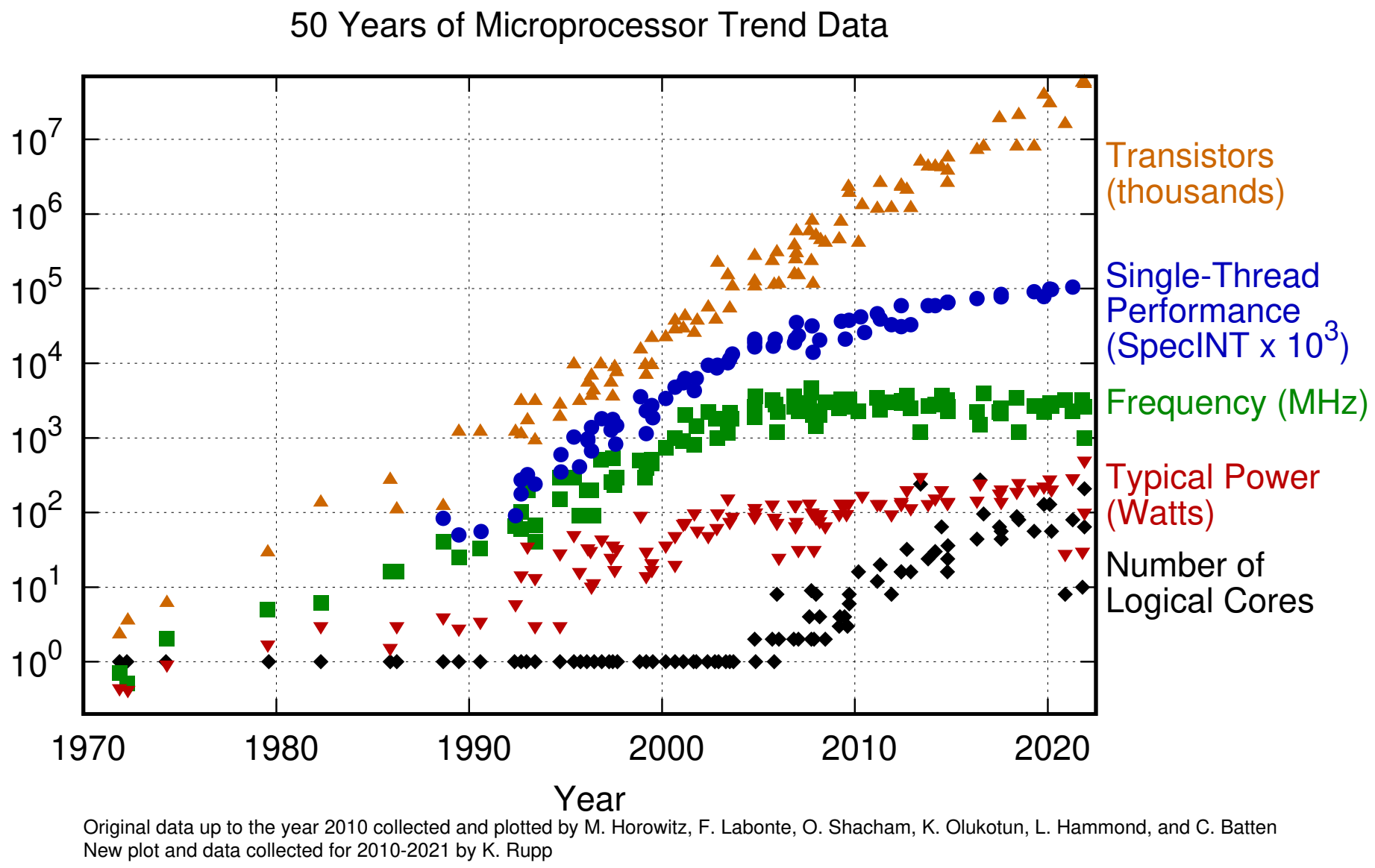

**Figure 5.1:** Displayed is the increase in computing performance over the last years. The Data displayed was collected by [65].

Through this increasing number of cores, the limit of the multiple execution methods is already reached or will be soon. In addition, it entirely neglects any benefits of alternative technologies. Changes to the fundamental programme structure and the essential algorithms are required to ensure future-proof growth and evolution to meet future needs. This section deals with algorithmic changes and utilizes the new development in the context of CORSIKA 8 [11] for underlying structural changes.

## 5.1 GPU Computing

Elemental parallel computation support is ubiquitous today, but in its early days, it was reserved for data centres. The first widespread internally parallel computing processors were graphics cards, for instance, the Sony Console GPU, in 1994, while the first two-core CPU was released by IBM only seven years later. The development of GPUs progressed fast, with the workload to colour thousands of pixels on the monitor in the early days and millions in modern times; the task suits the parallel execution model exceptionally well. Given the limited computational complexity of computing a three-dimensional representation for a monitor, a highly parallel architecture specialised in computing simple arithmetic operations quickly emerged. This specialisation allows the hardware complexity of each core to be dramatically reduced compared to that of a CPU core while significantly increasing the number of available data pipelines.

In the early 2000s, scientists began using GPUs to perform independent computations by representing problems as graphical problems [51, 22]. Some years later, the first **G**eneral **P**urpose **G**raphics **P**rocessing **U**nit "GPGPU" with programmable function pipeline based on "CUDA" was released [53].

Some early restrictions on computing on GPUs have been weakened or even removed, but the basic principles have mostly stayed the same. For efficient programming, it is therefore crucial to know these and to adapt the basic structure of the program to the existing limitations and available features:

1. Data Parallelism
2. Memory Size/Performance
3. Latency Hiding

4. Warp Level synchronisation
5. Tensor Cores
6. FP32 →FP64 Performance Penalty (lower on newer hardware)
7. In hardware acceleration (e.g. Textures, Intrinsics)

Random access patterns to data at the level of individual threads must be avoided as much as possible because the data transfer speed is slow compared to the time required for arithmetic operations. In addition, the memory transfer unit is physically shared between several threads, which results in an even higher wait time. The way to access data optimally is in a sequential fashion along the threads, where every thread is executing some code to load data into much faster memory:

```
id = threadID.x
blocksise = threadSize.x
localData = memory[id * blockSize]
doSomething( localData )
```

The compiler and shared memory controller optimise this access structure by coalescing the individual operations into a smaller number of large copy operations, reducing the overall blocking time. With a large enough number of threads, the GPU utilises an additional technique called latency hiding, where the currently waiting threadblock is swapped out with the next task to compute. The hardware handles memory transfers in the background; each process does not see a wait time in its perspective. On the other hand, to enable this functionality, there must be enough data to supply several thousand threads to overcommit the GPU by several factors.

Since each thread runs on highly simplified hardware, certain processor parts are shared between the neighbouring threads, called warps. This hardware reduction also includes the unit that determines which operations the Chip executes next, the instruction pipeline, with the consequence that divergent branches of execution, such as IF conditions, cannot be processed simultaneously but only one at a time, which in the worst case doubles the runtime. During the initial development of the GPU routines for CORSIKA 8, it was decided that, for now, complex and branch-heavy code, such as the interaction models for hadrons, would be kept on the CPU for the time being, similar to CORSIKA 7. From the above requirements on how to best structure the program, the following program flow can be derived:

1. Stage: Transfer particle track information from CPU to GPU
2. Stage: Filter tracks with domain knowledge, i.e., experiment locations
3. Stage: Generate Photons

4. Stage: Filter Photons
5. Stage: Propagate Photons

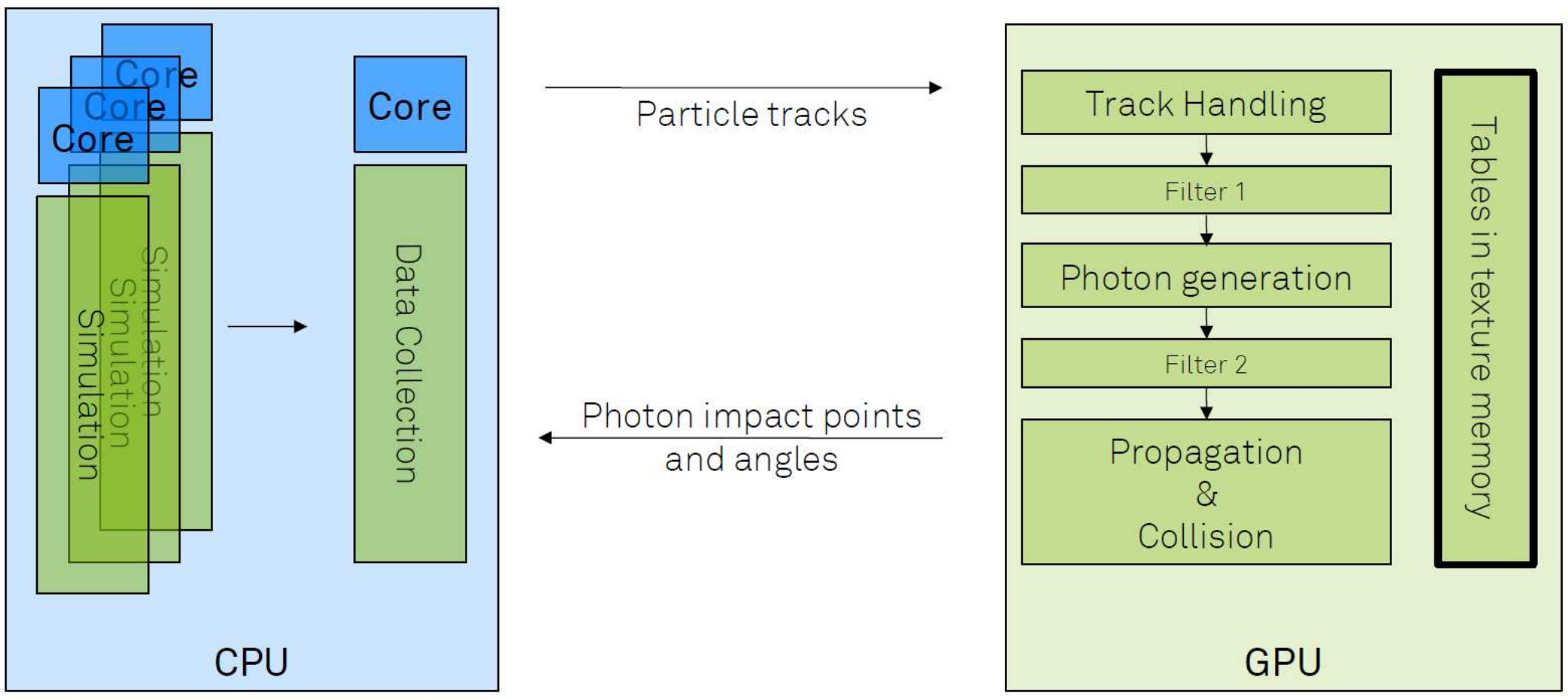

**Figure 5.2:** Overview of the programme structure used for GPU-accelerated photon propagation.

The program's control structure is displayed schematically in figure 5.2, with the CPU regime on the left side and the GPU regime on the right. Due to the overhead of transferring data from the CPU to the GPU and back, plus the need to provide enough data to the GPU device, it is necessary to first accumulate tracks in memory on the CPU side and then transfer them in a single step. To fully utilise the expensive acceleration hardware, only a collection of parallel-running CPU instances generates data fast enough to keep it busy. It necessitates a complete synchronisation and data aggregation stage described later in 5.4 to make this possible. The procedure works well for data centres with a high number of cores for every GPU available on a node. For more specialised clusters, such as AI Accelerators, the ratio between GPU and CPU is more heavily weighted to GPUs, which mandates the transfer of tracks from external nodes to be fully utilised.

## 5.2 Atmosphere and Interpolation

In addition to the limitation mentioned above, the current generation AI accelerators inherit some functionality of their early predecessors, the typical computer graphics card, specifically the texturing unit. Employed initially to efficiently calculate the images, called textures, used to colour polygons in graphics applications, it allows mapping and, here, more critical, linear interpolation between support points directly in hardware. It supports indexing in up to 3 Dimensions and up to four 32-bit floating point values per index or pixel. The number of entries depends on the GPU architecture but should be kept below 65536 elements to support a wide range of hardware.

Another limiting factor of using the texturing unit is the internal interpolation method, presumably using 16-bit floats(no complete documentation exists for the used Nvidia A100 GPUs), which introduces additional deviation from the original value. The less precise hardware method was compared with CPU-based interpolation methods to calculate the height-dependent refractive index distribution and check the results' quality. This is done by comparing the interpolation results of randomly sampled heights, where the exponential interpolation is used as a reference. An additional crosscheck based on the CPU Version with linear interpolation and identical support points allows for a direct comparison. The resulting curves are displayed in figure 5.3

The results show a predictable decrease in precision with increased support point distance. Overall the deviation for a reasonable number of support points, every $10\,\mathrm{m}$ to $100\,\mathrm{m}$, barely exceeds $1\,\%$ for the reduced refractive index $n-1$. Regarding the accuracy of the atmospheric measurements and the general application of $n$, not $n-1$ in the methods, the impact is negligible.

From a runtime point of view, the texture memory performs better than the usual access methods. This results from the "random" positions of the particles, which do not scale well for coalesced memory access within a warp. The texture benefits from a dedicated texture cache and optimisations when accessing nearby data. For comparison, the runtime of different sample sizes and, comparatively more importantly, different support distances were generated, and the results are displayed in figure 5.4.

This method of interpolation is not only used for the calculation of atmospheric properties but can also be used to accelerate access to the correction factor table, up to three dimensions, for the fast propagation according to chapter 4.4.1, further described in 5.3.4.

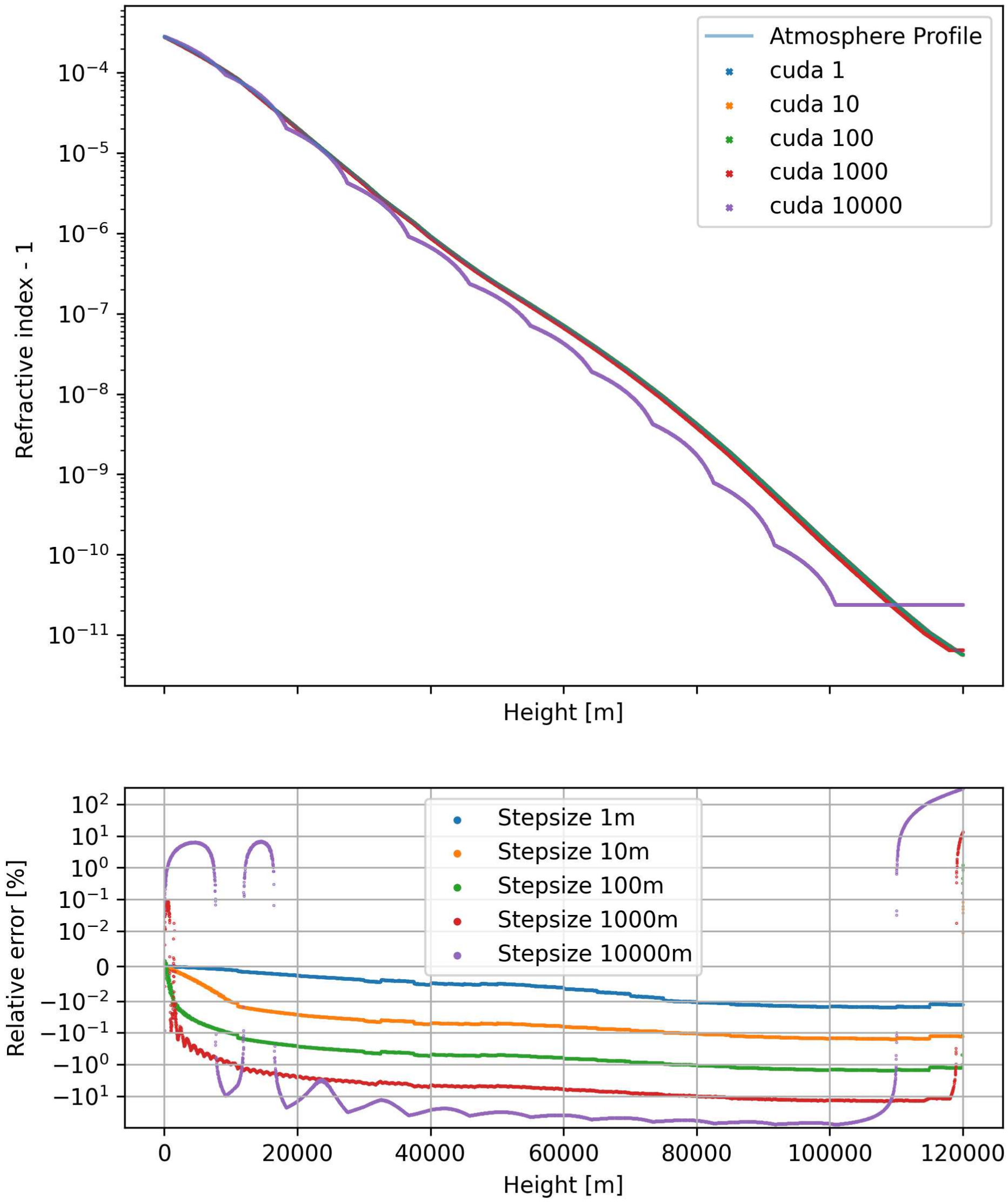

**Figure 5.3:** Displayed are the GPU interpolated values compared to the Exponential and CPU interpolation with varying counts of support points.

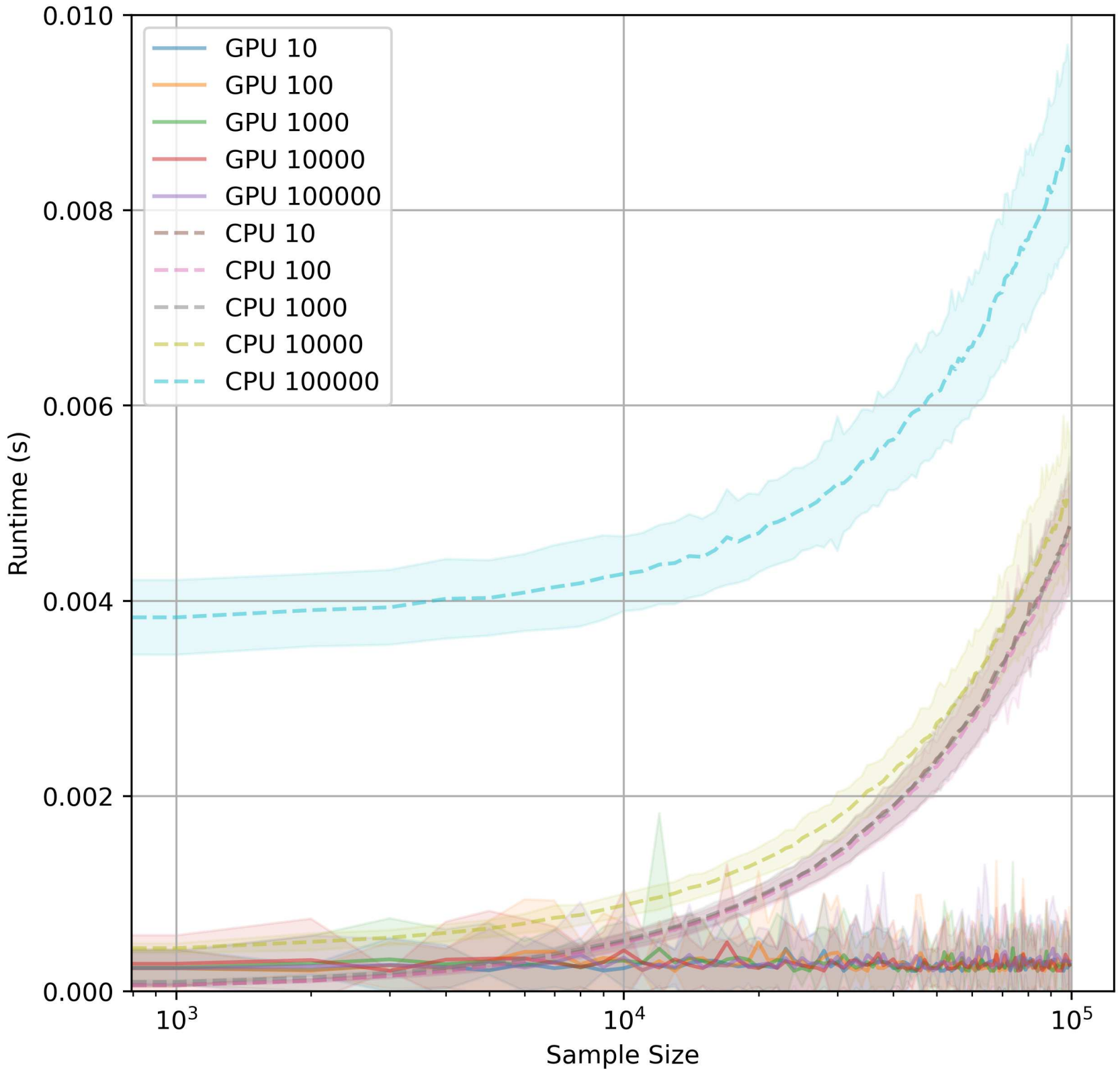

**Figure 5.4:** Runtime comparison of different sample sizes and support distances in m (see legend) for interpolating random positions. The provided curves include the overhead of memory allocation and transfer to the GPU.

## 5.3 Kernel Implementation

The GPU implementation is functionally divided into several distinct stages following the previously mentioned (see Image 5.2) structure. For performance reasons, the separation is only functional and implemented, e.g. through synchronisation barriers on the GPU, not a split into distinct kernel calls, which would need dedicated CPU time to schedule the function via the driver API. This increases the first memory transfer to initialise all data but enables the use of shared memory between the stages because individual calls can only share data via global memory allocations.

### 5.3.1 Import and Filter

The filter stage is the first working stage in the computing pipeline and must include all functionality to ingress the relevant information used. The particle information transferred to the GPU storage area is already simplified and contains only the minimal set of values to minimise data transfer times:

| Name | Datatype | Size in Byte |
|---|---|---|
| Position | double | $3\times$ 8 B |
| Direction | double | $3\times$ 8 B |
| Mass | double | $1\times$ 8 B |
| Energy | double | $1\times$ 8 B |
| Charge | float | $1\times$ 4 B |
| Age | float | $1\times$ 4 B |
| | | 76 B |

The position and direction data are transferred from global memory into the register working memory and checked for the possibility of measurable light production. In addition, mass, charge, and energy are combined into a more straightforward light yield parameter. The age of the photon, with the cascade start set as zero, can be propagated through the GPU or applied afterwards as an offset by bookkeeping the light-generating particles' ages. This value is essential for photon arrival time calculation. If all preliminary checks and filter stages are successful, the particle is stored in shared memory for further work.

### 5.3.2 Photon Generation

Photon generation is done in the same way as explained in 4.2 for CPU. Created photons are transferred to shared memory to be processed further. No detailed history is kept for information like photon origin to limit the bookkeeping workload.

The used A100 GPU has an L1 Cache size of 192 kB, which is shared between texture memory, shared memory and normal cache use. The drivers allows to choose 0 kB, 8 kB, 16 kB, 32 kB, 48 kB, 64 kB, 100 kB, 132 kB and 164 kB as reserved storage for shared memory exclusive use where 1 kB is reserved by the system. Unoptimised, the photon data structure has a size of 60 B, each consisting of the following data:

| Name | Datatype | Size in Byte |
|---|---|---|
| Position | float | $3\times$ 8 B |
| Direction | float | $3\times$ 8 B |
| Wavelength | float | $1\times$ 4 B |
| Age | float | $1\times$ 4 B |
| Weight | float | $1\times$ 4 B |
| | | 60 B |

The default value of 48 kB is sufficient to store ≈783 photons. With enough cache to spare, the size has been increased to 100 kB, which is supported by a wider range of accelerators, allowing up to 1650 photons to accumulate before the propagation routine is executed and the buffer drained.

### 5.3.3 Fluorescence Emission

The light yield is presently calculated for each particle on the CPU to provide a foundation for several experiments. One reason is that CORSIKA 8, in the utilised version, does not provide a standard interface to access the relevant ionisation loss information in the release branch. The solution, as of now, uses a very coarse approximation to test the existing implementation. The number of photons that should be created is then forwarded to the GPU in combination with the particle start and end position to generate and propagate the photons. The wavelength of each photon is sampled according to the probability distribution of each relaxation channel [16].

### 5.3.4 Photon Tracing

For the number of photons needed to be propagated, more than $7 \times 10^8$ Cherenkov photons alone for an average 1 TeV vertical EM-Shower (without filtering applied), the precise methods described in 3.4 are even on acceleration hardware computing intensive and provide too little benefits for the majority of propagation through the atmosphere. Therefore, the much faster interpolation method described in 4.4.1 transfers the photons to ground level or a bounding box around the experiment. In the first step, the tiny surviving fraction of upwards or, if inverted (e.g. balloon experiments), downwards-going photons are discarded. This removal avoids additional condition checks in the used algorithms later on. The remaining photons are then projected to the observation level via the straight-line methods and checked if they are inside the telescope array's bounding box.

## 5.4 Data Aggregation and synchronisation

With GPUs' high throughput, the propagated particles generated by a single-threaded CPU instance of CORSIKA are insufficient to keep the GPU fully saturated. With an internal bandwidth peeking at 2 TB/s and a minimal CPU - GPU data transfer rate of 32 GB/s, depending on the exact configuration, the GPU can, in theory, process up to several million track segments per second.

Even with all the calculations for photon generation and propagation, which is the limiting factor, the number of particles processed can still be much larger than what is currently being produced. With CORSIKA 7 being the fastest implementation, the number of particles generated per second, displayed in figure 5.5, is used as a benchmark.

As there is currently no widespread support for native multithreading in CORSIKA 8 [48] to generate enough particles, the classic "physicist parallelisation" of running the program multiple times has to be used. Given the technical problem of sharing the GPU between multiple instances (which is done with slow context switches) and the fact that several tens of thousands of tracks are needed for each kernel call, a separate program that collects everything and deploys the GPU code is the better solution. Another advantage is the possible use of GPU acceleration for CORSIKA 7 and the crosschecking between the two versions.

To transfer the track segments between the different programs several methods exist:

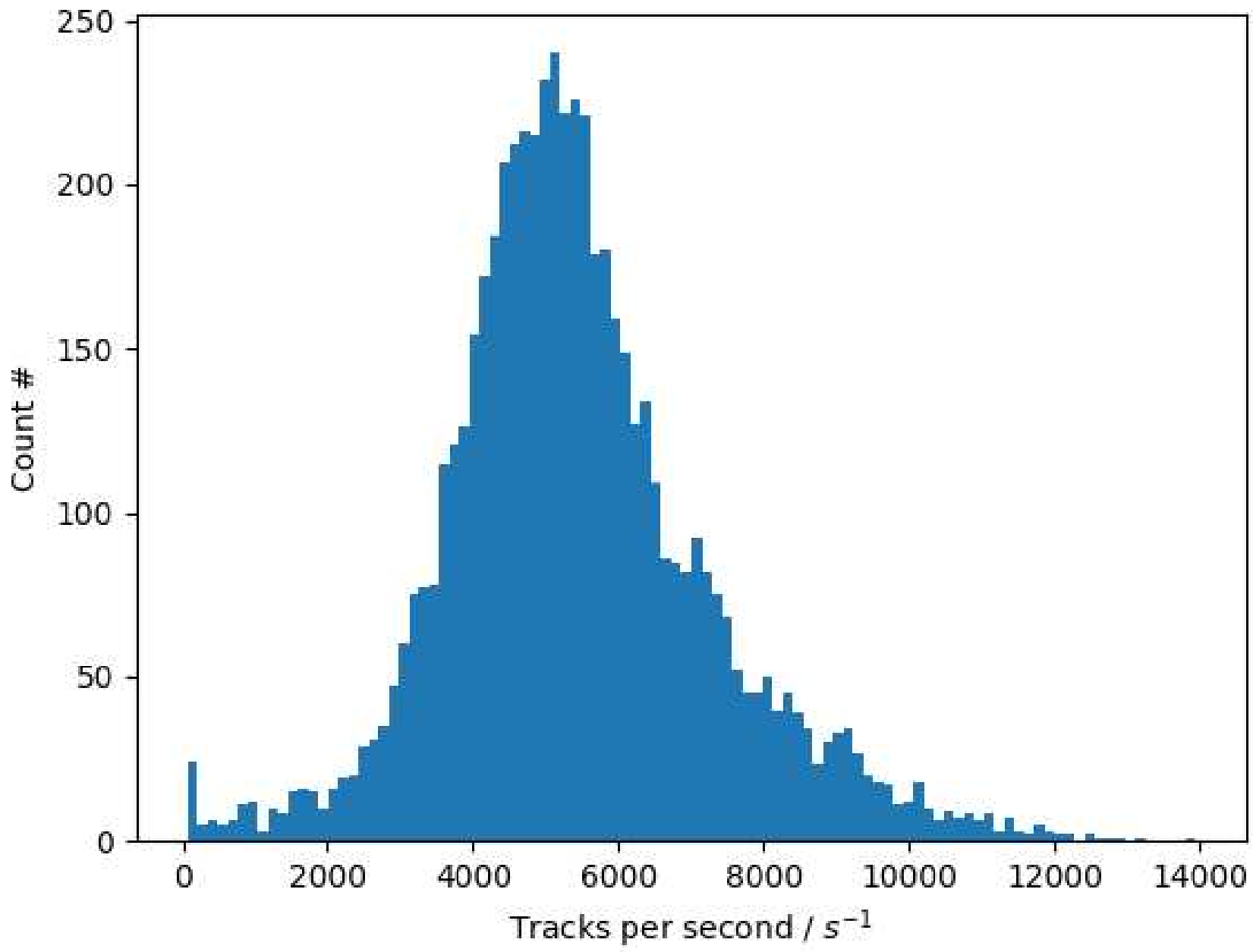

**Figure 5.5:** Displayed is the number of tracks per second provided by a single-threaded CORSIKA 7 instance. The displayed values are the average over 30 seconds.

- **Network Socket** - Transfer via IP Ethernet Network is the most flexible solution. With a theoretical maximum throughput of 40 Gbit/s to 400 Gbit/s for the available machines and latency around 100 µs date, production (e.g. CORSIKA) and the GPU Jobs can be placed on completely different machines. Keeping the machines in the same datacenter, or at least densely connected, is still mandatory to have low latency and high bandwidth. In theory, the data can be transferred over the internet, but it will significantly slow down the overall process so much as not to be feasible anymore.
- **Shared Memory** - The fastest way to transfer data between processes is to use shared memory. The data is directly transferred between the processes without copying it to the kernel. The downside is the need to keep the processes on the same machine and the limited memory available. This is the recommended way to transfer data for the GPU Machines available.
- **File System** - The slowest way to transfer data is to write it to the file system and read it back in the other process live or later. This method is the most flexible way to transfer data between processes and machines but is generally not recommended for GPU Jobs. The file system is orders of magnitude slower, and the storage used memory can exceed TB even for small runs.
- **Infiniband** - The fastest way to transfer data between independent machines, with throughput similar to Ethernet but much lower latency and overhead. The downside is the need for specialised hardware and the locality; the machines must be in the same data centre and often physically close.

The communication between programs was implemented with the help of ØMQ (pronounced Zero MQ), the open-source universal messaging library [41], [3]. With its different transportation methods and the availability of several hardware platforms, OS Types, and programming languages, the library is the perfect fit to provide an easy-to-use and flexible interface.

To provide flexibility over several use cases in CORSIKA 8, which includes the Cherenkov methods, Radio simulation, as well as custom modules outside of the simulation framework (e.g. in other programming languages for rapid prototyping), two distinct patterns, displayed in figure 5.6, were implemented. Both patterns provide the essential feature of automatic throttling to avoid data loss. Through this, the producer is slowed down in the case of a slower consumer to prevent memory overflow. The internal cache handles peak loads in case of higher complexity, e.g. storing checkpoints without slowing down the overall simulation.

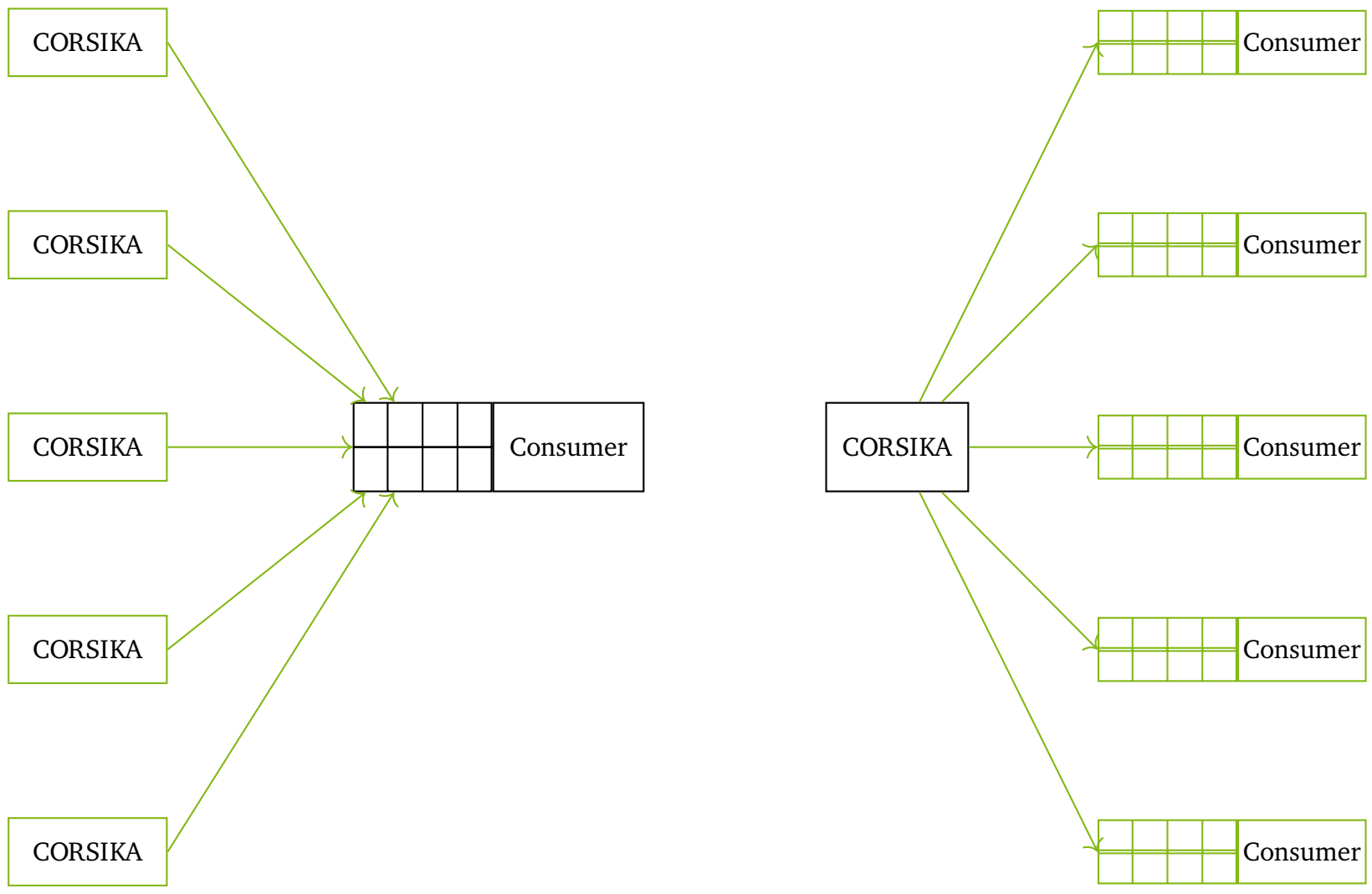

**Figure 5.6:** Displayed is the schematic representation of the two provided design patterns in the synchronisation modules available for CORSIKA 7 and CORSIKA 8. It represents the "Many-to-one" pattern, where multiple simulations send data to a single consumer, and the "One-to-many" pattern, where a single simulation sends data to multiple consumers.

# 6 Comparison and Results

## 6.1 Synchronisation

The synchronisation methods implemented in CORSIKA 7 and 8 make it possible to communicate with arbitrary programs as long as the interface is implemented. Sample implementations for the programming languages C++ and Python were developed and act as documentation or an introduction for new users. While the C++ implementation can directly use the ZMQ library and has access to the required binary conversation from the raw data into the implemented class structure, a pure Python implementation would experience additional overhead in converting incoming datatypes. As a solution, the data receiving and conversion is kept in C++ and exported through a pybind11 [46] interface exposing the relevant information.

The primary purpose of this implementation is ease of use and the broadest possible applicability for rapid prototyping of ideas. In the case of CORSIKA 7, this means that all usable data is exported:

- Particle tracks
- Cherenkov particle tracks
- Cherenkov track segments
- Interaction points
- Generated Cherenkov photons
- Event and simulation headers

Without a specific focus, the amount of information sent is large, creating a significant overhead. A single data segment, called a packet, consists of a character-based identifier and one of the above items as payload in varying sizes. It is created and sent when the data is available, resulting in very low latency. Measuring the effect on the simulation in the default configuration against an empty client, which immediately logs and discards all messages, shows that the raw packet throughput stabilises at $\approx 660\,000\,1/\mathrm{s}$ with a data transfer rate of $\approx 130\,\mathrm{MB/s}$. Disabling message sending increases performance to $\approx 4\,600\,000\,1/\mathrm{s}$ in packet generation rate, a factor of 7. A reduction to only export the required data can significantly increase the throughput, with photons and Cherenkov track segments having the most significant impact.

An impact study with CORSIK A8 is not relevant at this stage; with the much slower speed of the simulation, the impact will be quite small in percentage terms. The implementation provides basically the same top-level list of information as CORSIKA 7 but differs in some details. Hence, separate header and data packets are used. An additional feature is the inclusion of particle tracks, which allows direct comparison of secondary effect implementations such as Cherenkov or Radio on the same shower, completely avoiding the otherwise statistical comparison of distributions.

## 6.2 Cutting Strategies

The cutting strategies with their five different distant metrics explained in section 4.3.2 are listed here again:

1. Center of array
2. Closest to array outline
3. Center of the telescopes
4. Closest to 2D telescope
5. Closest to 3D telescope

To provide a, preferably, universal recipe or guide on efficient cutting strategies, different telescope constellations were tested with a modified Simulation, which labels if and how many photons from a track reach a bounding sphere around the telescope after they were propagated through the atmosphere. The resulting cuts are by no means optimal because angular acceptance is completely ignored, but they provide a good starting point for all experiments if further optimisation is needed. A CTA-North and -South-like example configuration, details are attached to the appendix B.3, is specifically selected for its ongoing importance.

### 6.2.1 Early particle removal

The first data reduction stage - particle removal - as explained in 4.3.2, is one of the biggest algorithmic changes and requires rigorous testing because of its potentially large negative influence on the simulation. For the three most useful options, angle to the array centre, angle to the individual telescope centres, and minimal angle to the telescope bounding spheres, the resulting distributions are displayed in the three figures 6.1. They display the number of sub-tracks with the corresponding angles that do or do not contribute to the detectable light emission. It becomes visible that no clear separation can be found for extended telescope arrays when only one parameter is considered. By adding the distance in a 2D Plot (Figure 6.2 and B.2) a clearer separation can be found:

$$y/x^{b_{\text{cut}}} < a_{\text{cut}}$$

Where $x$ is the angle and $y$ is the emission height, $b_{\text{cut}}$ the lines angle, and $a_{\text{cut}}$ the lines offset. The values must be determined for each experiment individually.

### 6.2.2 Photon removal

The removal of photons, the second data reduction stage, can be done after the propagation correction when the photon position is known precisely or with wider margins before the correction is applied. The second method significantly reduces the number of table lookups required for applying the correction. The three figures 6.3 show the effects of the removal process by displaying margins against the number of photons surviving or accidentally removed for the example production up to $60°$ inclination. With a removal rate of $\approx 80\,\%$ and more, all three options provide good results. More problematic is the low purity of detectable photons with $0.04$ and $0.06$ respectively, when used to determine which photons should be stored directly. As a result, the file size would increase significantly, causing problems in several places. Depending on the storage system and statistics required, a size increase of a factor of 2 can be reasonably accepted. However, only the best and slowest to calculate method with the 3D distance to the telescope's bounding sphere would provide a reasonable file size of this level.

Due to the higher storage usage, along with the additional time required for file writing and later analysis, these filters can not be recommended to determine which photon to store exclusively. A better solution is the addition of a third data reduction stage after the propagation correction, which directly stores the photon output in telescopes' specific file storage.

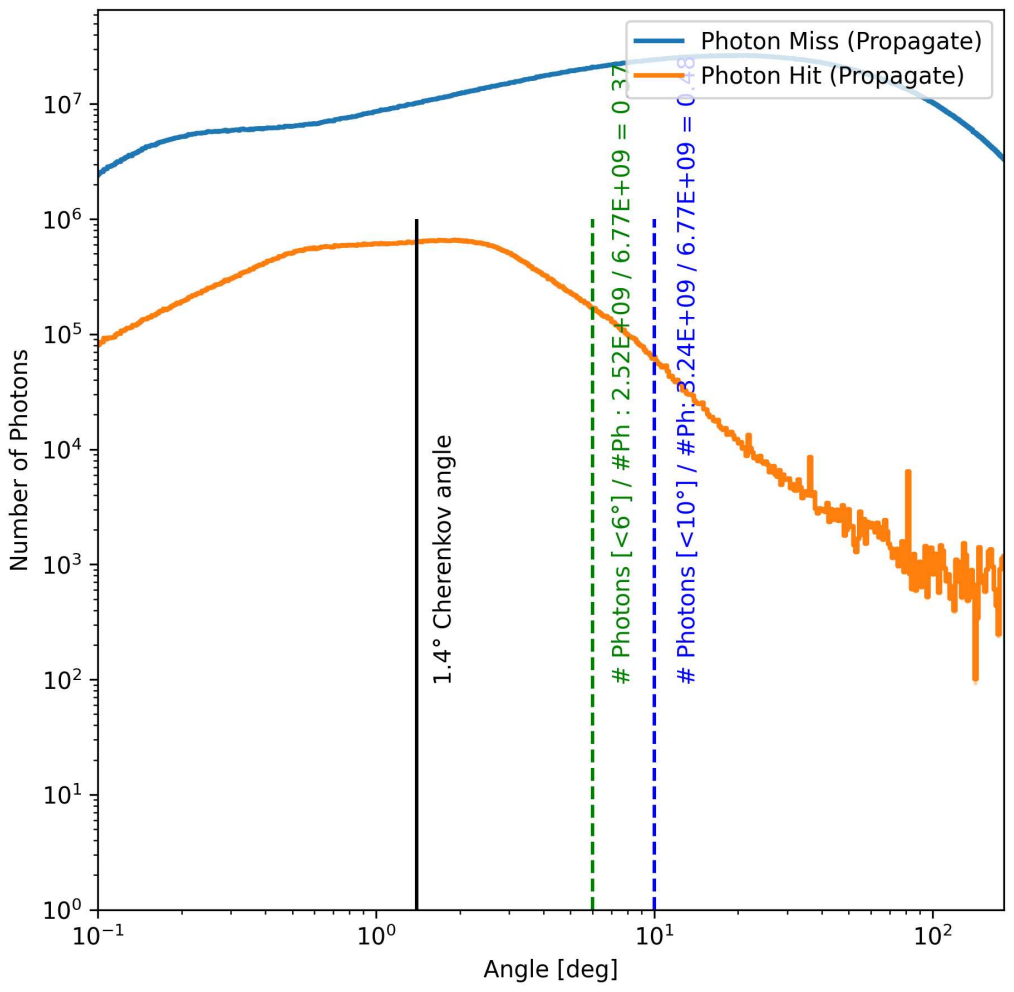

**a:** Center of array

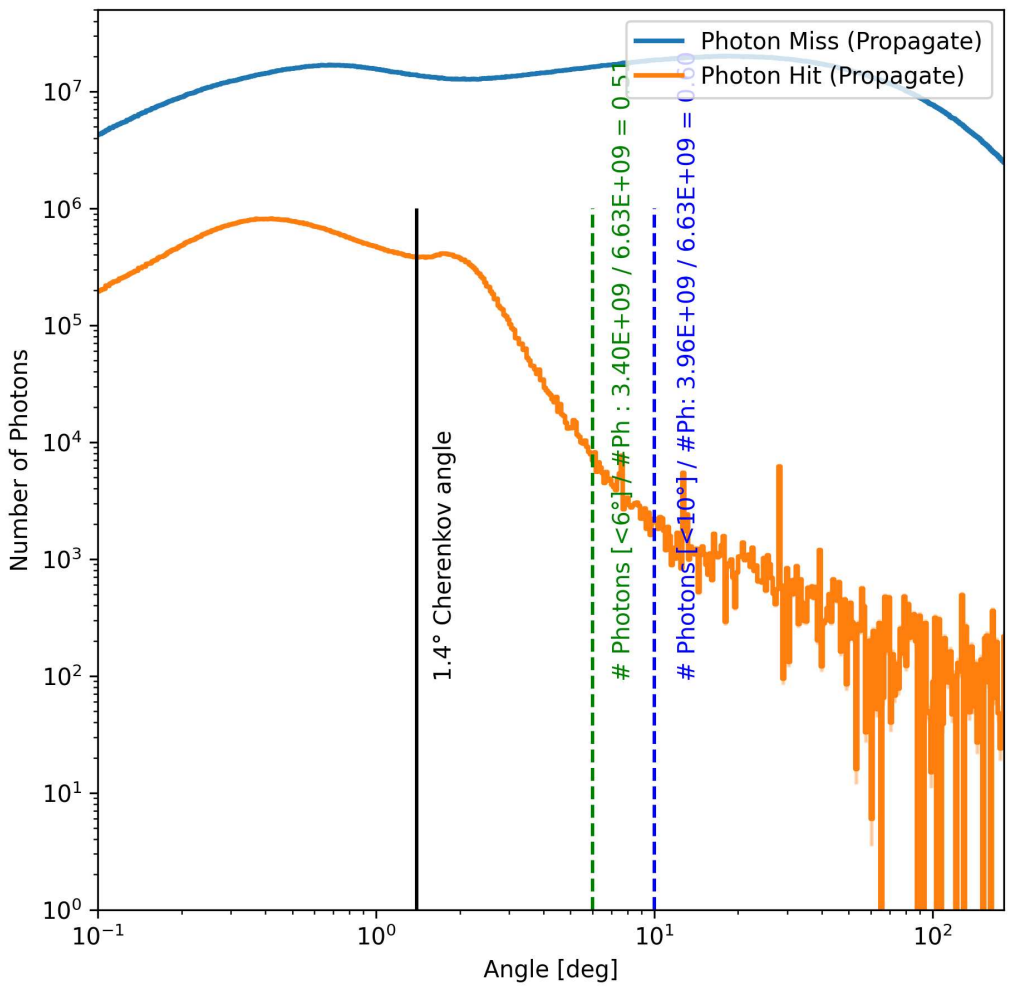

**b:** Center of telecopes

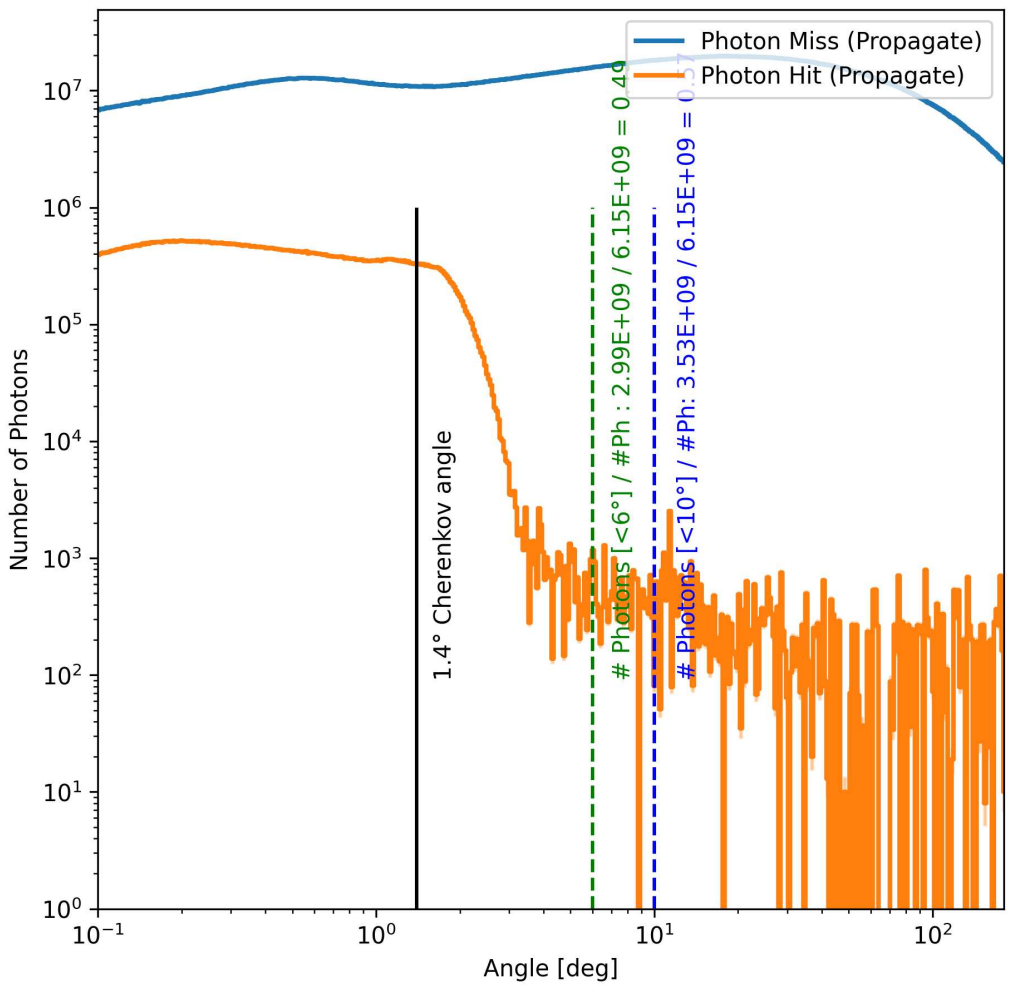

**c:** Bounding sphere

**d:** Vertical Photon Cascade

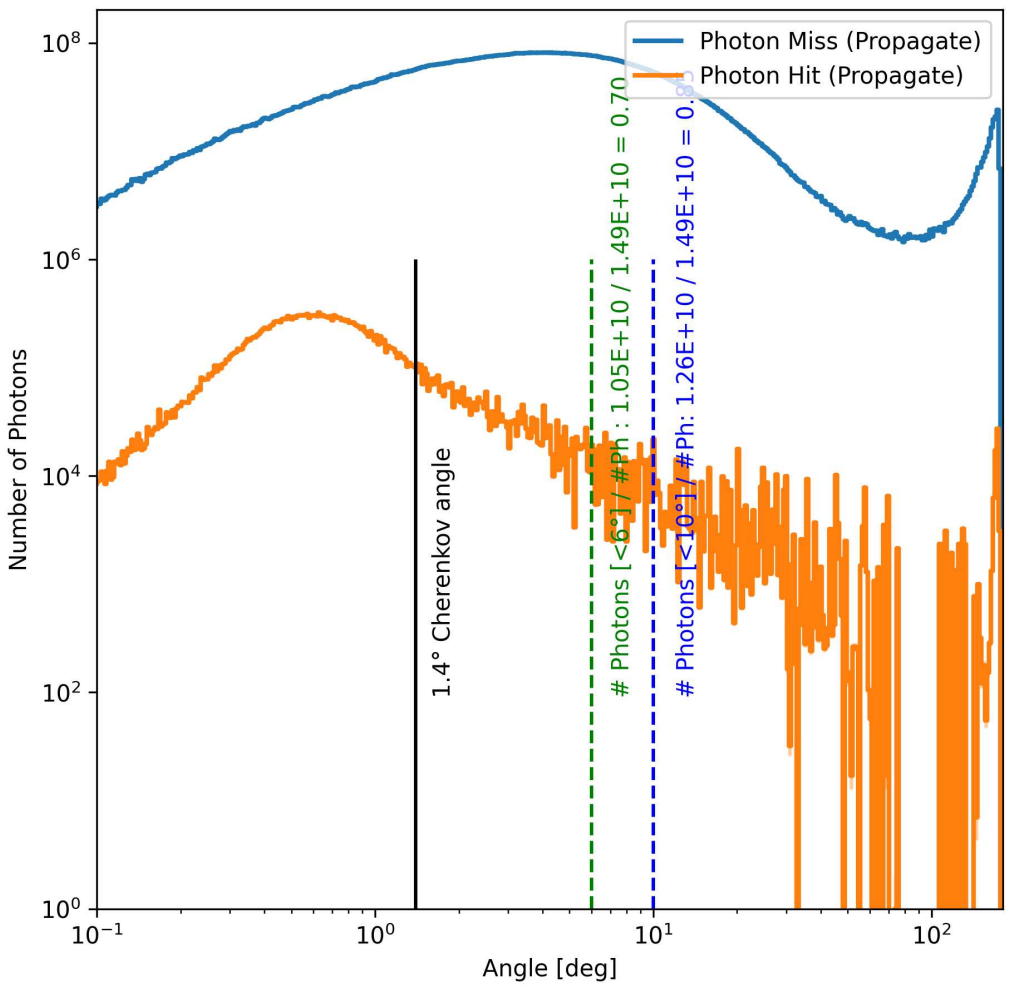

**e:** Center of array

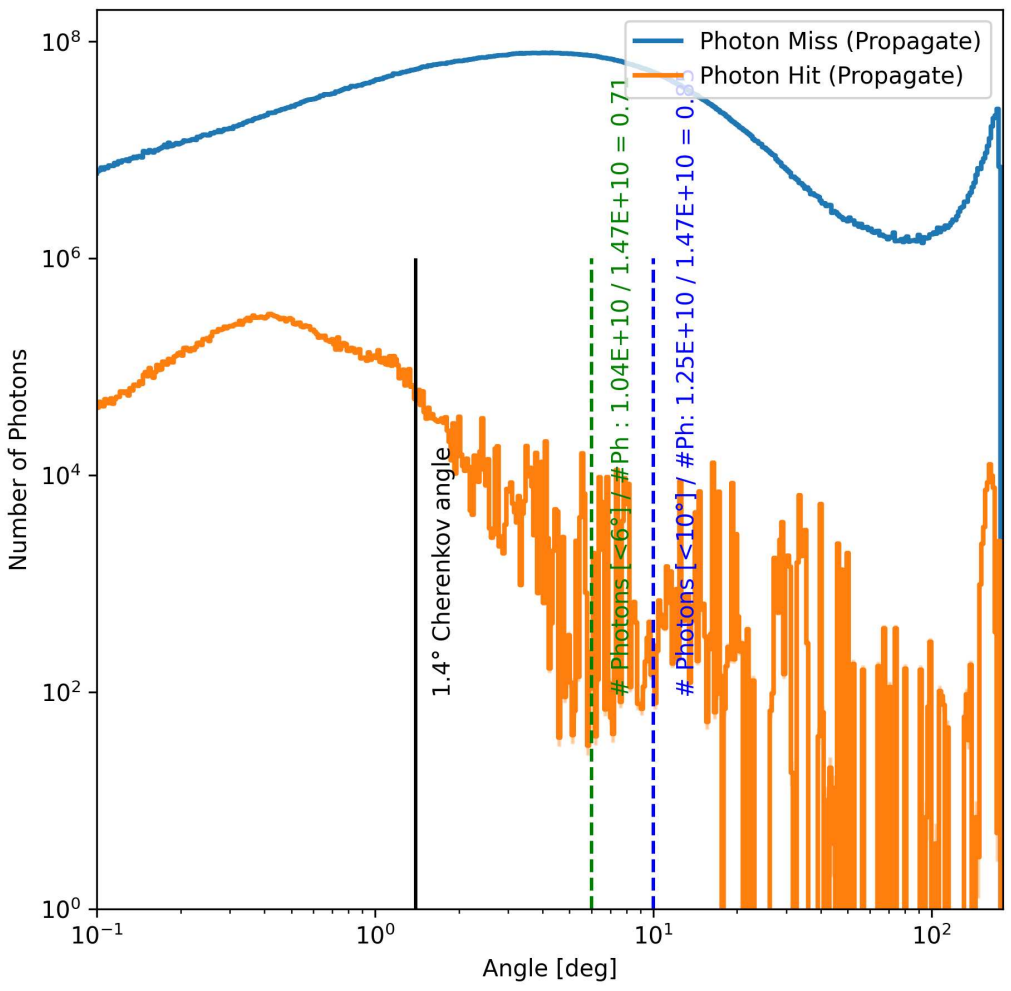

**f:** Center of telecopes

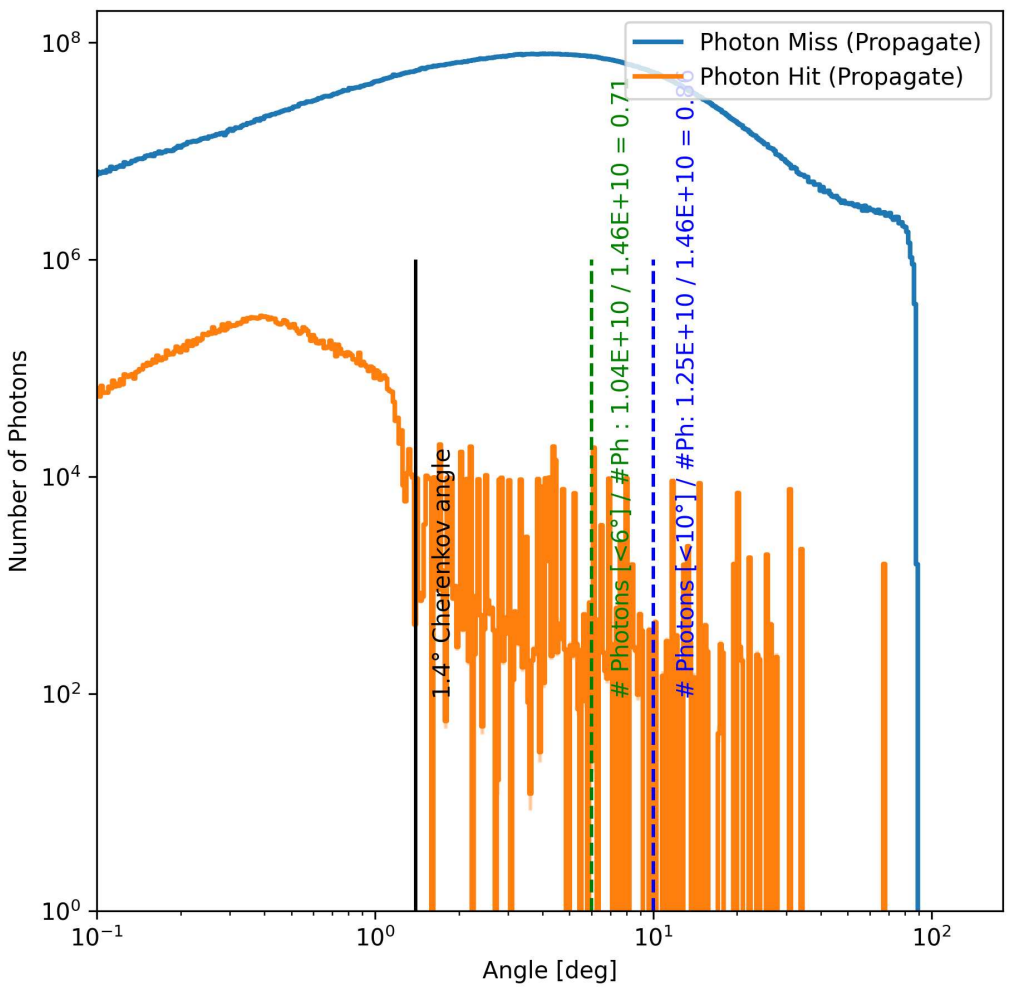

**g:** Bounding sphere

**h:** 0° to 60° Proton Cascade with $500\,\mathrm{m}$ Scattering

**Figure 6.1:** Displayed is the minimal angle between the particle's flight direction and the vector from the track centre to a specific telescope feature. Particles that contribute to the detectable light, defined as hitting the bounding sphere of the telescope, are displayed in blue, the others in orange. The plots were generated with 200 individual cascades of $500\,\mathrm{GeV}$ each.

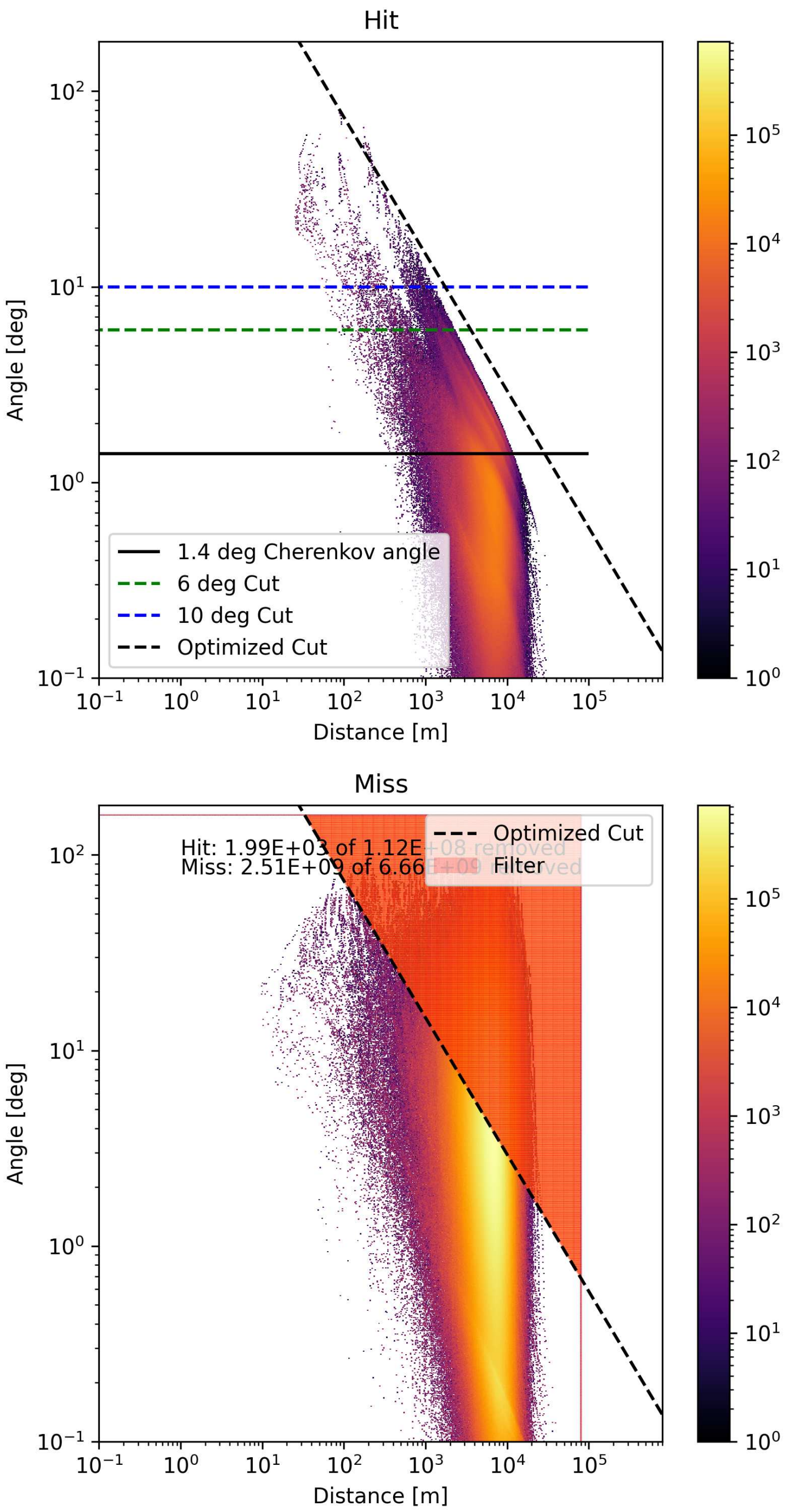

**a:** Center of telecopes

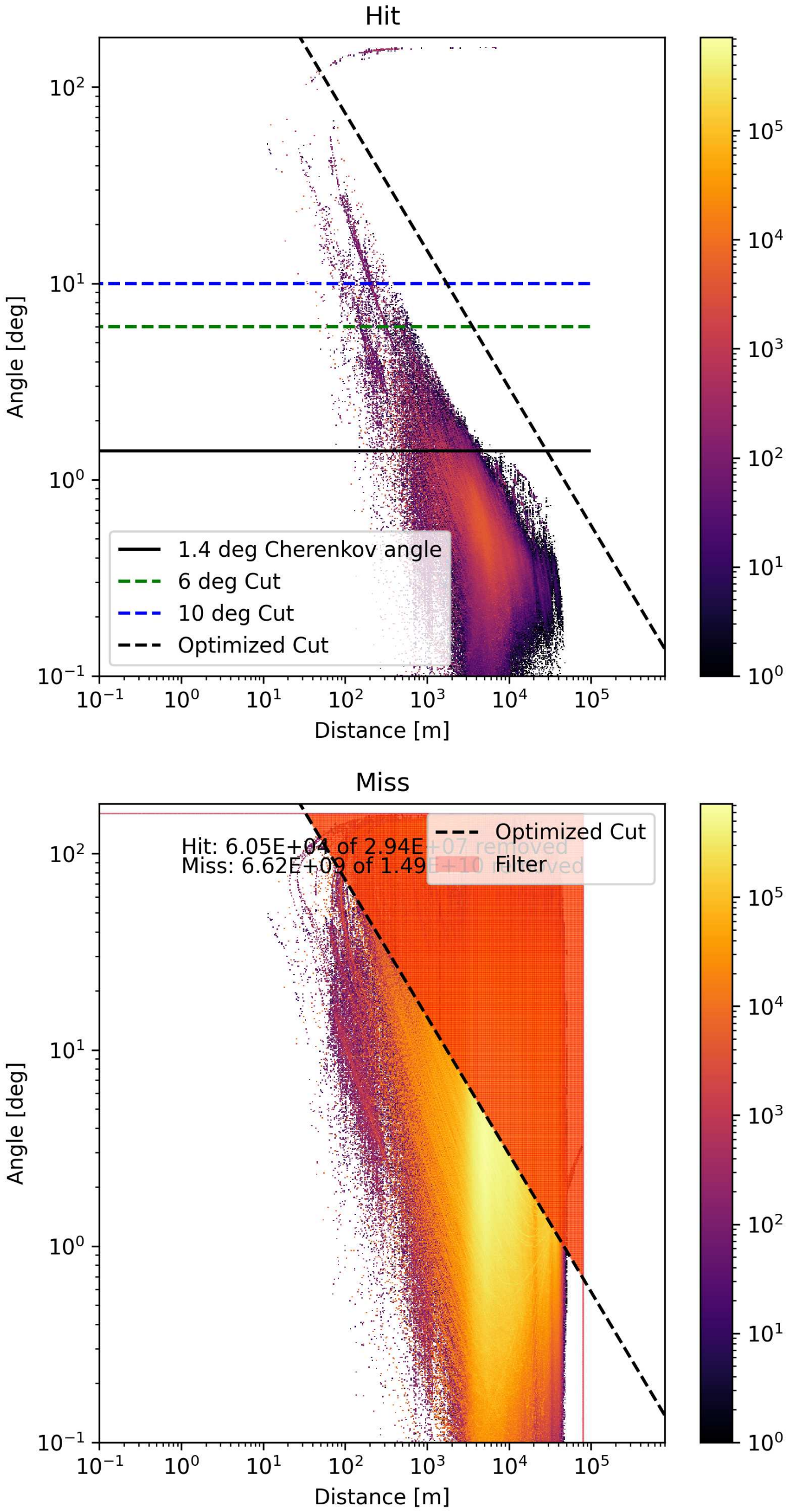

**b:** Center of telecopes

**Figure 6.2:** Displayed is the minimal angle between the particle's flight direction and the vector from the track middle to a specific telescope feature against the distance to the array centre. Particles that contribute to the detectable light, defined as hitting the bounding sphere of the telescope, are displayed in blue, the others in orange. The plots were generated with 200 individual cascades of 500 GeV each.

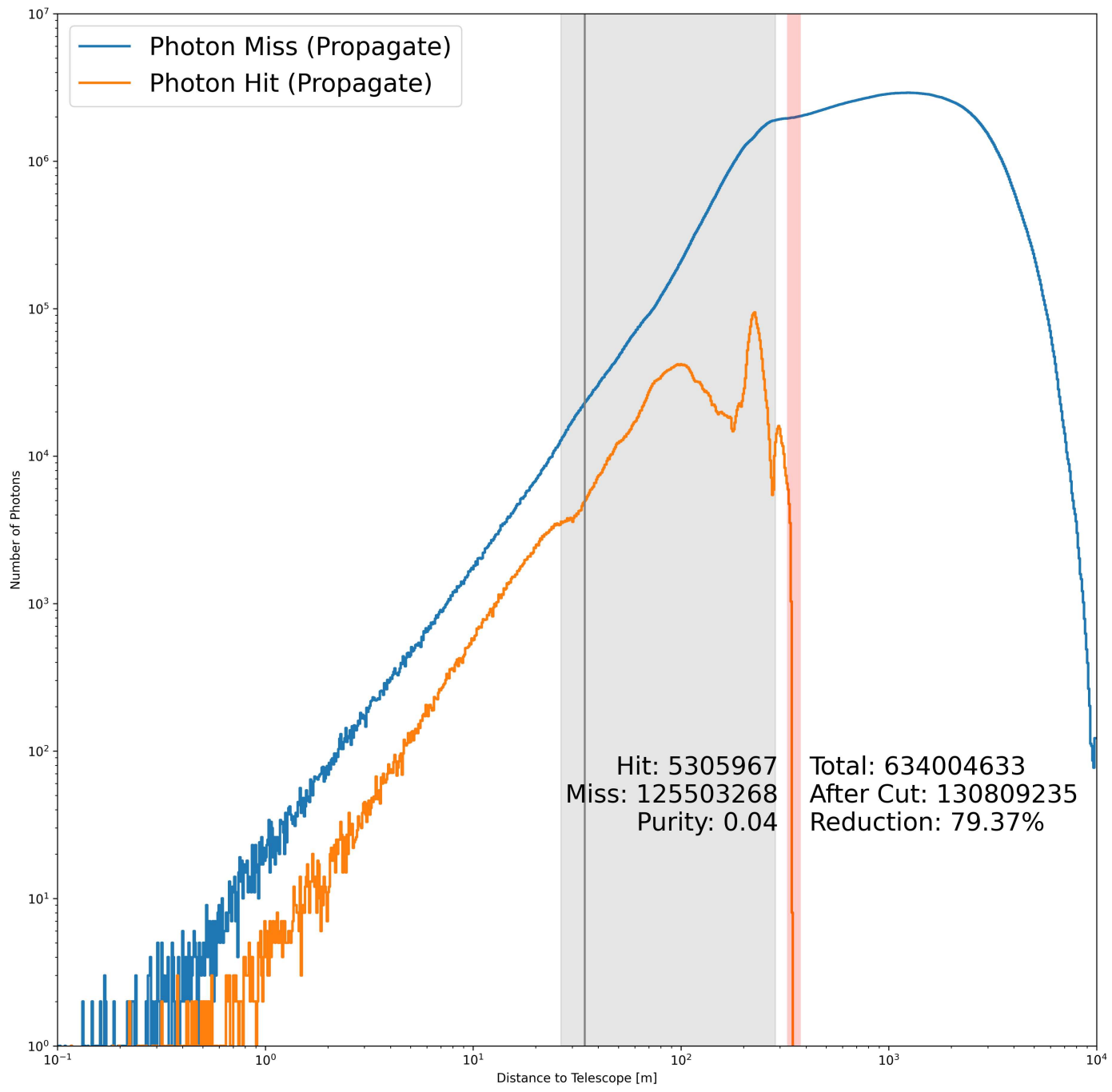

**a:** distance to array center

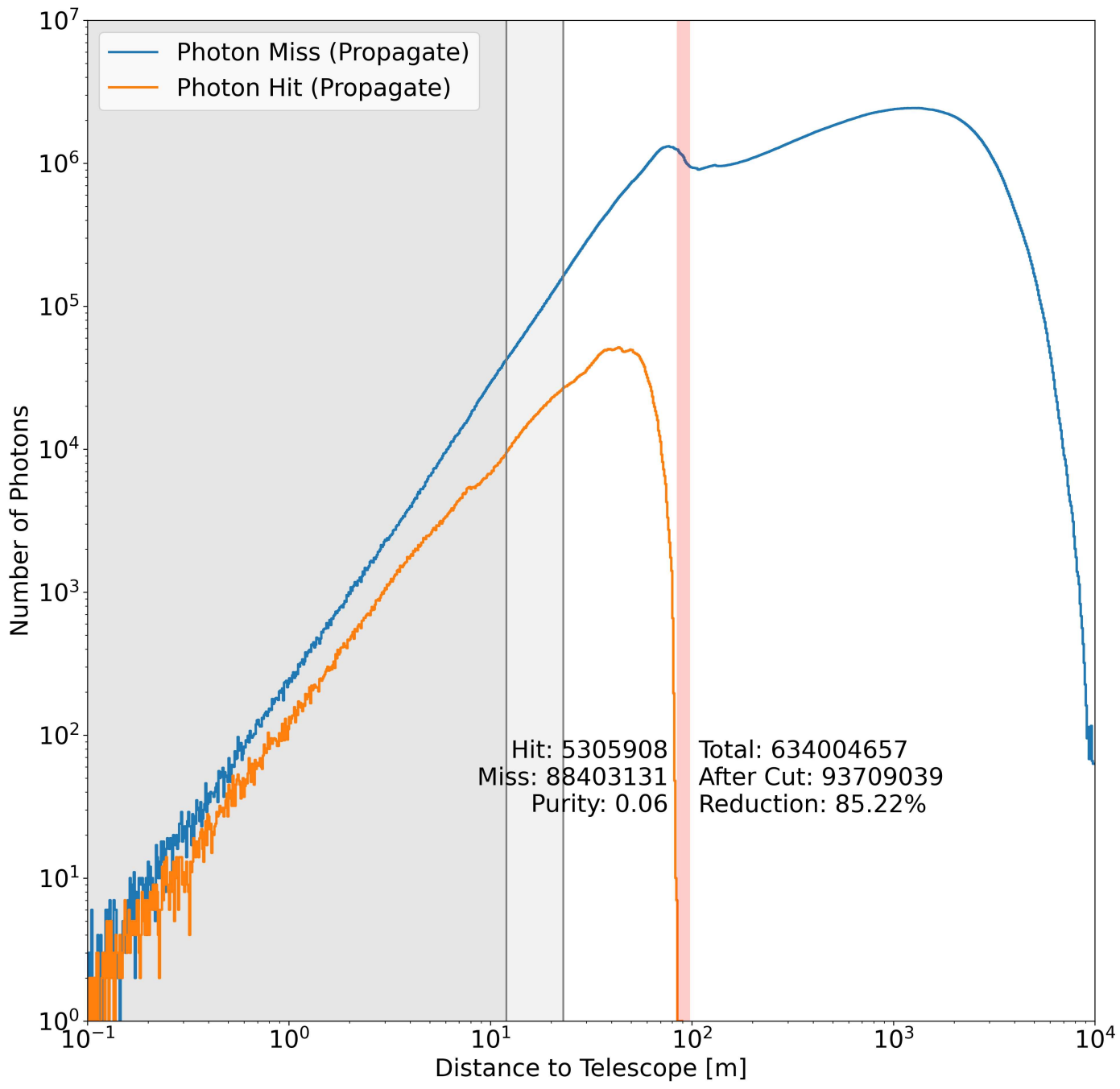

**b:** Minimal distance to telescope centers

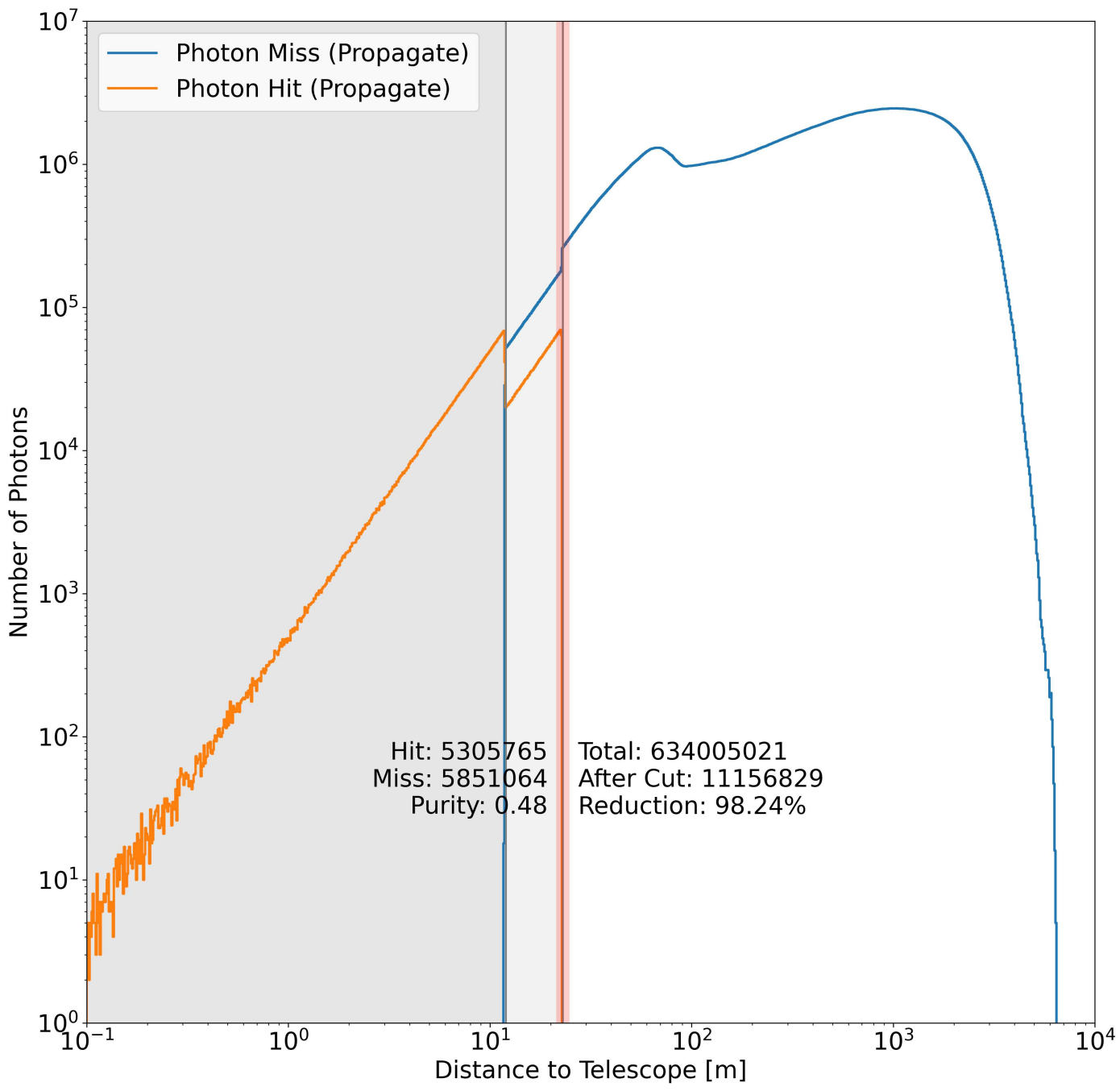

**c:** Minimal distance to sphere around telescopes with telescope size as radius

**Figure 6.3:** Displayed are different distance measurements of photons generated after the particle removal filter (with a simple $10°$ Telescope Center cut). The grey-shaded section displays the physical area where a photon can hit the telescopes if it were a 2D Circle on the ground. The red vertical line marks a possible cut position with the results written beside it.

## 6.3 Propagation crosscheck

All implemented propagation algorithms are cross-checked with the theoretical model for a linear atmosphere to ensure that they produce physically correct results with the required accuracy. Furthermore, the findings are compared to the CORSIKA 7 simulation, validated by its extensive use throughout the research field and further systematic studies [59, 64].

### 6.3.1 Theory

The analytical solution previously calculated, which is fully described by the formula derived theoretically in section 3.4.3, is applied to a, with height decreasing, linear refractive index profile. The numerical methods are applied to the same atmosphere and compared to the theoretical ”true” light path. The refractive index values used, $n_1 = 1.5$ at ground level to $n_0 = 1.0$ at the height of $100\,\mathrm{km}$, are higher compared to the natural atmosphere to amplify potential deviations. The method allows the direct observation of variations in the result of the numerical method originating from, e.g. step length or changes in the algorithm. The figures 6.4 display a selection of the obtained results. It is visible that the numerical models agree with a high relative accuracy with the theoretical model. The deviation does not exceed $0.1\,\%$ even for high Zenith angles. Extrapolating the results to arbitrary atmospheric models should work as long as the scale of variation exceeds the steplength by at least an order of magnitude.

The steplength resolution shows only a slight impact in precision depending on the algorithm used. The deviation is below $1\,\%$. With quasi-linear scaling in runtime, 4.2, and only a single execution at the beginning of the simulation, the adverse effects of a small step length are not too significant; therefore, a higher choice is often the more secure method.

### 6.3.2 CORSIKA 7 - Replay

Comparisons with CORSIKA 7 are not done on an individual photon basis but on the cumulative Cherenkov light pool on the ground level. Here, not only the direct comparison between CORSIKA 7 and CORSIKA 8 is important, but isolating differences between the Cherenkov Codes is required to avoid systematic differences between the fundamental simulations. This is possible by injecting CORSIKA 7-generated particle subtracks via the synchronisation module (see 5.4 for details) in CORSIKA 8. The results, called CORSIKA Replay, work on binary identical particle tracks, and any differences in the light output can directly be attributed to differences in the module. Differences between both simulations do not directly imply a bug in any of the two versions but can originate from different simplifications and algorithms used in the propagation. CORSIKA 7, for example, utilises a planar atmospheric model for most parts of the Cherenkov light propagation.

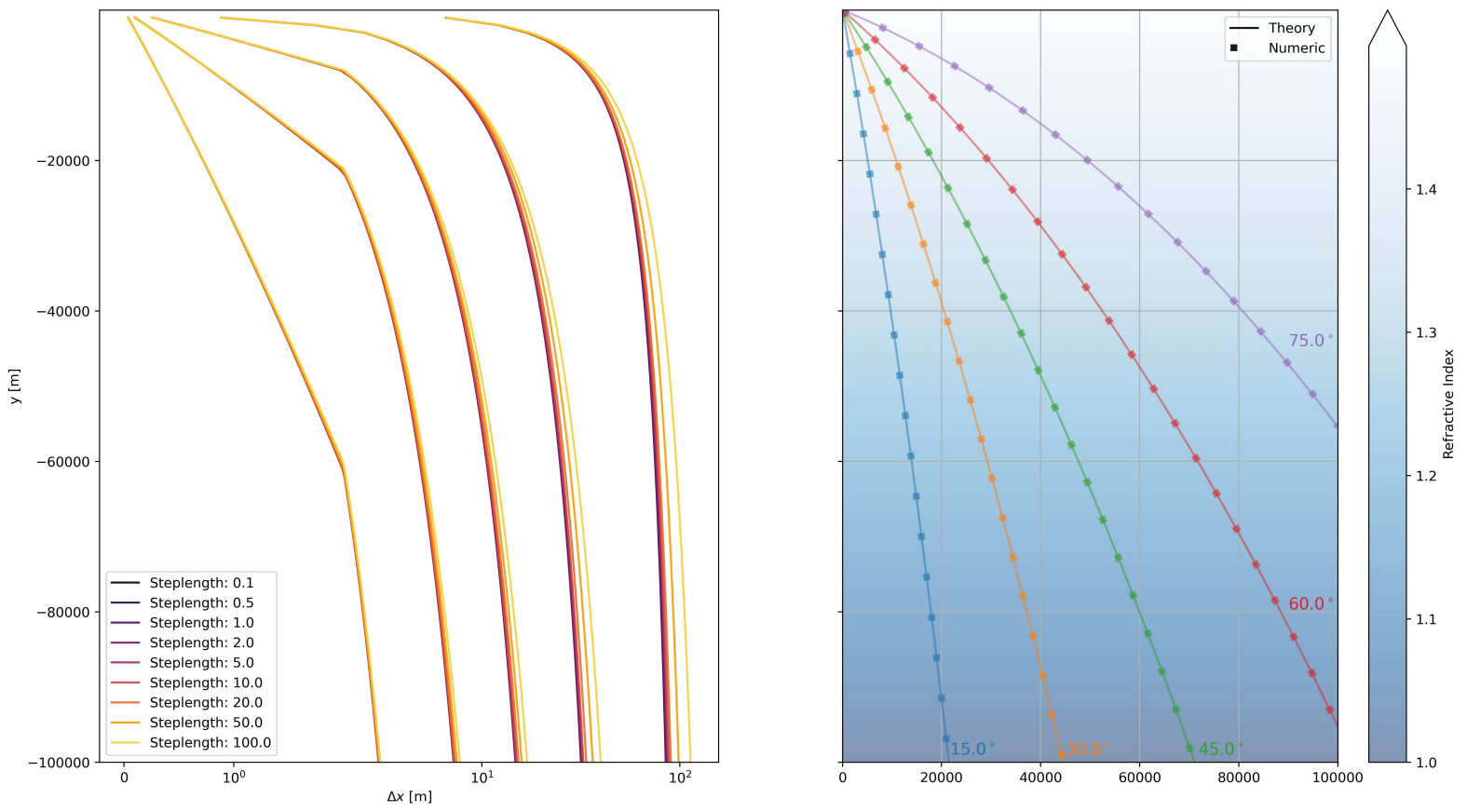

**Figure 6.4:** Displayed is the direct comparison of the numerical propagation against the theoretical derived for a linear refractive index medium. The direct path is displayed in the upper figure, and the relative variation at different heights is in the lower plots. For a better comparison, different steplengths were used.

Several values can be compared, with the highest importance being the intensity distribution on the ground level and the arrival time. If the magnetic field is deactivated, the lightcone displays nearly rotational symmetric behaviour for vertical showers. This enables the reduction to radial coordinates and a direct comparison as done in image 6.5a. For the arrival time points, the averages are compared to each other. Here, radial and 2D distribution allow similar conclusions and are displayed in 6.5b and 6.5c respectively.

### 6.3.3 CORSIKA 8

A comparison of the complete CORSIKA 8 simulation stack with CORSIKA 7 is included in the figures 6.5. An absolute comparison of the arrival times is impossible due to internal variations in the CORSIKA 7 and 8 timing calculations. However,

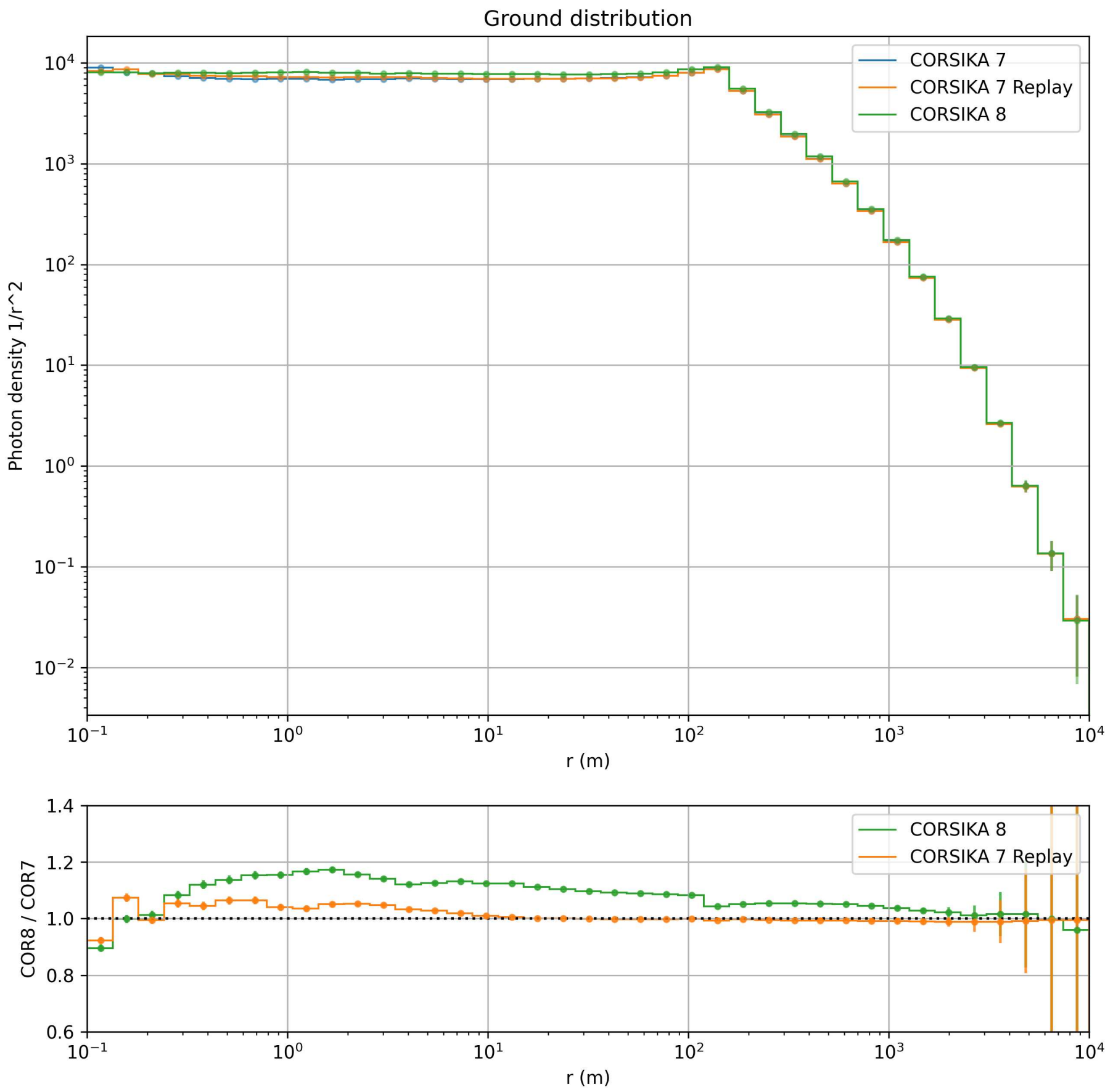

**a:** Radial distribution of the Cherenkov light pool

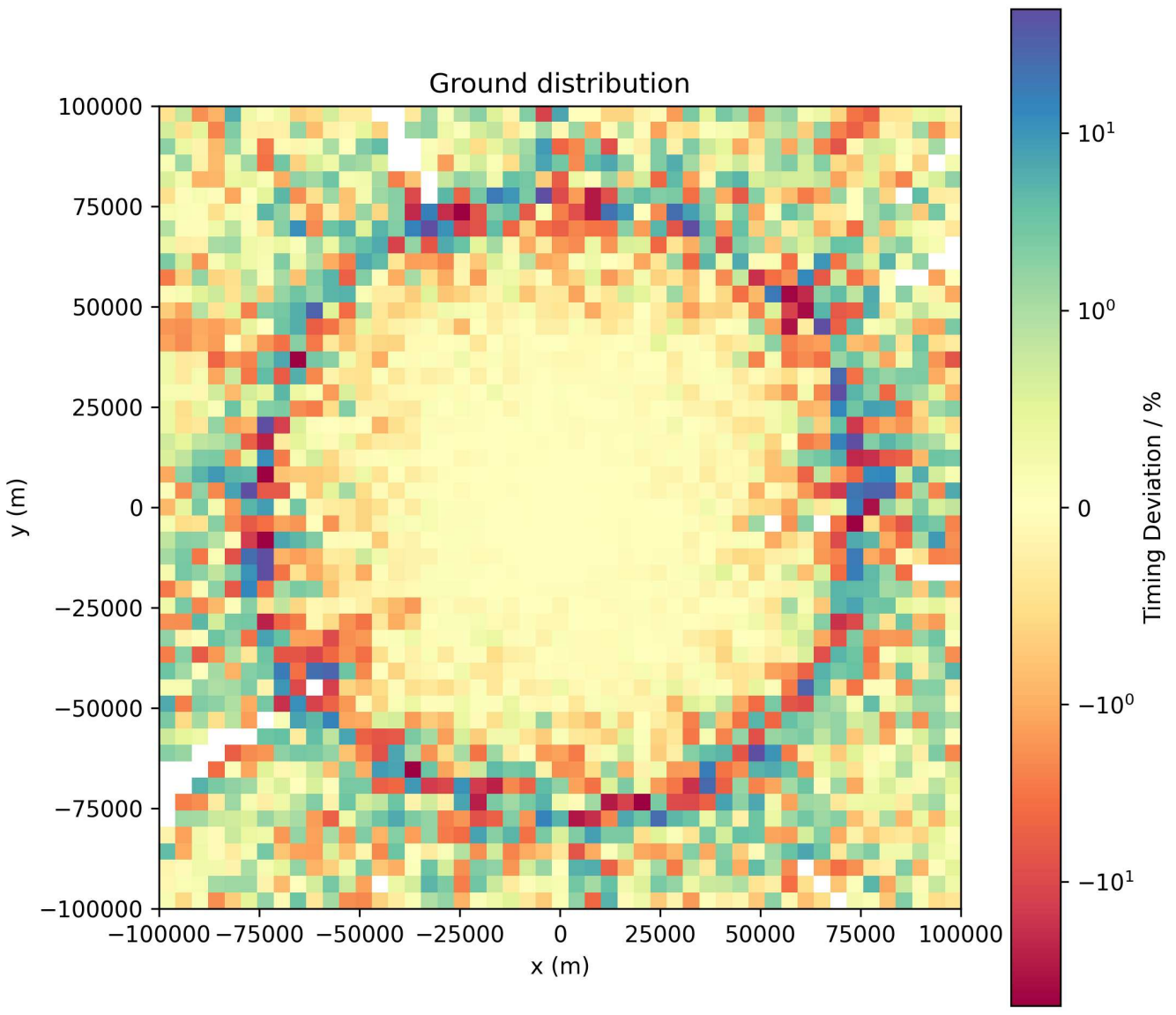

**b:** CORSIKA 7 Replay / CORSIKA 7

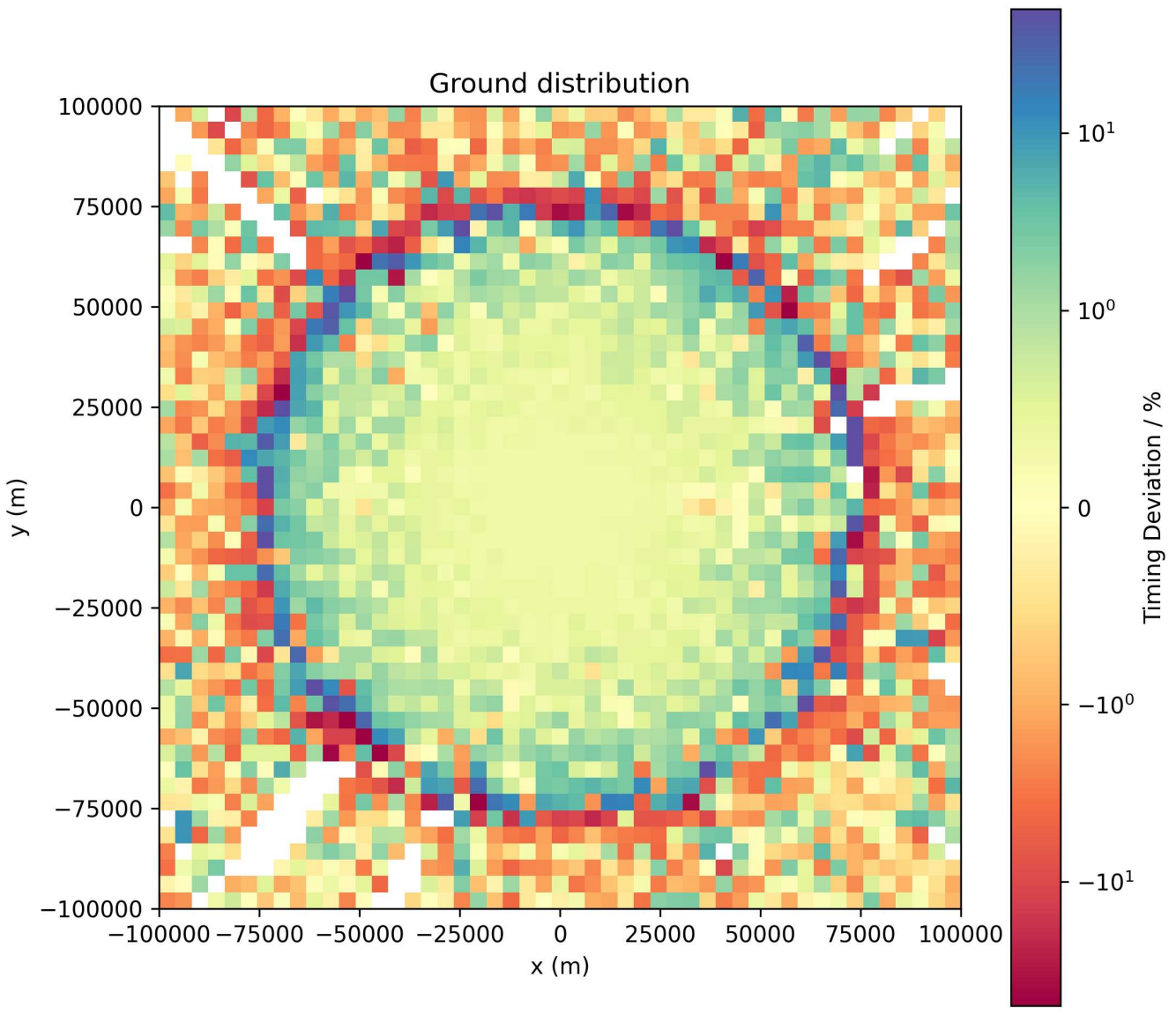

**c:** CORSIKA 8 / CORSIKA 7

**Figure 6.5:** Comparison of the Cherenkov light pool of the CORSIKA 7 simulation and the replay of the same data with the different propagation algorithms used in CORSIKA 8. The upper figure displays the radial distribution of the Cherenkov light pool. The second and third figure displays the deviation in average arrival time.

to make a comparison, the average arrival time of all photons is taken as the zero baseline and deviations are shown relative to this value.

The overall differences in time and radial profile are minor, but compared to the replay results, they are larger than can be explained statistically. The deviation can be attributed to the different propagation algorithms used. However, the difference does not exceed $20\,\%$ in the radial profile and is under $5\,\%$ for the central shower region.

CORSIKA 8 is still under development; with each release, changes may affect the distributions generated by the Cherenkov modules. Therefore, an automated test setup has been used to trigger a re-run of a pre-defined sample to check for excessive differences between software versions and CORSIKA 7 to detect introduced bugs early.

## 6.4 Last Mile Simulation

In previous releases of CORSIKA 7, the photon impact positions on the ground were used as the simulation output. Experiment-specific code was then used to read the photon data from the storage and simulate hardware-dependent properties. A typical step for most is the ray tracing of the photon path through the detector hardware. This often includes shadowing from the experiment's truss structure or camera, reflection from optical mirrors and, in rare cases, bending due to lenses or light collection structures. A raytracer can replace the photon propagator in a defined area around the experiment to avoid re-implementation and decrease the number of photons stored at the end of the simulation. Currently, the following features are natively supported in this photon raytracing method:

- Reflection on a plane (e.g. first surface mirror)
- Light bending at a boundary layer (e.g. lens or Winston cone)
- Light bending through an inhomogeneous medium (e.g. atmosphere)
- Absorption by objects

The figures 6.6 show the results of the implemented features for a light guide and a MAGIC like telescope with $17\,\mathrm{m}$ diameter parabola and focal point distance of $17\,\mathrm{m}$ as well. The camera has a diameter of $1.47\,\mathrm{m}$ where angular acceptance is not considered.

The supported geometry is currently limited to parametrically defined surfaces or simple mesh objects, as the underlying acceleration structure (KD tree) has yet to support all features fully. Rarely used effects such as polarization or Fresnel reflection are not yet implemented but can be added in the future if required.

This implementation reduces the overhead introduced through storing and loading of photons. If GPU support was already utilized, the photons could be directly processed there, reducing an additional transfer. The resulting advantage is the reduction in file size to a few individual photons hitting the camera surface, depending on the efficiency of the telescope. The disadvantage of this implementation is that the detector geometry is directly incorporated into the simulation results and cannot be changed afterwards.

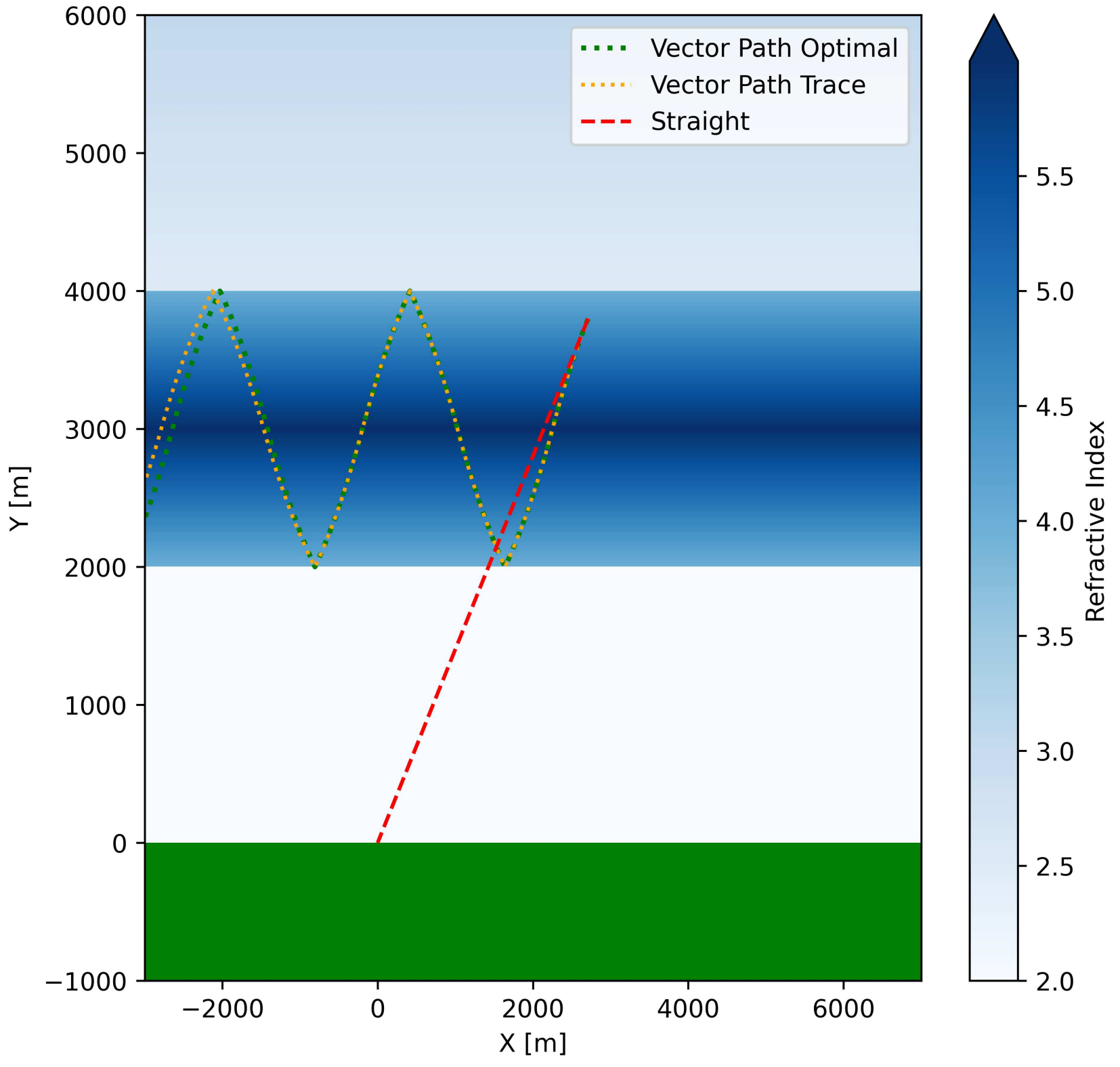

**a:** Total internal reflection in a custom layered refractive index media.

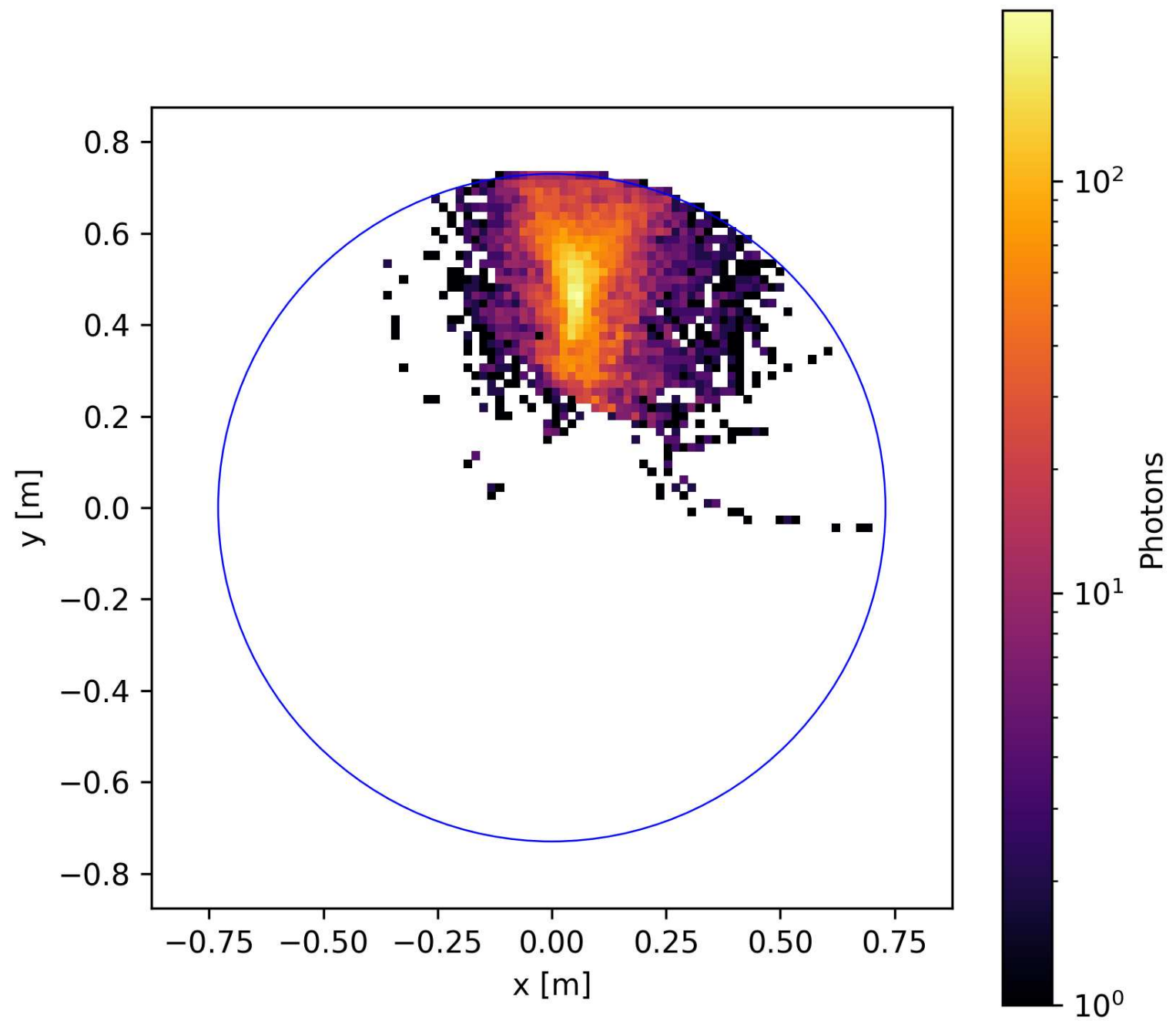

**b:** Photon on a camera surface of a MAGIC like telescope.

**Figure 6.6:** Additional simulation features in an area close to the experiment.

# 7 Conclusion and Outlook

## 7.1 Conclusion

The goal to develop a new implementation for the simulation of atmospheric Cherenkov light in CORSIKA 8 has been achieved. Two independent implementations are available: a CPU-based version directly integrated into a CORSIKA 8 branch and a GPU implementation available as a secondary package for compatibility reasons. The branch is planned to be merged into the first release version of CORSIKA 8 planned for this year (2023).

Several algorithmic optimizations in the form of data reduction and interpolation methods have been introduced and tested to allow faster and more efficient calculation of Cherenkov light emissions explicitly tailored to individual experiments. Depending on the overall configuration, up to $90\,\%$ of the specific workload for IACT simulation can be avoided without introducing noteworthy deviations from the full simulations. For fluorescence emission, “beamforming” options were explored to produce photons only in the required direction. For individual telescopes, a significant reduction in otherwise undetectable photons could be achieved. For the more commonly deployed experimental setup with multiple detectors, the resulting overhead proved too computationally intensive to be feasible.

The atmospheric photon transport matched the results of the existing CORSIKA 7 implementation while allowing more flexible changes through its clear and modular structure. Additionally, several iterative implementations were tested and integrated. These enable the photon path tracing through exotic environments or the last mile simulation of the entire optical detector model.

## 7.2 Outlook

With quality control against CORSIKA 7 done, the entire CORSIKA 8 pipeline needs to be compared over a longer timeframe with experimental data to estimate and compare if the existing mismatch between simulation and data changed. With the first steps taken for the GPU integration into CORSIKA 8, the next steps would be to extend the current interfaces to other aspects of the simulation which are equally or more computing intensive. One of the critical topics would be the integration of photon propagation for optical dense media like liquid or solid water and handling radio emission inside the shower. Booth topics could heavily benefit from the high processing performance of modern GPU platforms.

# A Appendix

## A.1 Integration for refractive index formula

### A.1.1 Linear Refractive Index Profile

$$
\begin{aligned}
\int \frac{1}{f}\,\mathrm{d}x &= \int \frac{\mathrm{d}y}{\sqrt{(c-y)^2 - f^2}} \\
&\text{Substitute } z = c - y \text{ and } \mathrm{d}y = -\,\mathrm{d}z \\
&= \int -\frac{\mathrm{d}z}{\sqrt{z^2 - f^2}} \\
&\text{Substitute } \frac{f}{\cos(\theta)} = z \text{ and } \frac{f \cdot \tan(\theta)}{\cos(\theta)}\,\mathrm{d}\theta = \mathrm{d}z \\
&= \int -\frac{1}{\sqrt{\left(\frac{f}{\cos(\theta)}\right)^2 - f^2}} \cdot \frac{f \cdot \tan(\theta)}{\cos(\theta)}\,\mathrm{d}\theta \\
&= \int -\frac{1}{f \cdot \sqrt{\left(\frac{1}{\cos(\theta)}\right)^2 - 1}} \cdot \frac{f \cdot \tan(\theta)}{\cos(\theta)}\,\mathrm{d}\theta \\
&\text{with } \sqrt{\left(\frac{1}{\cos(\theta)}\right)^2 - 1} = \tan(\theta) \\
&= \int -\frac{1}{\cos(\theta)}\,\mathrm{d}\theta = \begin{cases} \frac{1}{2}\log\left|\frac{1+\sin(\theta)}{1-\sin(\theta)}\right| + \text{const.} \\ \log\left|\frac{1}{\cos(\theta)} + \tan(\theta)\right| + \text{const.} \\ \log\left|\tan\left(\frac{\theta}{2} + \frac{\pi}{4}\right)\right| + \text{const.} \end{cases}
\end{aligned} \tag{A.1}
$$

The second case works for all resubstitutions

$$\text{Reverse substitution } \frac{1}{\cos(\theta)} = \frac{z}{f}$$

$$\text{and } \tan(\theta) = \sqrt{\frac{1}{\cos(\theta)^2} - 1} = \frac{\sqrt{z^2 - f^2}}{f}$$

$$= -\log\left(\frac{z}{f} + \frac{\sqrt{z^2 - f^2}}{f}\right) + \text{const.}$$

$$= -\log\left(\frac{z + \sqrt{z^2 - f^2}}{f}\right) + \text{const.}$$

$$\text{Reverse substitution } z = c - y$$

$$= -\log\left(\frac{(k-y) + \sqrt{(k-y)^2 - f^2}}{f}\right) + \text{const.}$$

# B Appendix Plots & Tables

## B.1 Atmospheric Interpolation Plots

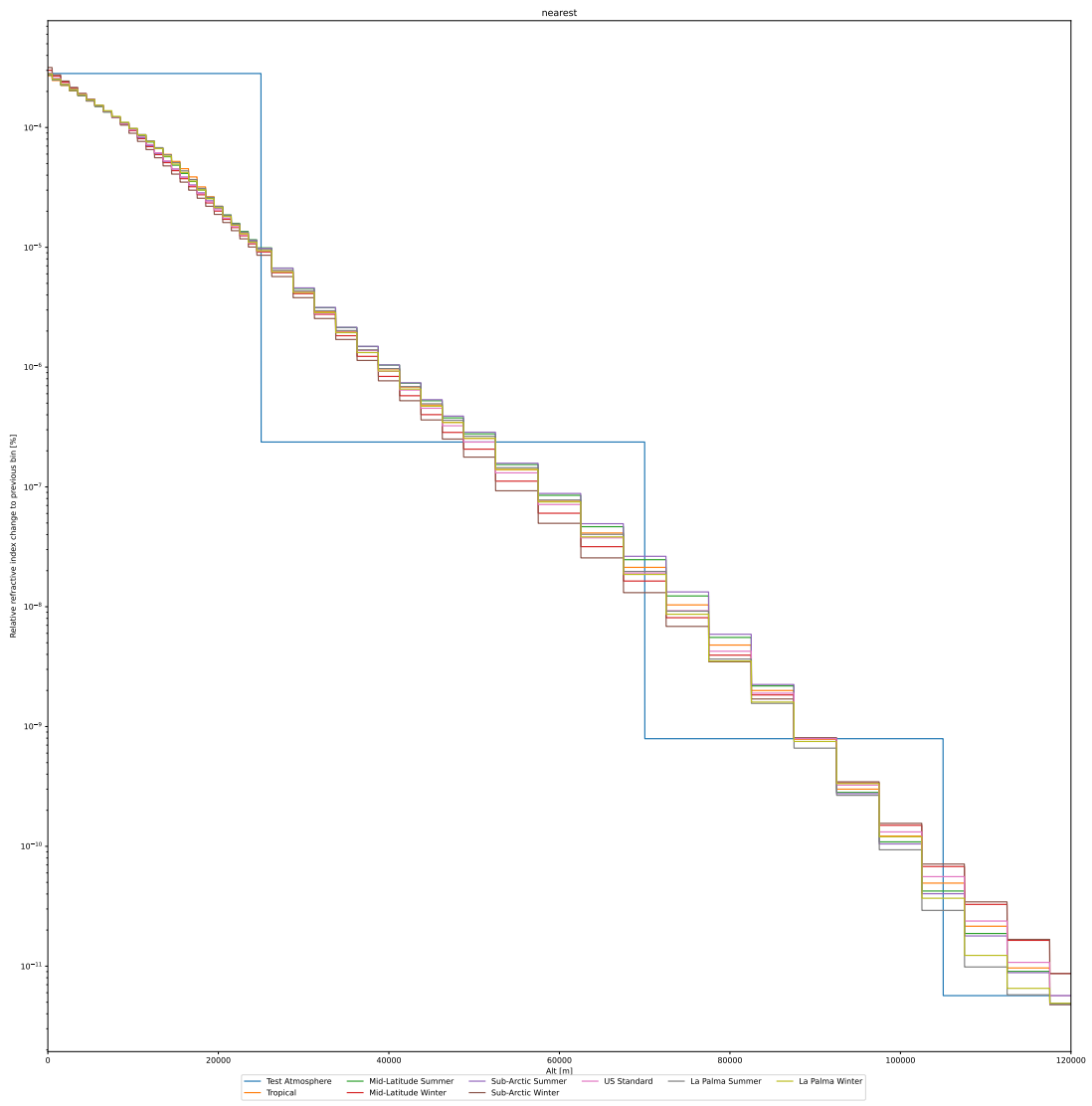

**a:** Displayed is the nearest neighbor interpolation.

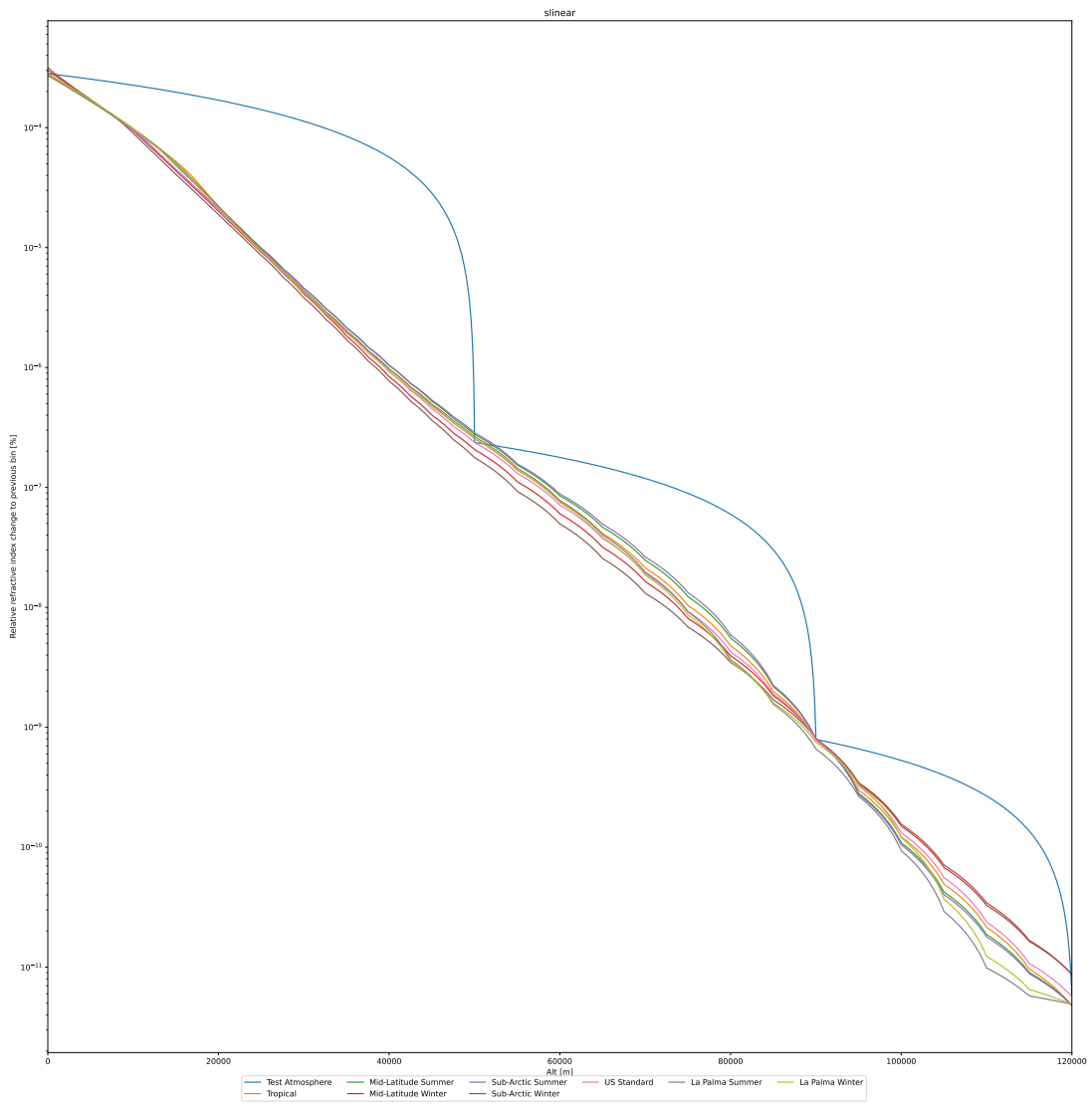

**b:** Displayed is the linear interpolation between support points.

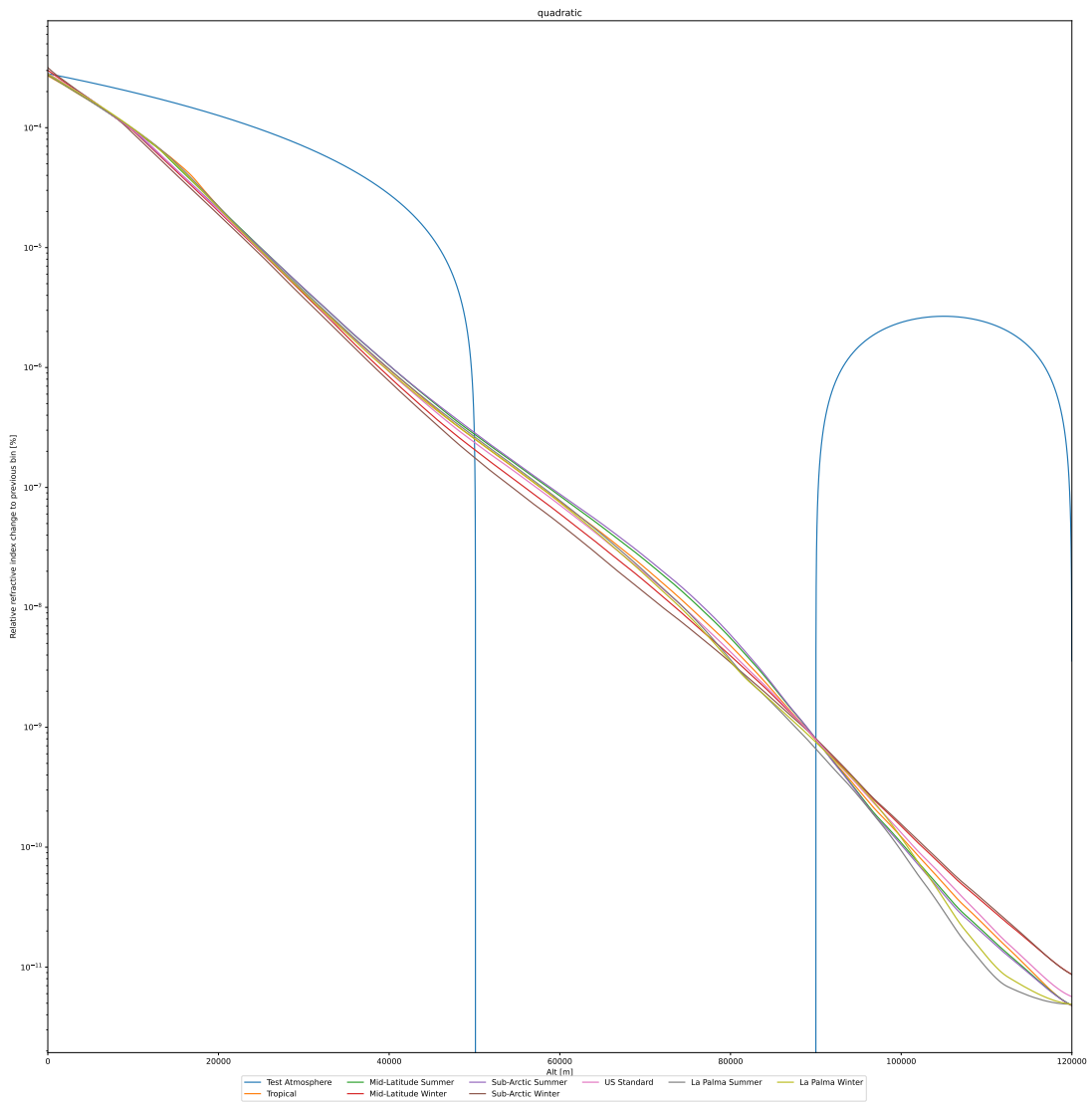

**c:** Displayed is the Quadratic Spline interpolation between support points.

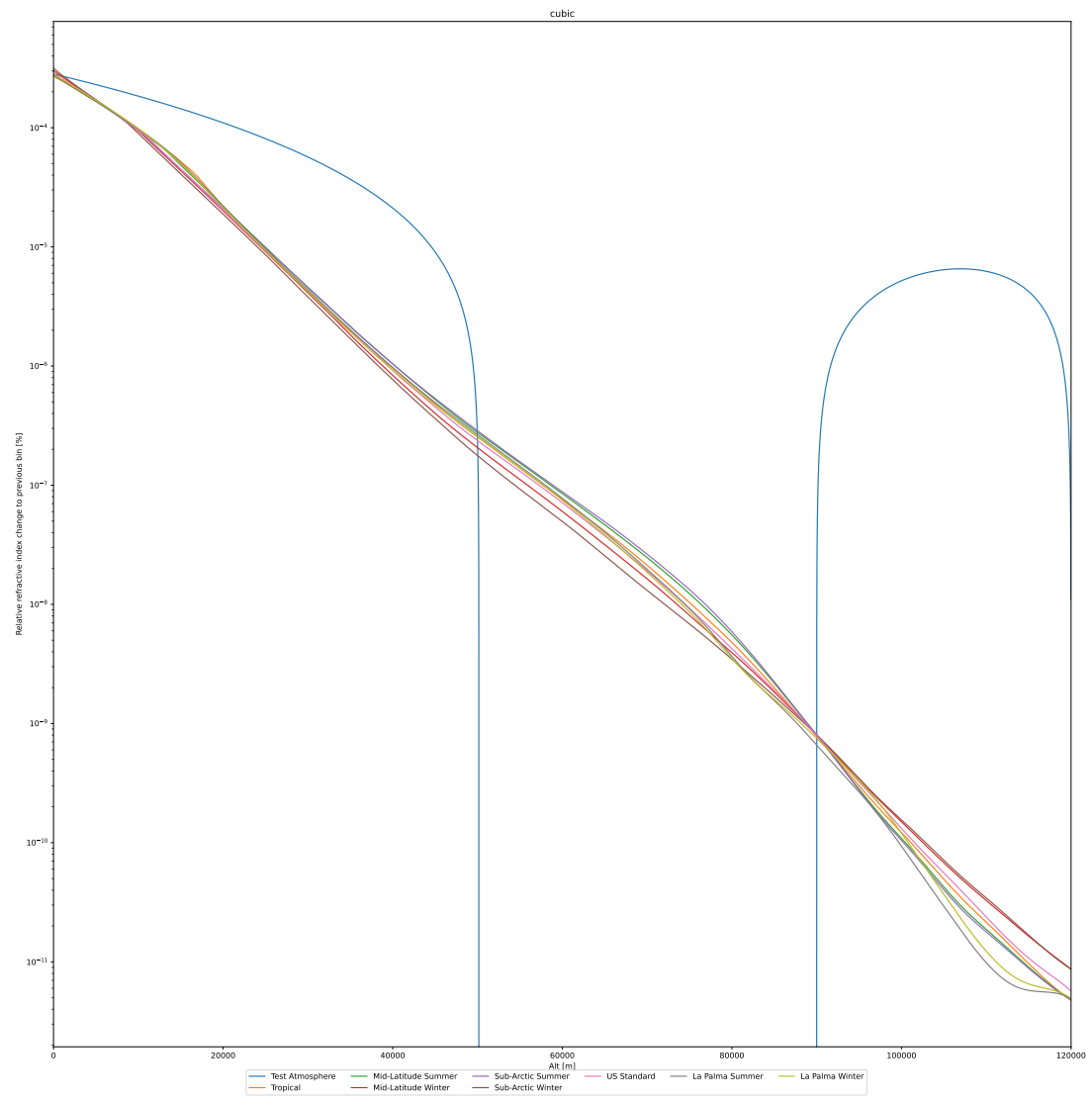

**d:** Displayed is the Cubic Spline interpolation between support points.

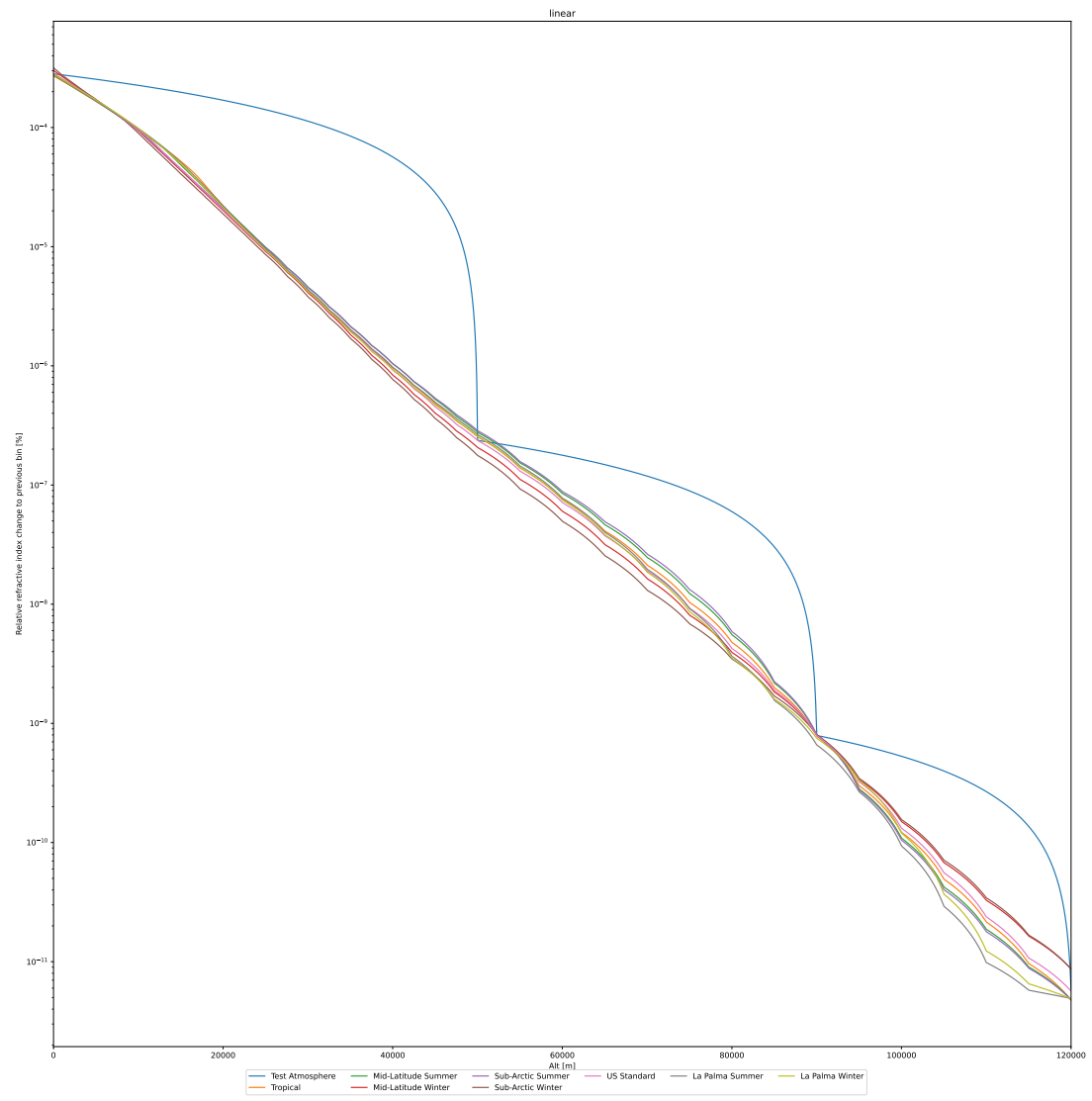

**e:** Displayed is the exponential interpolation between support points.

**Figure B.1:** Collection of different atmospheric interpolation plots with varying degrees of precision. The strong outlier in some plots is the US Standard Atmosphere with massively reduced support points, which is used as a performance test.

## B.2 Cutting Strategies

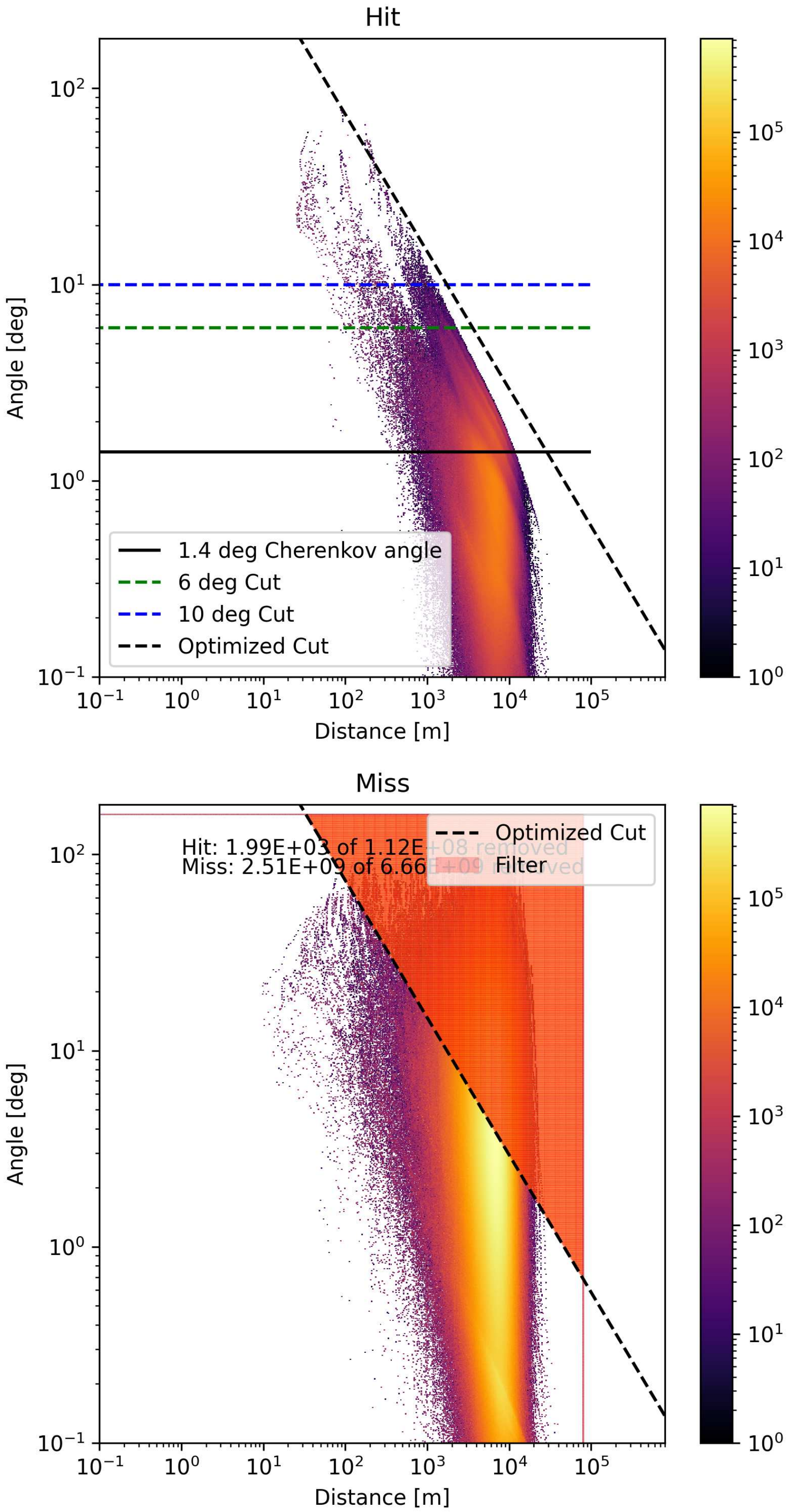

**a:** Center of array

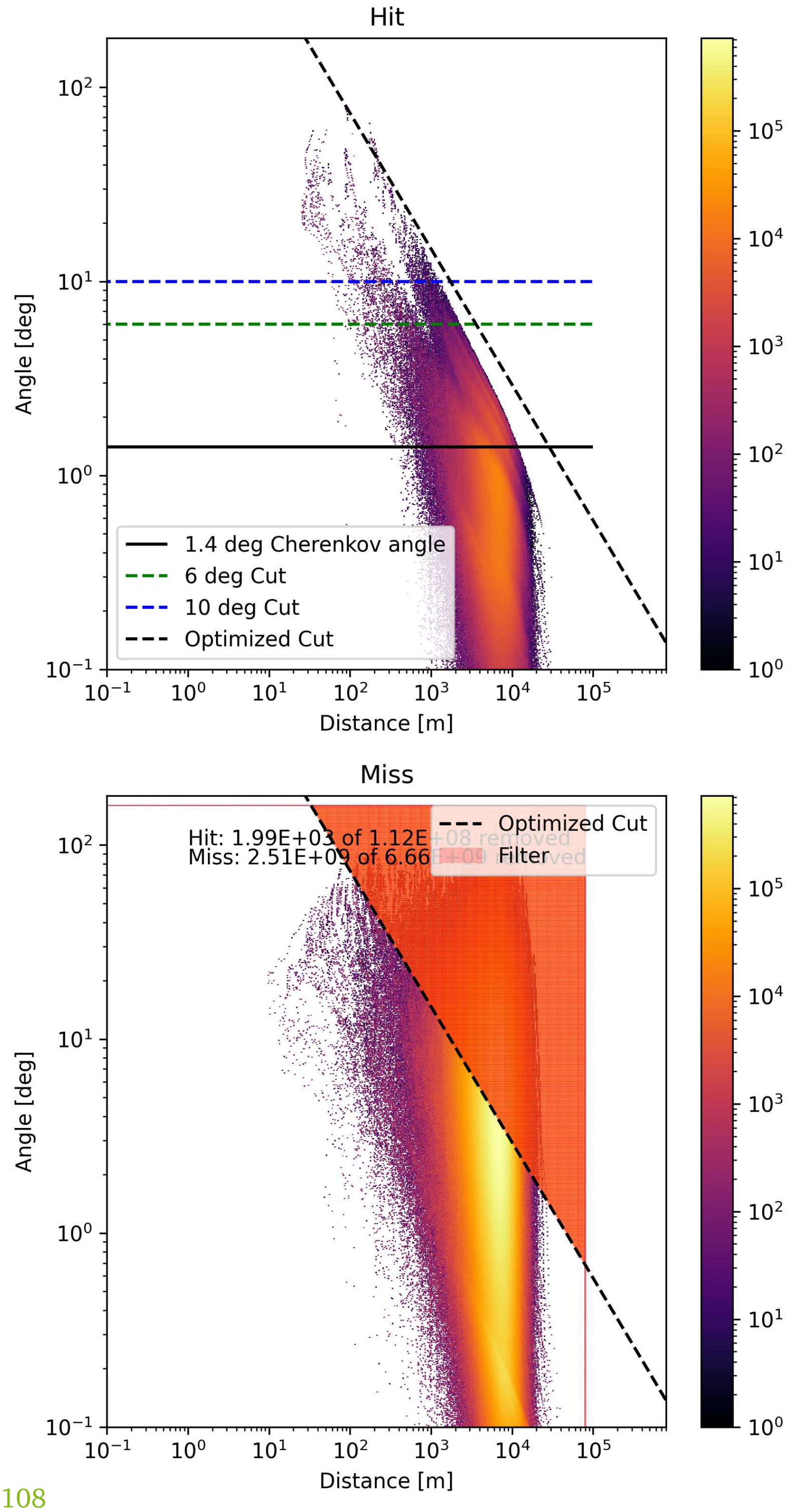

**b:** Center of telecopes

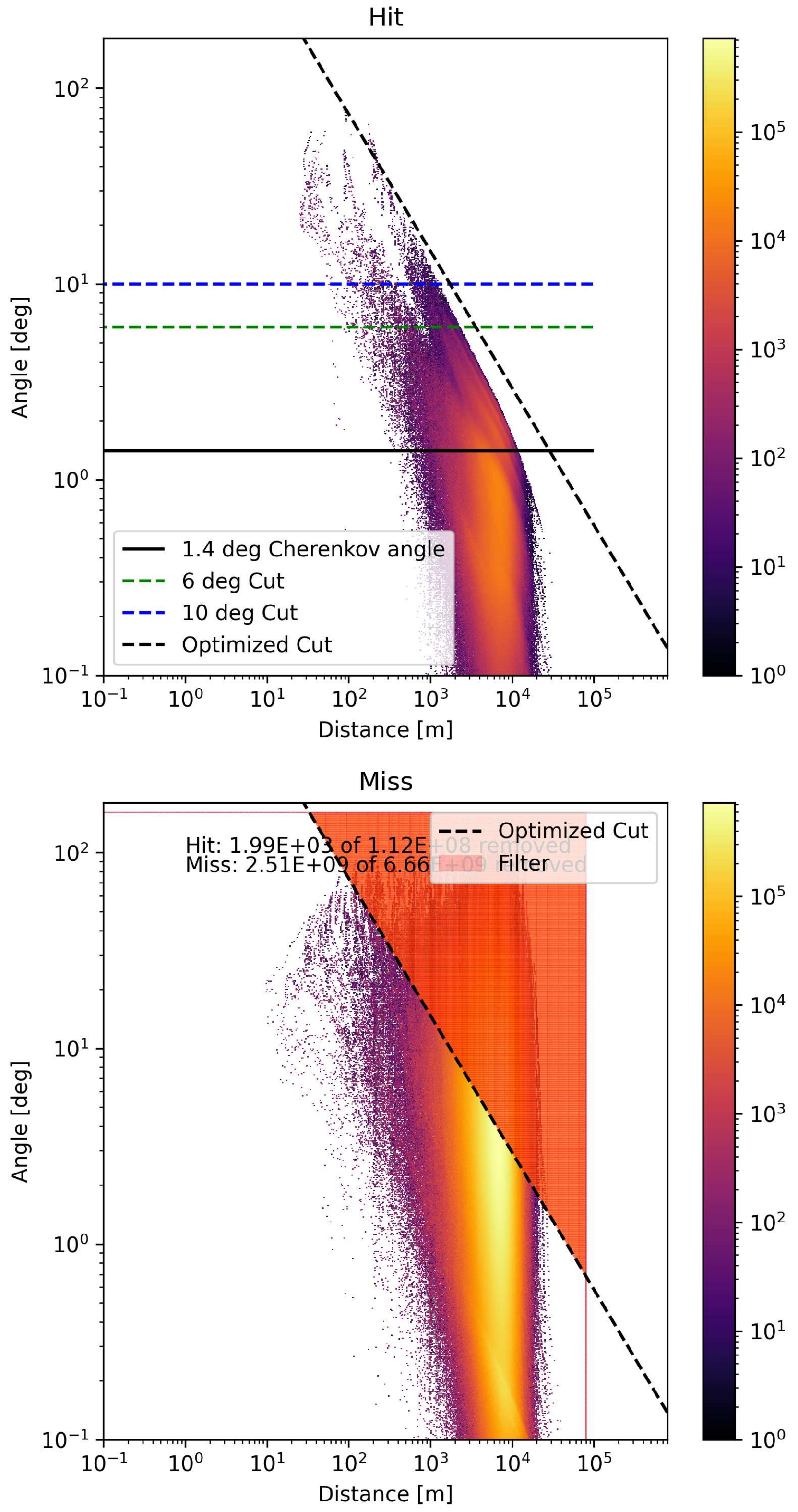

**c:** Bounding sphere

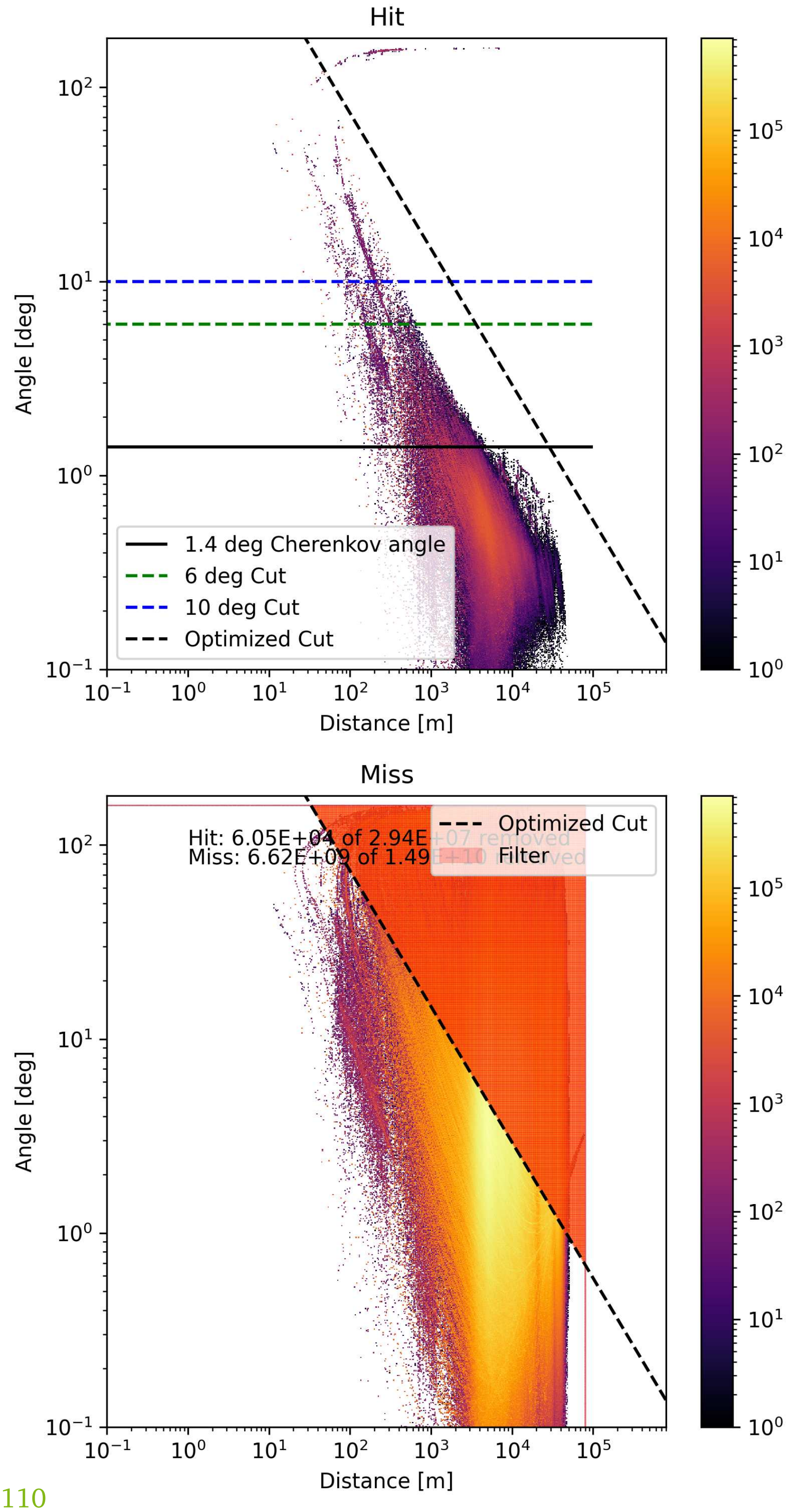

**d:** Center of array

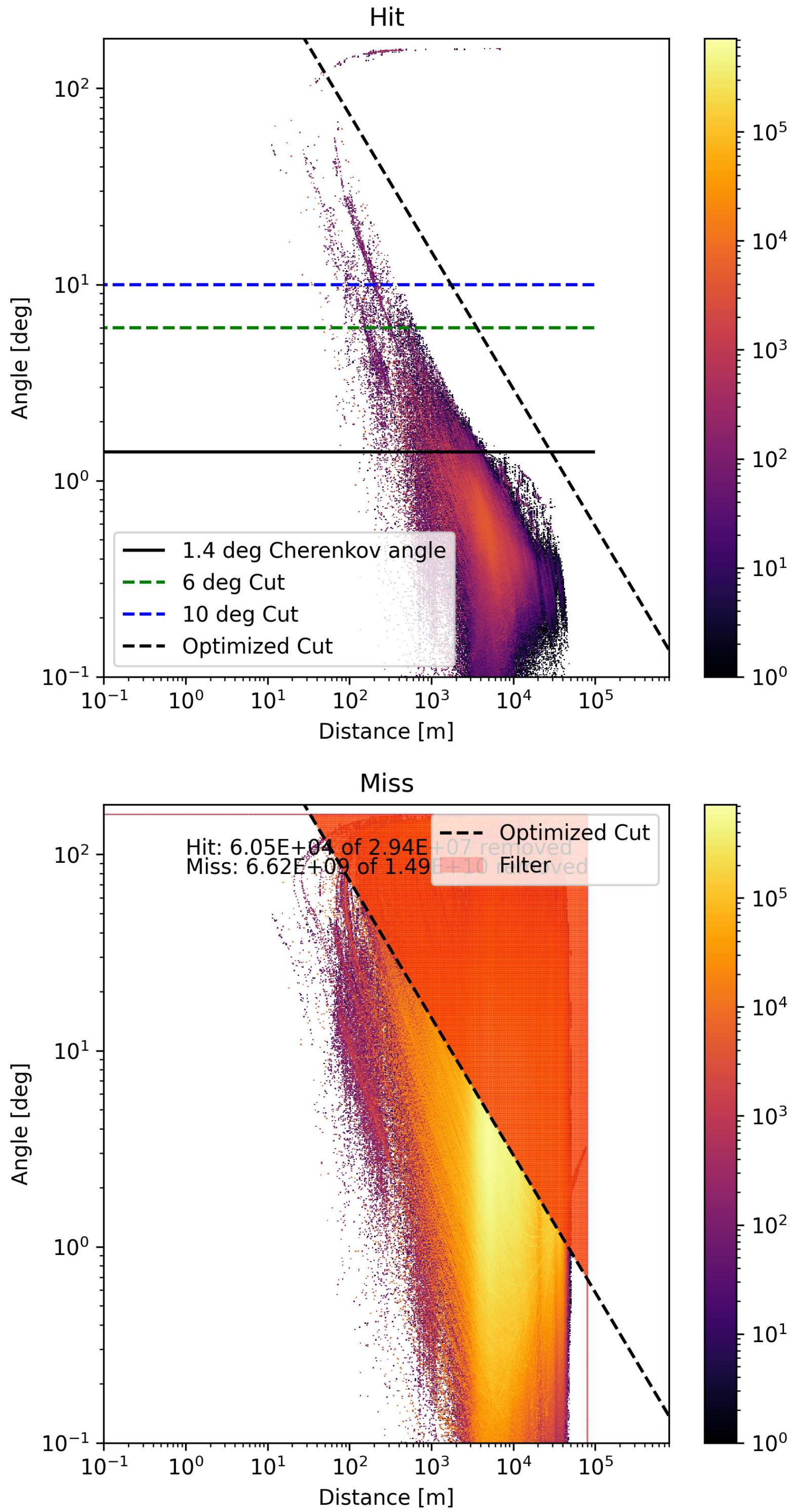

**e:** Center of telecopes

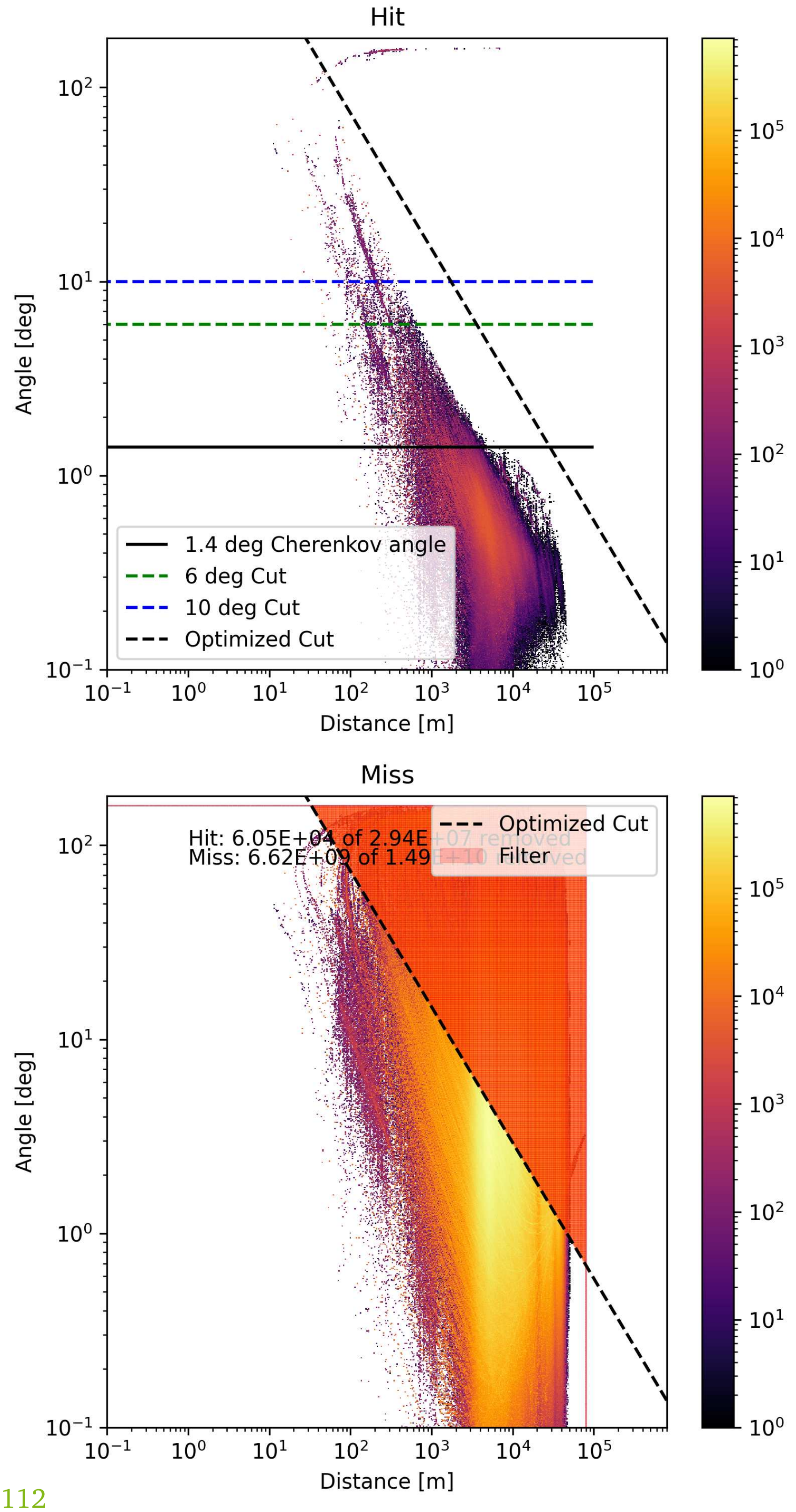

**f:** Bounding sphere

**Figure B.2:** Displayed is the minimal angle between the particle's flight direction and the vector from the track center to a specific telescope feature against the distance to the array center. Particles that contribute to the detectable light, defined as hitting the bounding sphere of the telescope, are displayed in blue, the others in orange. The plots were generated with 200 individual cascades of 500 GeV each.

## B.3 Model Experiments

### B.3.1 Telescope

### B.3.2 Telescope Array 1

| ID | x [m] | y [m] | z [m] | radius [m] |
|---|---|---|---|---|
| - 71 | -52 | 43 | 23 | |
| - 35 | 66 | 32 | 23 | |
| 75 | 50 | 28 | 23 | |
| 31 | -65 | 32 | 23 | |
| -212 | 6 | 50 | 12 | |
| -153 | 169 | 24 | 12 | |
| 27 | 199 | 12 | 12 | |
| 176 | 73 | 9 | 12 | |
| 140 | -189 | 29 | 12 | |
| - 76 | -189 | 42 | 12 | |
| -216 | -167 | 63 | 12 | |
| 0 | - 0 | 29 | 12 | |
| 20 | -300 | 53 | 12 | |

### B.3.3 Telescope Array 2

Table B.1:

| ID | (x, y, z) [m] | Radius [m] | ID | (x, y, z) [m] | Radius |
|---|---|---|---|---|---|
| 1 | $-21, -65, 34$ | 11.1 m | 2 | $80, -1, 29$ | 11.1 m |
| 3 | $-19, 65, 31$ | 11.1 m | 4 | $-120, 1, 33$ | 11.1 m |
| 5 | $-0, -0, 24$ | 5.8 m | 6 | $-1, -151, 31$ | 5.8 m |
| 7 | $-3, -325, 39$ | 5.8 m | 8 | $1, 151, 25$ | 5.8 m |
| 9 | $3, 325, 24$ | 5.8 m | 10 | $156, 237, 24$ | 5.8 m |

Continued on next page

Table B.1: (Continued)

| | | | | | |
|---|---|---|---|---|---|
| 11 | $147, 74, 21$ | 5.8 m | 12 | $146, -77, 26$ | 5.8 m |
| 13 | $152, -240, 30$ | 5.8 m | 14 | $-152, 240, 27$ | 5.8 m |
| 15 | $-146, 77, 28$ | 5.8 m | 16 | $-147, -74, 28$ | 5.8 m |
| 17 | $-157, -237, 38$ | 5.8 m | 18 | $295, 447, 25$ | 5.8 m |
| 19 | $317, 160, 18$ | 5.8 m | 20 | $308, -3, 20$ | 5.8 m |
| 21 | $314, -166, 22$ | 5.8 m | 22 | $287, -453, 38$ | 5.8 m |
| 23 | $-287, 453, 17$ | 5.8 m | 24 | $-314, 166, 30$ | 5.8 m |
| 25 | $-308, 3, 30$ | 5.8 m | 26 | $-317, -160, 35$ | 5.8 m |
| 27 | $-295, -447, 52$ | 5.8 m | 28 | $582, -6, 14$ | 5.8 m |
| 29 | $-582, 6, 36$ | 5.8 m | 30 | $207, 157, 14$ | 2.1 m |
| 31 | $204, -161, 20$ | 2.1 m | 32 | $-204, 161, 22$ | 2.1 m |
| 33 | $-207, -157, 28$ | 2.1 m | 34 | $169, 423, 18$ | 2.1 m |
| 35 | $161, -426, 33$ | 2.1 m | 36 | $-161, 426, 10$ | 2.1 m |
| 37 | $-169, -423, 42$ | 2.1 m | 38 | $5, 520, 12$ | 2.1 m |
| 39 | $-5, -520, 41$ | 2.1 m | 40 | $396, 400, 11$ | 2.1 m |
| 41 | $388, -408, 28$ | 2.1 m | 42 | $-388, 408, 13$ | 2.1 m |
| 43 | $-396, -400, 50$ | 2.1 m | 44 | $496, 105, 9$ | 2.1 m |
| 45 | $493, -115, 12$ | 2.1 m | 46 | $-494, 115, 28$ | 2.1 m |
| 47 | $-496, -105, 30$ | 2.1 m | 48 | $7, 724, 12$ | 2.1 m |
| 49 | $-7, -724, 60$ | 2.1 m | 50 | $621, 313, 7$ | 2.1 m |
| 51 | $615, -324, 19$ | 2.1 m | 52 | $-615, 325, 20$ | 2.1 m |
| 53 | $-621, -313, 49$ | 2.1 m | 54 | $442, 669, 30$ | 2.1 m |
| 55 | $429, -677, 47$ | 2.1 m | 56 | $-429, 677, 7$ | 2.1 m |
| 57 | $-442, -669, 70$ | 2.1 m | 58 | $820, -8, 4$ | 2.1 m |
| 59 | $-820, 8, 31$ | 2.1 m | 60 | $228, 795, 25$ | 2.1 m |
| 61 | $213, -799, 56$ | 2.1 m | 62 | $-213, 799, 10$ | 2.1 m |
| 63 | $-228, -795, 68$ | 2.1 m | 64 | $9, 944, 27$ | 2.1 m |
| 65 | $-9, -944, 75$ | 2.1 m | 66 | $668, 563, 13$ | 2.1 m |
| 67 | $657, -576, 43$ | 2.1 m | 68 | $-657, 576, 8$ | 2.1 m |

Continued on next page

Table B.1: (Continued)

| | | | | | |
|---|---|---|---|---|---|
| 69 | $-668, -563, 66$ | 2.1 m | 70 | $886, 219, 6$ | 2.1 m |
| 71 | $881, -236, 9$ | 2.1 m | 72 | $-881, 236, 24$ | 2.1 m |
| 73 | $-886, -219, 42$ | 2.1 m | 74 | $921, 463, 13$ | 2.1 m |
| 75 | $911, -481, 26$ | 2.1 m | 76 | $-912, 481, 11$ | 2.1 m |
| 77 | $-921, -463, 59$ | 2.1 m | 78 | $480, 967, 56$ | 2.1 m |
| 79 | $462, -976, 66$ | 2.1 m | 80 | $-462, 976, 17$ | 2.1 m |
| 81 | $-480, -967, 87$ | 2.1 m | 82 | $715, 843, 49$ | 2.1 m |
| 83 | $698, -857, 54$ | 2.1 m | 84 | $-698, 857, 15$ | 2.1 m |
| 85 | $-715, -843, 85$ | 2.1 m | 86 | $1100, -11, 4$ | 2.1 m |
| 87 | $-1100, 11, 28$ | 2.1 m | 88 | $250, 1108, 55$ | 2.1 m |
| 89 | $229, -1112, 78$ | 2.1 m | 90 | $-227, 1112, 25$ | 2.1 m |
| 91 | $-250, -1108, 89$ | 2.1 m | 92 | $964, 731, 27$ | 2.1 m |
| 93 | $950, -749, 46$ | 2.1 m | 94 | $-950, 749, 17$ | 2.1 m |
| 95 | $-964, -731, 83$ | 2.1 m | 96 | $1200, 358, 10$ | 2.1 m |
| 97 | $1193, -381, 19$ | 2.1 m | 98 | $-1193, 381, 13$ | 2.1 m |
| 99 | $-1200, -358, 47$ | 2.1 m | 100 | $-0, -0, 24$ | 5.8 m |
| 101 | $-1, -151, 31$ | 5.8 m | 102 | $-3, -325, 39$ | 5.8 m |
| 103 | $1, 151, 25$ | 5.8 m | 104 | $3, 325, 24$ | 5.8 m |
| 105 | $156, 237, 24$ | 5.8 m | 106 | $147, 74, 21$ | 5.8 m |
| 107 | $146, -77, 26$ | 5.8 m | 108 | $152, -240, 30$ | 5.8 m |
| 109 | $-152, 240, 27$ | 5.8 m | 110 | $-146, 77, 28$ | 5.8 m |
| 111 | $-147, -74, 28$ | 5.8 m | 112 | $-157, -237, 38$ | 5.8 m |
| 113 | $295, 447, 25$ | 5.8 m | 114 | $317, 160, 18$ | 5.8 m |
| 115 | $308, -3, 20$ | 5.8 m | 116 | $314, -166, 22$ | 5.8 m |
| 117 | $287, -453, 38$ | 5.8 m | 118 | $-287, 453, 17$ | 5.8 m |
| 119 | $-314, 166, 30$ | 5.8 m | 120 | $-308, 3, 30$ | 5.8 m |
| 121 | $-317, -160, 35$ | 5.8 m | 122 | $-295, -447, 52$ | 5.8 m |
| 123 | $582, -6, 14$ | 5.8 m | 124 | $-582, 6, 36$ | 5.8 m |
| 125 | $-157, -213, 38$ | 5.8 m | 126 | $-121, 77, 28$ | 5.8 m |

Continued on next page

Table B.1: (Continued)

| | | | | | |
|---|---|---|---|---|---|
| 127 | $-138, 59, 28$ | 5.8 m | 128 | $-157, -213, 38$ | 5.8 m |
| 129 | $-121, 77, 28$ | 5.8 m | 130 | $-138, 59, 28$ | 5.8 m |
| 131 | $1100, -20, 4$ | 2.1 m | 132 | $880, -8, 4$ | 2.1 m |
| 133 | $785, -43, 4$ | 2.1 m | 134 | $-228, -779, 68$ | 2.1 m |
| 135 | $910, 471, 13$ | 2.1 m | 136 | $228, 810, 25$ | 2.1 m |
| 137 | $-0, 350, 21$ | 2.1 m | 138 | $0, -350, 39$ | 2.1 m |
| 139 | $-270, 320, 22$ | 2.1 m | 140 | $-270, -320, 40$ | 2.1 m |
| 141 | $270, 320, 20$ | 2.1 m | 142 | $270, -320, 30$ | 2.1 m |
| 143 | $-280, 575, 10$ | 2.1 m | 144 | $-280, -575, 50$ | 2.1 m |
| 145 | $280, 575, 20$ | 2.1 m | 146 | $280, -575, 40$ | 2.1 m |
| 147 | $-685, 175, 25$ | 2.1 m | 148 | $-685, -175, 35$ | 2.1 m |
| 149 | $685, 175, 10$ | 2.1 m | 150 | $685, -175, 18$ | 2.1 m |
| 151 | $-1070, 250, 20$ | 2.1 m | 152 | $-1070, -250, 42$ | 2.1 m |
| 153 | $1070, 250, 5$ | 2.1 m | 154 | $1070, -250, 12$ | 2.1 m |
| 155 | $-970, -0, 30$ | 2.1 m | 156 | $970, -0, 5$ | 2.1 m |
| 157 | $120, -590, 45$ | 2.1 m | 158 | $-120, -590, 49$ | 2.1 m |
| 159 | $120, 590, 13$ | 2.1 m | 160 | $-120, 590, 11$ | 2.1 m |
| 161 | $-500, 465, 10$ | 2.1 m | 162 | $-500, -465, 52$ | 2.1 m |
| 163 | $500, 465, 15$ | 2.1 m | 164 | $500, -465, 30$ | 2.1 m |
| 165 | $-770, 360, 20$ | 2.1 m | 166 | $-770, -360, 53$ | 2.1 m |
| 167 | $770, 360, 10$ | 2.1 m | 168 | $770, -360, 17$ | 2.1 m |
| 169 | $-260, 920, 25$ | 2.1 m | 170 | $-260, -920, 75$ | 2.1 m |
| 171 | $260, 920, 45$ | 2.1 m | 172 | $260, -920, 65$ | 2.1 m |
| 173 | $-500, 815, 15$ | 2.1 m | 174 | $-500, -815, 75$ | 2.1 m |
| 175 | $500, 815, 45$ | 2.1 m | 176 | $500, -815, 53$ | 2.1 m |
| 177 | $-810, 655, 12$ | 2.1 m | 178 | $-810, -655, 68$ | 2.1 m |
| 179 | $810, 655, 20$ | 2.1 m | 180 | $810, -655, 41$ | 2.1 m |